\documentclass[10pt]{article}

\usepackage[dvips]{epsfig} 
\usepackage{amssymb,amsmath,amsthm,mathrsfs} 
\usepackage{graphicx}
\usepackage{lineno}

\usepackage[authoryear,sort&compress]{natbib}
\usepackage{url}
\usepackage{enumerate}
\usepackage{colordvi,color,pspicture}
\usepackage{psfrag}

\newcommand{\X}{\mathcal{X}}
\newcommand{\U}{\mathcal{U}}
\newcommand{\I}{\mathbf{I}}

\newcommand{\E}{\mathbf{E}}
\newcommand{\Var}{\mathbf{Var}}
\newcommand{\Cov}{\mathbf{Cov}}

\newcommand{\al}{\alpha}

\newcommand{\ah}{\widehat{\alpha}}
\newcommand{\bh}{\widehat{\beta}}
\newcommand{\pasy}{p_{\text{asy}}}
\newcommand{\pmc}{p_{\text{mc}}}
\newcommand{\prand}{p_{\text{rand}}}

\theoremstyle{plain}
\newtheorem{theorem}{Theorem}[section]

\theoremstyle{definition}

\theoremstyle{remark}

\newtheorem{remark}[theorem]{Remark}

\begin{document}


\title{Technical Report \# KU-EC-09-1:\\
Directional Clustering Tests Based on Nearest Neighbor Contingency Tables}
\author{
Elvan Ceyhan
\thanks{Address:
Department of Mathematics, Ko\c{c} University, 34450 Sar{\i}yer, Istanbul, Turkey.
e-mail: elceyhan@ku.edu.tr, tel:+90 (212) 338-1845, fax: +90 (212) 338-1559.
}
}

\date{\today}
\maketitle

\begin{abstract}
\noindent
Spatial interaction between two or more classes or species has important implications
in various fields and causes multivariate patterns such as segregation or association.
Segregation occurs when members of
a class or species are more likely to be found near members of the same class or conspecifics;
while association occurs when members of a class or species are more likely to be
found near members of another class or species.
The null patterns considered are random labeling (RL)
and complete spatial randomness (CSR) of points from two or more classes,
which is called \emph{CSR independence}, henceforth.
The clustering tests based on nearest neighbor contingency tables (NNCTs)
that are in use in literature are two-sided tests.
In this article, we consider the directional (i.e., one-sided) versions
of the cell-specific NNCT-tests and introduce
new directional NNCT-tests for the two-class case.
We analyze the distributional properties; compare the empirical significant levels and
empirical power estimates of the tests using extensive Monte Carlo simulations.
We demonstrate that the new directional tests have
comparable performance with the currently available NNCT-tests
in terms of empirical size and power.
We use four example data sets for illustrative purposes
and provide guidelines for using these NNCT-tests.
\end{abstract}





\newpage

\section{Introduction}
\label{sec:intro}
Spatial point patterns have important implications
in epidemiology, population biology, ecology, and other fields,
and have been extensively studied.
Most of the research on spatial patterns from the early
days on pertains to patterns of one type of points; i.e., to spatial pattern of a
type of points with respect to the ground (e.g., density, clumpiness, etc.).
These patterns for only one type of points usually fall under the pattern
category called {\em spatial aggregation} (\cite{coomes:1999}),
\emph{clustering}, or \emph{regularity}.
However, it is also of practical interest to
investigate the spatial interaction of one type of points
with other types (\cite{pielou:1961}).
The spatial relationships among two or more
types of points have interesting consequences, especially for plant species.
See, for example, \cite{pielou:1961}, \cite{pacala:1986}, and
\cite{dixon:1994,dixon:NNCTEco2002}.
For convenience and generality, we refer to the different types of points as ``classes",
but \emph{class} can stand for any characteristic of an individual at a
particular location.
For example, the spatial segregation pattern
has been investigated for \emph{species} (\cite{pielou:1961},
\cite{whipple:1980}, and \cite{diggle:2003}), \emph{age classes} of
plant species (\cite{hamill:1986}), \emph{fish species}
(\cite{herler:2005}), and \emph{sexes} of dioecious plants (\cite{nanami:1999}).
Many of the epidemiologic applications are for
a two-class system of case and control labels (\cite{waller:2004}).

Many univariate and multivariate (i.e., one-class and multi-class)
tests have been proposed for testing segregation of two classes
in statistical and other literature (\cite{kulldorff:2006}).
These include comparison of Ripley's $K(t)$
functions (\cite{diggle:1991}), comparison of NN distances
(\cite{diggle:2003}), and NNCTs (\cite{pielou:1961} and \cite{dixon:1994}).
\cite{pielou:1961} proposed various tests based on NNCTs for the two-class case only and
\cite{dixon:1994} introduced an overall test of segregation and
class-specific tests based on NNCTs for
the two-class case and extended his tests to multi-class case (\cite{dixon:NNCTEco2002}).
For the two-class case,
\cite{ceyhan:overall} discussed these tests and demonstrated that
Pielou's test is liberal under CSR independence or RL and
is only appropriate for a random sample of (base, NN) pairs.
If $v$ is a NN of point $u$,
then $u$ is called the \emph{base point} and $v$ is called the \emph{NN point}.
He also suggested the use of Fisher's exact test for NNCTs and evaluated its
variants and the exact version of Pearson's test in (\cite{ceyhan:exact-NNCT}).
Furthermore, \cite{ceyhan:cell2008} proposed new cell-specific and overall segregation tests
which are more robust to the differences in the relative abundance of classes
and have better performance in terms of size and power.

In literature,
most segregation tests are two-sided tests for the two-class case
or against a general alternative for the multi-class case.
In particular, the NNCT-tests in literature are not directional tests.
In this article, we discuss the directional (i.e., one-sided) versions of the
cell-specific tests of \cite{dixon:1994,dixon:NNCTEco2002} and \cite{ceyhan:cell2008}
and propose new directional segregation tests.
We compare these tests in terms of distributional properties,
and empirical size and power through extensive Monte Carlo simulations.
We also compare these tests with Ripley's $K$ or $L$-functions (\cite{ripley:2004})
and pair correlation function $g(t)$ (\cite{stoyan:1994}),
which are methods for second-order analysis of point patterns.
We only consider \emph{completely mapped data};
i.e., for our data sets, the locations of all events in a defined area are observed.
We show through simulation that
the newly proposed directional tests perform similar
(only slightly better in power) to the cell-specific tests of \cite{ceyhan:cell2008}
but perform better (in terms of empirical size and power) than Dixon's cell-specific tests.
Furthermore,
we demonstrate that our tests and Ripley's $L$-function and related methods
(i.e., second-order analysis) answer different questions about the pattern of interest.

``Dependence" in this article refers to the dependence
in the probabilistic sense between the cell counts which results from the
spatial dependence between the points from spatial point patterns.
Spatial dependence between points in a particular pattern is a well known phenomenon
and has been extensively studied.
In mathematical statistics, ``spatial dependence" is used
as a measure of the degree of
spatial interaction between independently measured
observations from a temporally or spatially ordered set of points.
The cause of spatial dependence is not the sample selection, nor sample preparation,
nor the measurement order (\cite{hald:1952}).
However, in practice, the spatial data show dependence
due to the order they occupy in space and time.
Modeling such dependence usually causes problems in the statistical analysis
because time is unidirectional,
and space is omnidirectional (see, for example, \cite{pace:2000a}).
Nearest neighbor spatial dependence is a consequence of this spatial dependence (\cite{pace:2000b}).
Just as the spatial patterns have been mostly analyzed for one-class patterns,
spatial dependence is usually investigated for one class only.
For example, indices of dependence (for clustering or regularity) are
proposed by \cite{lieshout:1996} and are extended for multi-type
point patterns in (\cite{lieshout:1999}).
In the latter article, the
authors introduce a dependence index whose values, if larger than 1,
indicate inhibition (similar to what we call segregation in this article),
and if smaller than 1, indicate positive association
(similar to what we call association in this article)
and equals to 1 for Poisson point patterns.
Hence, segregation and association can
be viewed as two opposite types of spatial dependence between two or
more classes.

For simplicity,
we discuss the spatial clustering patterns between two classes only;
the extension to the case with more classes is straightforward
for the cell-specific tests.
We discuss the null and alternative patterns in Section \ref{sec:null-and-alt},
describe the construction of the NNCTs in Section \ref{sec:nnct},
discuss Dixon's and Ceyhan's cell-specific segregation tests
in Sections \ref{sec:dixon-cell-spec} and \ref{sec:cell-spec-CSDA},
and introduce directional version of Pielou's test of segregation in Section \ref{sec:directional-k=2},
and new directional tests in Section \ref{sec:new-directional}.
We provide the empirical significance level analysis under
CSR independence in Section \ref{sec:empirical-size},
empirical power analysis under segregation and association alternatives in Section \ref{sec:emp-power},
and illustrate the tests in example data sets in Section \ref{sec:examples},
provide discussion and guidelines for using the tests in Section \ref{sec:disc-conc}.

\section{Null and Alternative Patterns}
\label{sec:null-and-alt}
The null hypothesis in the univariate spatial point pattern analysis
is usually \emph{complete spatial randomness} (\emph{CSR}) (\cite{diggle:2003}).
There are two benchmark hypotheses
to investigate the spatial interaction between multiple classes
in a multivariate process:
(i) \emph{independence}, which implies the classes of points are generated by independent
univariate processes and
(ii) \emph{random labeling} (RL), which implies that the class labels
are randomly assigned to a given set of locations in the region of interest (\cite{diggle:2003}).
In this article,
our null hypothesis is
$$H_o: \text{randomness in the NN structure}$$
which might result from two random pattern types:
CSR of points from two classes (this pattern will be called the \emph{CSR independence},
henceforth) or RL.
In the CSR independence pattern,
points from each of the two classes independently satisfy the CSR pattern
in the region of interest.

Although CSR independence and RL are not same,
they lead to the same null model for NNCT-tests,
since a NNCT does not require spatially-explicit information.
That is, when the points from two classes are assumed to be independently uniformly
distributed over the region of interest, i.e., under the CSR independence pattern,
or
when only the labeling (or marking) of a set of fixed points
(where the allocation of the points might be regular, aggregated, or clustered, or of lattice type)
is considered, i.e., under the RL pattern,
there is randomness in the NN structure.
We discuss the differences in practice and theory for either case.
The distinction between CSR independence and RL is very important
when defining the appropriate null model in practice;
i.e., the null model depends on the particular ecological context.
\cite{goreaud:2003} state that CSR independence implies that
the two classes are \emph{a priori}
the result of different processes (e.g., individuals of different species or age cohorts),
whereas RL implies that some processes affect \emph{a posteriori}
the individuals of a single population
(e.g., diseased vs. non-diseased individuals of a single species).
We provide the differences in the proposed tests
under these two patterns.
For a more detailed discussion of CSR independence and RL patterns,
see (\cite{ECarXivCellSpec:2008}).

As clustering alternatives, we consider two major types of spatial patterns:
\emph{segregation} and \emph{association}.
{\em Segregation} occurs if the
NN of an individual is more likely to be of the same
class as the individual than to be from a different class;
i.e., the members of the same class tend to be clumped or clustered
(see, e.g., \cite{pielou:1961}; \cite{dixon:1994}; and \cite{coomes:1999}).
For instance, one type of plant might not grow well
around another type of plant and vice versa.
In plant biology, one class of points might represent
the coordinates of trees from a species with large canopy,
so that other plants (whose coordinates are the other class of points)
that need light cannot grow (well or at all) around these trees.
In epidemiology, one class of points might
be the geographical coordinates of residences of cases
and the other class of points might be the coordinates of the residences of controls.
Furthermore, social and ethnic segregation of residential areas can be viewed
as a special type of segregation.
Given the locations of the residences,
the ethnic identity or social status of the residents can be viewed as class labels assigned randomly or not.
For example, the residents of similar social status or same ethnic identity might tend to gather
in certain neighborhoods which is an example of segregation
as opposed to RL of the residences.

{\em Association} occurs if the NN of an individual is more
likely to be from another class than to be of the same class as the individual.
For example, in plant biology, the two classes of points
might represent the coordinates of mutualistic plant species,
so the species depend on each other to survive.
As another example, the points from one class
might be the geometric coordinates of
parasitic plants exploiting  the other plant whose coordinates are the points of the other class.

The patterns of segregation and association do not
only result from multivariate interaction between the classes.
It is also conceivable to have either of these patterns
without any interaction between the point processes;
for example,
consider the case where species happen to have the same or different fine-scale habitat preferences.
Each of the two patterns of segregation and association are not symmetric in the sense that,
when two classes are segregated (or associated), they do not necessarily
exhibit the same degree of segregation (or association).
For example, when points from each of two classes labeled as $X$ and $Y$
are clustered at different locations,
but class $X$ is loosely clustered (i.e., its point intensity in the clusters is smaller)
compared to class $Y$
so that classes $X$ and $Y$ are segregated but class $Y$
is more segregated than class $X$.
Similarly, when class $Y$ points are clustered around class $X$ points
but not vice versa,
classes $Y$ and $X$ are associated, but class $Y$ is more associated with class $X$
compared to the other way around.
Although it is not possible to list all of the many different
types of segregation (and association),
its existence can be tested by an analysis of the
NN relationships between the classes (\cite{pielou:1961}).

\section{Nearest Neighbor Contingency Tables}
\label{sec:nnct}
NNCTs are constructed using the NN frequencies of classes.
The construction of NNCTs for two classes is described here; extension to multi-class
case is straightforward.
Consider two classes with labels $1,2$ which stand for classes $X$ and $Y$,
respectively.
Let $n_i$ be the number of points from class $i$ for $i \in \{1,2\}$ and
$n$ be the total sample size.
If the class of each point and the class of its NN were recorded,
the NN relationships fall into four distinct categories:
$(1,1),\,(1,2);\,(2,1),\,(2,2)$ where in cell $(i,j)$, class $i$ is the \emph{base class},
while class $j$ is the class of its \emph{NN}.
That is, the $n$ points constitute $n$ (base,NN) pairs.
Then each pair can be categorized with respect to
the base label (row categories) and NN label (column categories).
Denoting $N_{ij}$ as the frequency of cell $(i,j)$
(i.e., the count of all (base,NN) pairs each of which has label $(i,j)$) for $i,j \in
\{1,2\}$ yields the NNCT in Table \ref{tab:NNCT-2x2}
where the column sum $C_j$ is the number of times class $j$ points
serve as NNs for $j \in \{1,2\}$.
Furthermore, $N_{ij}$ is the cell count for
cell $(i,j)$ that is the sum of all (base,NN) pairs each of which
has label $(i,j)$.
Notice that $n=\sum_{i,j}N_{ij}$; $n_i=\sum_{j=1}^2\, N_{ij}$; and $C_j=\sum_{i=1}^2\, N_{ij}$.
By construction, if $N_{ij}$ is larger (smaller) than expected,
then class $j$ serves as NN more (less) to class $i$ than expected,
which implies (lack of) segregation if $i=j$ and (lack of) association of class $j$
with class $i$ if $i\not=j$.
Furthermore, we adopt the convention that variables denoted by upper (lower) case letters are random (fixed) quantities.

\begin{table}[ht]
\centering
\begin{tabular}{cc|cc|c}
\multicolumn{2}{c}{}& \multicolumn{2}{c}{NN class}& \\
\multicolumn{2}{c|}{}& class 1 &  class 2 & sum  \\
\hline
&class 1 &    $N_{11}$  &   $N_{12}$  &   $n_1$  \\
\raisebox{1.5ex}[0pt]{base class}
&class 2 &    $N_{21}$ &  $N_{22}$    &   $n_2$  \\
\hline
& sum    &    $C_1$   & $C_2$         &   $n$  \\
\end{tabular}
\caption{
\label{tab:NNCT-2x2}
NNCT for two classes.}
\end{table}

Observe that column sums and cell counts are random, while row sums and the overall sum are fixed
quantities in a NNCT.
Under segregation, the diagonal entries,
$N_{ii}$ for $i=1,2$, tend to be larger than expected; under
association, the off-diagonals tend to be larger than expected.
The general alternative is that some cell counts are different
than expected under CSR independence or RL.

(\cite{pielou:1961}) suggested the use of the Pearson's $\chi^2$ test of
independence for NNCTs,
but her test has been shown to be inappropriate in literature (see, e.g., \cite{meagher:1980}).
The main problem with her test is that sampling distribution of the
cell counts in the NNCTs is not correct.
The assumption for the use of $\chi^2$ test for NNCTs is the independence between
cell-counts (and rows and columns also), which is violated for CSR independence or RL
data (see \cite{dixon:1994} and \cite{ceyhan:overall}).
\cite{dixon:1994} derived the appropriate (asymptotic) sampling
distribution of cell counts under RL, hence his test is
appropriate for CSR independence (\cite{ceyhan:overall}).
Nevertheless, \cite{ceyhan:overall} demonstrated that
all these tests are consistent, in the sense that under any
alternative (of segregation or association), the power tends to one,
as sample sizes tend to infinity.
While Dixon's test has the
appropriate nominal size under CSR independence,
Pielou's test is liberal (\cite{ceyhan:overall}).
\cite{ceyhan:exact-NNCT} also suggested
the use of Fisher's exact test for NNCTs
and evaluated its variants and the exact version of Pearson's test.

\section{Directional Segregation Tests Based on NNCTs}
\label{sec:directional}
In this section, we describe the sampling distribution of cell counts for NNCTs,
discuss directional versions of the cell-specific test for cell $(1,1)$,
directional version of Pielou's test of segregation,
and introduce new directional tests of segregation.

\subsection{Dixon's Cell-Specific Test of Segregation}
\label{sec:dixon-cell-spec}
\cite{dixon:1994} proposed a series of tests for segregation based on NNCTs.
In Dixon's framework, the probability of a class $j$ point
serving as a NN of a class $i$ point depends only on
the class sizes (row sums), but not the total number of times class
$j$ serves as a NN (column sums).
The level of segregation is tested by
comparing the observed cell counts to the expected cell counts
under RL of points that are fixed or a realization of points from CSR independence.
Dixon demonstrated that under RL,
one can write down the cell frequencies as Moran join count statistics (\cite{moran:1948}).
He then derived the means, variances, and covariances of
the cell counts (frequencies) (\cite{dixon:1994, dixon:NNCTEco2002}).

The null hypothesis under CSR independence or RL is given by
\begin{equation}
\label{eqn:Exp[Nij]}
H_o:\,\E[N_{ij}]=
\begin{cases}
n_i(n_i-1)/(n-1) & \text{if $i=j$,}\\
n_i\,n_j/(n-1)     & \text{if $i \not= j$,}
\end{cases}
\end{equation}
where $n_i$ is the sample size for  class $i$.
Observe that the expected cell counts depend only on the size of each class
(i.e., row sums), but not on column sums.
Furthermore,
\begin{equation}
\label{eqn:VarNij}
\Var[N_{ij}]=
\begin{cases}
(n+R)\,p_{ii}+(2\,n-2\,R+Q)\,p_{iii}+(n^2-3\,n-Q+R)\,p_{iiii}-(n\,p_{ii})^2 & \text{if $i=j$,}\\
n\,p_{ij}+Q\,p_{iij}+(n^2-3\,n-Q+R)\,p_{iijj} -(n\,p_{ij})^2               & \text{if $i \not= j$,}
\end{cases}
\end{equation}
with $p_{xx}$, $p_{xxx}$, and $p_{xxxx}$ are the probabilities that
a randomly picked pair, triplet, or quartet of points, respectively,
are the indicated classes and are given by
\begin{align}
\label{eqn:probs}
p_{ii}&=\frac{n_i\,(n_i-1)}{n\,(n-1)},  & p_{ij}&=\frac{n_i\,n_j}{n\,(n-1)},\nonumber\\
p_{iii}&=\frac{n_i\,(n_i-1)\,(n_i-2)}{n\,(n-1)\,(n-2)}, &
p_{iij}&=\frac{n_i\,(n_i-1)\,n_j}{n\,(n-1)\,(n-2)},\\
 p_{iijj}&=\frac{n_i\,(n_i-1)\,n_j\,(n_j-1)}{n\,(n-1)\,(n-2)\,(n-3)},&
p_{iiii}&=\frac{n_i\,(n_i-1)\,(n_i-2)\,(n_i-3)}{n\,(n-1)\,(n-2)\,(n-3)}.\nonumber
\end{align}
Furthermore, $Q$ is the number of points with shared NNs,
which occur when two or more points share a NN and
$R$ is twice the number of reflexive pairs.
A (base,NN) pair $(u,v)$ is \emph{reflexive} if $(v,u)$ is also a (base,NN) pair.
Then $Q=2\,(Q_2+3\,Q_3+6\,Q_4+10\,Q_5+15\,Q_6)$ where $Q_k$
is the number of points that serve as a NN to other
points $k$ times.

The test statistic suggested by Dixon is given by
\begin{equation}
\label{eqn:dixon-Zij}
Z^D_{ij}=\frac{N_{ij}-\E[N_{ij}]}{\sqrt{\Var[N_{ij}]}}
\end{equation}
where $\E[N_{ij}]$ is given in Equation \eqref{eqn:Exp[Nij]}
and $\Var[N_{ij}]$ is given in Equation \eqref{eqn:VarNij}.
One-sided and two-sided tests are possible for each cell $(i,j)$
using the asymptotic normality of $Z^D_{ij}$ given in Equation
\eqref{eqn:dixon-Zij} (\cite{dixon:1994}).
In fact, asymptotic
normality of the cell counts are rigorously proved by
\cite{dixon:1994} using the technique proposed by \cite{cuzick:1990}
for the diagonal entries.
For the two-class case, the asymptotic
normality of the off-diagonal entries follow trivially; for the
multi-class case, normality does only generalize to the diagonal
entries, but Monte Carlo simulations suggest the normality of
off-diagonal entries also (\cite{dixon:NNCTEco2002}).

\begin{remark}
\label{rem:asymp-structure}
There are two major types of asymptotic structures for spatial data (\cite{lahiri:1996}).
In the first, any two observations are required to be at least a fixed distance apart,
hence as the number of observations increase, the region on which the process
is observed eventually becomes unbounded.
This type of sampling structure is called ``increasing domain asymptotics".
In the second type, the region of interest is a fixed
bounded region and more or more points are observed in this region.
Hence the minimum distance between data points tends to zero
as the sample size tends to infinity.
This type of structure is called ``infill asymptotics", due to \cite{cressie:1993}.
The sampling structure in our asymptotic sampling distribution
could be either one of infill or increasing domain asymptotics,
as we only consider the class sizes and the total sample size
tending to infinity regardless of the size of the study region. $\square$
\end{remark}

\subsection{Cell-Specific Tests of Segregation in \cite{ceyhan:cell2008}}
\label{sec:cell-spec-CSDA}
In standard cases like multinomial sampling with fixed
row totals and conditioning on the column totals,
the expected cell count for cell $(i,j)$ in contingency tables
is $\E[N_{ij}]=N_i\,C_j/n$.
We first consider the difference $\Delta_{ij}:=N_{ij}-N_i\,C_j/n$ for cell $(i,j)$.
Notice that under RL, $N_i=n_i$ are fixed,
but $C_j$ are random quantities and $C_j=\sum_{i=1}^q N_{ij}$,
hence $\Delta_{ij}=N_{ij}-\frac{n_i\,C_j}{n}.$ Then under RL,
$\displaystyle \E[\Delta_{ij}]= \frac{n_i(n_i-1)}{(n-1)}-\frac{n_i^2}{n} \, \I(i=j)+
\frac{n_i\,n_j}{(n-1)}-\frac{n_i\,n_j}{n}  \, \I(i \not= j)$ (\cite{ECarXivCellSpec:2008}).
\begin{equation}
\E[\Delta_{ij}]=
\begin{cases}
\frac{n_i(n_i-1)}{(n-1)}-\frac{n_i}{n}\,\E[C_j] & \text{if $i=j$,}\\
\frac{n_i\,n_j}{(n-1)}-\frac{n_i}{n}\,\E[C_j]  & \text{if $i \not= j$.}
\end{cases}
\end{equation}

\noindent
For all $j$, $\E[C_j]=n_j$, since
\begin{multline*}
\E[C_j]=\sum_{i=1}^q \E[N_{ij}]=
\frac{n_j(n_j-1)}{(n-1)}+\sum_{i\neq j} \frac{n_i n_j}{(n-1)}
=\frac{n_j(n_j-1)}{(n-1)}+\frac{n_j}{(n-1)}\sum_{i\neq j} n_i\\
=\frac{n_j(n_j-1)}{(n-1)}+\frac{n_j}{(n-1)}(n-n_j)=n_j.
\end{multline*}

\noindent
Therefore,
\begin{equation}
\label{eqn:Exp[Deltaij]}
\E[\Delta_{ij}]=
\begin{cases}
\frac{n_i(n_i-1)}{(n-1)}-\frac{n_i^2}{n} & \text{if $i=j$,}\\
\frac{n_i\,n_j}{(n-1)}-\frac{n_i\,n_j}{n}  & \text{if $i \not= j$.}
\end{cases}
\end{equation}

\noindent
For all $i=j=1$,
\begin{eqnarray*}
\E[\Delta_{11}]&=&\frac{n_1(n_1-1)}{(n-1)}-\frac{n_1}{n}\left( \E[N_{11}]+\E[N_{21}] \right)\\
&=&\frac{n_1(n_1-1)}{(n-1)}-\frac{n_1}{n}
\left(\frac{n_1(n_1-1)}{(n-1)}+\frac{n_1\,n_2}{(n-1)}\right)\\
&=&\frac{n_1(n_1-1)}{(n-1)}-\frac{n_1^2}{n}.
\end{eqnarray*}
For $i=1$ and $j=2$,
\begin{eqnarray*}
\E[\Delta_{12}]&=&\frac{n_1\,n_2}{(n-1)}-\frac{n_1}{n}\left(\E[N_{12}]+\E[N_{22}]\right)\\
&=&\frac{n_1\,n_2}{(n-1)}-\frac{n_1}{n}
\left(\frac{n_1\,n_2}{(n-1)}+\frac{n_2\,(n_2-1)}{(n-1)}\right)\\
&=&\frac{n_1\,n_2}{(n-1)}-\frac{n_1\,n_2}{n}.
\end{eqnarray*}
Similarly,
$$\E[\Delta_{21}]=\frac{n_1\,n_2}{(n-1)}-\frac{n_1\,n_2}{n}~~~ \text{  and  }~~~
\E[\Delta_{22}]=\frac{n_2\,(n_2-1)}{(n-1)}-\frac{n_2^2}{n}.$$
\noindent
Notice that the expected value of $\Delta_{ij}$ is not zero under RL.
Hence, instead of $\Delta_{ij}$, 
we suggest the following test statistic:
\begin{equation}
\label{eqn:Tij}
T_{ij}=
\begin{cases}
N_{ij}-\frac{(n_i-1)}{(n-1)}C_j & \text{if $i=j$,}\\
N_{ij}-\frac{n_i}{(n-1)}C_j     & \text{if $i \not= j$.}
\end{cases}
\end{equation}
Then $\E[T_{ij}]=0$, since for $i=j$,
$$
\E[T_{ii}]=\E[N_{ii}]-\frac{(n_i-1)}{(n-1)}\E[C_i]=
\frac{n_i(n_i-1)}{(n-1)}-\frac{(n_i-1)}{(n-1)}n_i=0,
$$
and for $i \neq j$,
$$
\E[T_{ij}]=\E[N_{ij}]-\frac{(n_i-1)}{(n-1)}\E[C_j]=
\frac{n_i\,n_j}{(n-1)}-\frac{(n_i-1)}{(n-1)}n_j=0.
$$
Furthermore,
for $i=1$ and $j=2$,
\begin{eqnarray*}
\E[T_{12}]&=&\frac{n_1\,n_2)}{(n-1)}-\frac{n_1}{(n-1)}\left(\E[N_{12}]+\E[N_{22}]\right)\\
&=&\frac{n_1\,n_2}{(n-1)}-\frac{n_1}{(n-1)}
\left(\frac{n_1\,n_2}{(n-1)}+\frac{n_2\,(n_2-1)}{(n-1)}\right)\\
&=&\frac{n_1\,n_2}{(n-1)}-\frac{n_1\,n_2}{(n-1)}=0.
\end{eqnarray*}
Likewise $\E[T_{21}]=0$ and $\E[T_{22}]=0$.
For the variance of $T_{ij}$, we have
\begin{equation}
\label{eqn:Var[Tij]}
\Var[T_{ij}]=
\begin{cases}
\Var[N_{ij}]+\frac{(n_i-1)^2}{(n-1)^2}\Var[C_j]-2\frac{(n_i-1)}{(n-1)}\Cov[N_{ij},C_j]
& \text{if $i=j$,} \vspace{.05 in}\\
\Var[N_{ij}]+\frac{n_i^2}{(n-1)^2}\Var[C_j]-2\frac{n_i}{(n-1)}\Cov[N_{ij},C_j] & \text{if $i \not= j$,}
\end{cases}
\end{equation}
where
$\Var[N_{ij}]$ are as in Equation \eqref{eqn:VarNij},
$\Var[C_j]=\sum_{i=1}^q \Var[N_{ij}]+\sum_{k \neq i}\sum_i \Cov[N_{ij},N_{kj}]$
and
$\Cov[N_{ij},C_j]=\sum_{k=1}^q \Cov[N_{ij},N_{kj}]$
with $\Cov[N_{ij},N_{kl}]$ are as in Equations (4)-(12) of \cite{dixon:NNCTEco2002}.
The proposed cell-specific test in standardized form is
\begin{equation}
\label{eqn:new-Zij}
Z^C_{ij}=\frac{T_{ij}}{\sqrt{\Var[T_{ij}]}}.
\end{equation}

Recall that in the two-class case,
each cell count $N_{ij}$ has asymptotic normal distribution (\cite{cuzick:1990}).
Hence, $Z^C_{ij}$ also converges in law to $N(0,1)$ as $n\rightarrow \infty$.
Moreover, one and two-sided versions of this test are also possible.

Under CSR independence, the distribution of the test statistics above
is similar to the RL case.
The only difference is that $Z^C_{ij}$ asymptotically has
$N(0,1)$ distribution conditional on $Q$ and $R$.

Dixon's cell-specific test in Equation \eqref{eqn:dixon-Zij} depends
on the frequencies of (base, NN) pairs (i.e., cell counts),
and measures deviations from expected cell counts.
On the other hand, Ceyhan's cell-specific test in Equation \eqref{eqn:new-Zij}
is the difference of cell counts and column and row sums;
in fact, it can be seen as a difference of two statistics and has zero expected
value for each cell.

In the two-class case, segregation of class $i$ from class $j$ implies lack of association
between classes $i$ and $j$ ($i \not= j$)
association between classes $i$ and $j$ implies lack of segregation between them ($i \not= j$),
since $Z^D_{i1}=-Z^D_{i2}$ for $i=1,2$.
The same holds for the new cell-specific tests,
since $Z^C_{1j}=-Z^C_{2j}$ for $j=1,2$.

\subsection{Directional Version of Pielou's Test of Segregation}
\label{sec:directional-k=2}
\cite{pielou:1961} constructed NNCTs based on NN frequencies which
yield tests that are independent of quadrat size
(see also \cite{krebs:1972}) for two classes.
She used Pearson's $\chi^2$ test of independence to detect any
deviation from randomness in NN structure.
The corresponding test statistic is given by
\begin{equation}
\label{eqn:piel-seg-2x2}
\X_P^2=\sum_{i=1}^2\sum_{j=1}^2\frac{(N_{ij}-\E[N_{ij}])^2}{\E[N_{ij}]}
\end{equation}
where $\E[N_{ij}]=(n_i\,c_j)/n$ with $c_j$ being the observed sum
for column $j$.
Under CSR independence or RL,
this test is liberal, i.e., has larger Type I error rate
than the desired level (\cite{ceyhan:overall}).

Pielou's test, when used for a NNCT based on a random sample of (base, NN) pairs,
measures deviations from the independence of cell counts,
but does not indicate the direction of the deviation (e.g., segregation or association).
To determine the direction, one needs to check the NNCT.
Since $\X_P^2 \stackrel{\text{approx}}{\sim}
\chi^2_1$, for large $n$, we can write $\X_P^2=Z_P^2$ where
$Z_P \stackrel{\text{approx}}{\sim} N(0,1)$,
where $N(0,1)$ stands for the standard normal distribution.
By some algebraic manipulations, among other possibilities,
 $Z_P$ can be written as
\begin{equation}
\label{eqn:2x2-Zform}
Z_P=\left(\frac{N_{11}}{n_1}-\frac{N_{21}}{n_2} \right)\sqrt{\frac{n_1\,n_2\,n}{C_1\,C_2}}.
\end{equation}
See (\cite{bickel:1977}) for the sketch of the derivation.
For example, $Z_P$ could also be written as
$\displaystyle Z_P=\left(N_{11}-\frac{n_1\,C_1}{n}\right)
\left[\frac{n_1\,n_2\,C_1\,C_2}{n^3}\right]^{-1/2}$.

We point out that these directional tests are not
appropriate for testing CSR independence or RL, due to inherent
dependence of cell counts in NNCTs for such patterns,
but only appropriate for NNCTs based on a random sample
of (base,NN) pairs.

\subsubsection{Empirical Correction of Directional Version of Pielou's Test}
\label{sec:MC-correction}
We demonstrate in Section \ref{sec:empirical-size} that the directional and two-sided tests
based on $Z_P$ (i.e., the directional version of Pielou's test) are both liberal
in rejecting the null hypothesis.
We adjust this test for location and scale based on Monte Carlo simulations
to render it have the desired level, $\al$.
For the null case, we simulate the CSR independence pattern only, with
classes 1 and 2 of sizes $n_1$ and $n_2$, respectively.
At each of $N_{mc}=10000$ replicates, under $H_o$, we generate data for the
pairs of $(n_1,n_2) \in \{(10,10),(10,30),(10,50),(30,30),(30,50),(50,50),
(100,100),(200,200)\}$
points iid from $\U((0,1)\times (0,1))$, the uniform distribution on
the unit square.
These sample size combinations are chosen so that we can see the
influence of small and large samples, and
of the differences in the relative abundances on the tests.

\begin{figure}[ht]
 \rotatebox{-90}{ \resizebox{3.25 in}{!}{\includegraphics{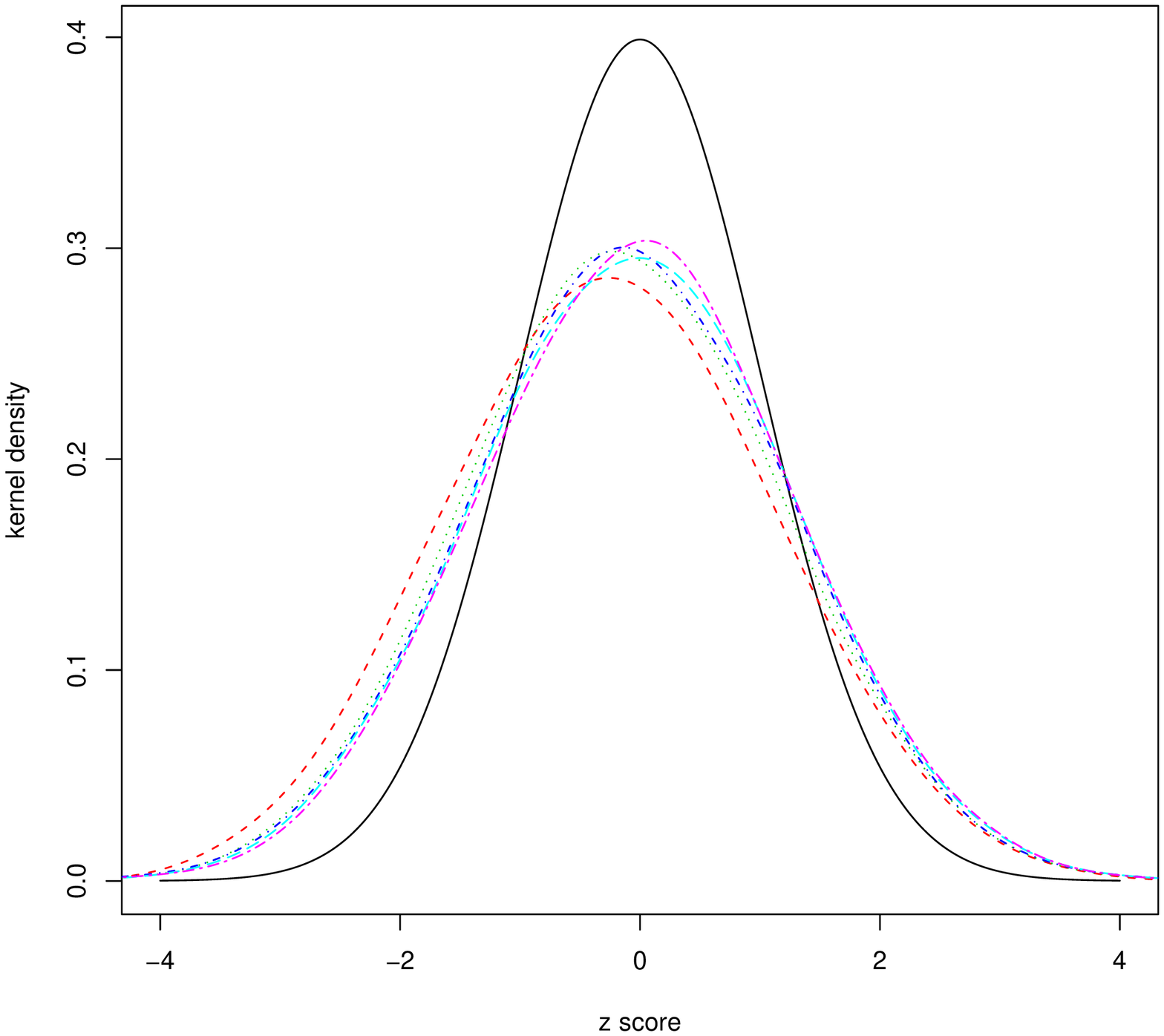} }}
 \rotatebox{-90}{ \resizebox{3.25 in}{!}{\includegraphics{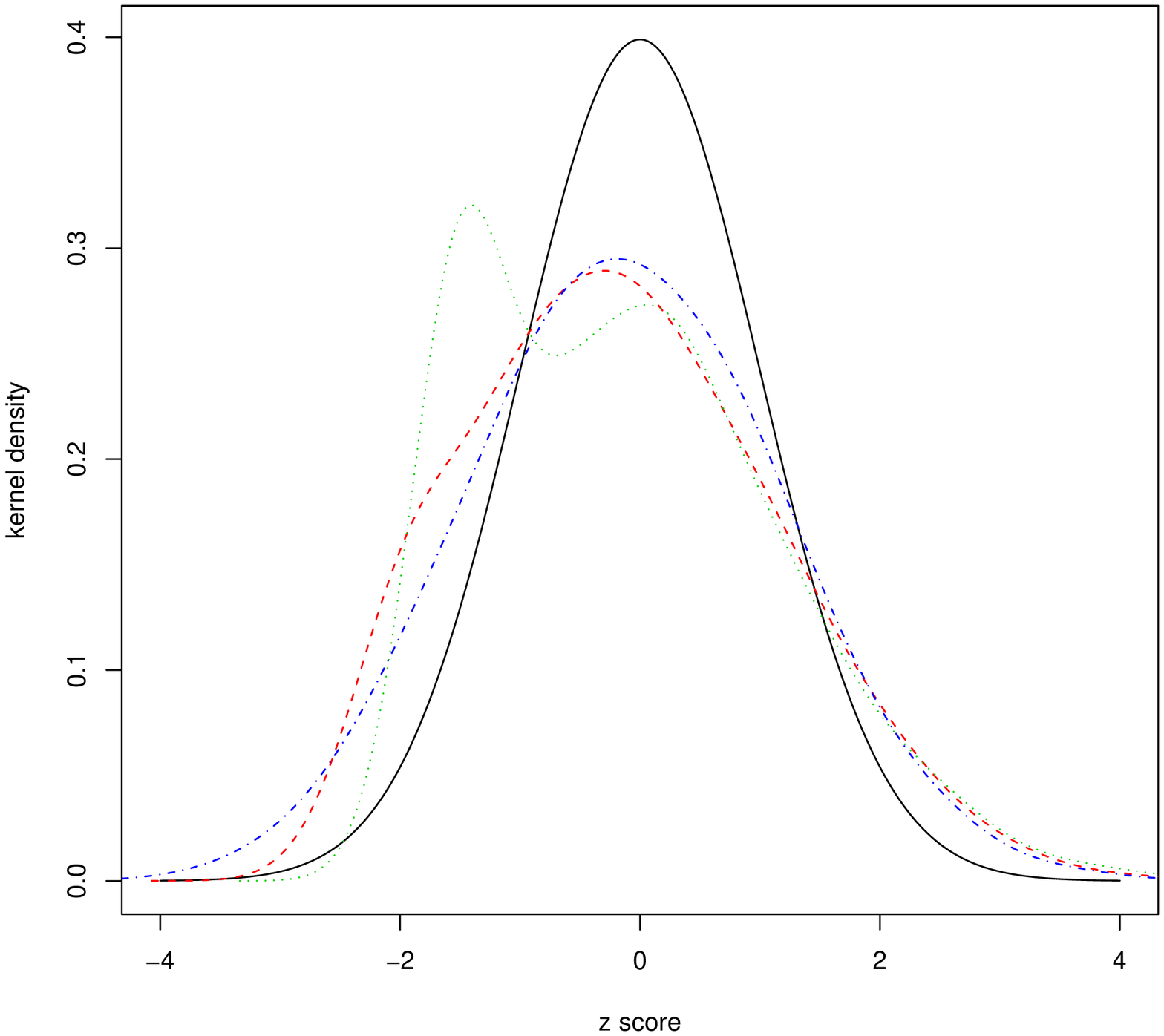} }}
\caption{
\label{fig:z-scores}
The density plot of standard normal distribution (solid line) and
the kernel density estimates of the directional z-scores for similar
(left) and very different (right) sample size combinations.}
\end{figure}

We record $Z_P$ values at each Monte Carlo
replication for each sample size combination.
In Figure \ref{fig:z-scores}, we present the kernel
density estimates for directional $Z$-tests and the standard normal
density (solid line) in
order to make distributional comparisons.
Observe that for balanced sample size
combinations, there seems to be a need for scaling, and a mild
adjustment in location; while for unbalanced sample size combinations,
the location discrepancy between standard normal and kernel density
estimates seem to be larger.

We tabulate the sample means and variances of the test statistics in
Table \ref{tab:means-variances}, where $Z_P$ is for the statistic defined in
Equation \eqref{eqn:2x2-Zform}.
Since the empirical mean for $Z_P$, although tending to zero for
large $n$, is negative for small samples, we also consider
$Z^a_P=Z_P\,\I(Z_P \leq 0)$ for the association, and $Z^s_P=Z_P\,\I(Z_P \geq 0)$
for the segregation alternatives separately,
where $\I(\cdot)$ stands for the indicator function.

\begin{table}[ht]
\centering
\begin{tabular}{|c|c|c|c||c|c|c|}
\hline
\multicolumn{7}{|c|}{Empirical Means and Variances of the Test Statistics} \\
\hline
&\multicolumn{3}{|c||}{Means}&\multicolumn{3}{|c|}{Variances} \\
\hline
$(n_1,n_2)$ & $Z_P$ & $Z^a_P$ & $Z^s_P$ & $Z_P$ & $Z^a_P$ & $Z^s_P$  \\
\hline
 (10,10) &  -.243 & -1.026 & .835 & 1.734 & .744 & .611 \\
 (10,30) &  -.155 & -1.004 & .960 & 1.623 & .508 & .671 \\
 (10,50) &  -.131 & -1.000 & .985 & 1.559 & .343 & .749 \\
 (30,30) &  -.135 & -.996 & .910 &  1.636 & .645 & .595 \\
 (30,50) &  -.128 & -1.049 & .971 & 1.637 & .613 & .576 \\
 (50,50) &  -.091 & -.990 & .931 &  1.639 & .648 & .593 \\
 (100,100) &  -.058 & -1.007 & .956 & 1.643 & .624 & .597 \\
 (200,200) &  -.036 & -1.006 & .962 & 1.627 & .623 & .591 \\
\hline
\end{tabular}
\caption{ \label{tab:means-variances} The empirical means and
variances for the test statistics.}
\end{table}

Let $M(Z_P)$ be the sample mean and $V(Z_P)$ be the sample
variance of $Z_P$ values. Let $M(Z^a_P)$ and $V(Z^a_P)$,
$M(Z^s_P)$ and $V(Z^s_P)$ be similarly
defined for $Z^a_P$  and $Z^s_P$ values, respectively.
Then Table \ref{tab:means-variances}
suggests that $M(Z^a_P)=-1.00$ and $V(Z^a_P)=0.62$; while
$M(Z^s_P)=0.96$ and $V(Z^s_P)=0.59$.
On the other hand, $M(Z_P)$ does not stabilize at a value, but tends to zero;
and $V(Z_P)$ is about 1.63.
Since, we use the critical
values based on standard normal distribution, after adjusting we want
$Z_P \sim Z$, $Z^a_P \sim Z^-$ and $Z^s_P \sim Z^+$, where $Z^-=Z\,\I(Z \leq 0)$ and
$Z^+=Z\,\I(Z \geq 0)$ with $Z \sim N(0,1)$.
Therefore $Z_P$ does not need adjusting for location (we take the mean of $Z_P$ to be 0),
but needs an adjustment in scale for variance.
We transform the $Z_P$ scores as $Z_{mc}:= Z_P/\beta_n$
so that $\Var[Z_{mc}]=\Var[Z]=1$.
By numerical integration, we find the mean and variances of $Z^-$ and $Z^+$ as
$\E[Z^+]=-\E[Z^-]=0.798$ and $\Var[Z^+]=\Var[Z^-]=0.363$ (by symmetry).
We transform the $Z^a_P$ and $Z^s_P$ scores by adjusting for
location and scale as $Z^a_{mc}:=\left(Z^a_P-\alpha_a\right)/\beta_a$
and $Z^s_{mc}:=\left(Z^s_P-\alpha_s\right)/\beta_s$ so that
$\E[Z^a_{mc}] \approx \E[Z^-]=-0.798$ and $\Var[Z^a_{mc}] \approx \Var[Z^-]=0.363$; and
$\E[Z^s_{mc}] \approx \E[Z^+]=0.798$ and $\Var[Z^s_{mc}] \approx \Var[Z^+]=0.363$ would hold.
Such transformations will convert the $Z_P$ into a standard normal (approximately);
and $Z^a_P$ and $Z^s_P$ values into restricted standard normal (approximately),
since the Figure \ref{fig:z-scores}
suggests that $Z_P$ scores are approximately normal, but not standard normal.
Using the sample estimate $V(Z_P)$ for $\Var(Z_P)$,
solving for $\beta_n$ yields $\beta_n=\sqrt{1.63}=1.277$.
Similarly, using the sample estimates $M(Z^a_P)$ and $V(Z^a_P)$ for
$\E(Z^a_P)$ and $\Var(Z^a_P)$, solving for $\alpha_a$ and $\beta_a$
yields $\beta_a=\sqrt{V(Z^a_P)/0.363}=1.307$ and $\alpha_a=0.043$.
Likewise, we find $\beta_s=\sqrt{V(Z^s_P)/0.363}=1.275$ and $\alpha_s=-0.057$.

\subsection{New Directional Tests of Segregation}
\label{sec:new-directional}
In the directional version of Pielou's test
using $Z_P$ in Equation \eqref{eqn:2x2-Zform}, the quantities
$n_1,\,n_2$ and $n$ are fixed, while $C_1,\,C_2$, and
$N_{11},\,N_{21}$  are random quantities.
Because of the fact that
the product $C_1\,C_2$ is in the denominator under the square root,
$\E[Z_P]$ and $\Var[Z_P]$ are not analytically tractable, hence the
correct (asymptotic) distribution of $Z_P$ is not
available. Therefore, one way to fix $Z_P$ to have the appropriate
Type I error rate is as in Section \ref{sec:MC-correction} or by
randomization test.
\cite{meagher:1980} propose and illustrate using Monte Carlo simulation
to calculate critical values of Pielou's test statistic under RL.
This is the same concept as our randomized (Monte Carlo corrected) test,
but the details are not the same:
\cite{meagher:1980} use a Monte-Carlo computation of the critical value
while we use a Monte Carlo-based moment adjustment that is intended for use
when the study region is rectangular and sample sizes are similar.

Alternatively we modify $Z_P$ to make the distribution of it to be analytically tractable.
To this end,
let $\displaystyle T_n:=\frac{N_{11}}{n_1}-\frac{N_{21}}{n_2}$ and
$\displaystyle U_n:=\sqrt{\frac{n_1\,n_2}{C_1\,C_2}}$, then $\displaystyle Z_P=\sqrt{n}\,U_n\,T_n$.
Note that $\displaystyle \E[T_n]=\left(\frac{\E[N_{11}]}{n_1}-\frac{\E[N_{21}]}{n_2}\right)=\frac{-1}{(n-1)}$.
Using the asymptotic normality of cell counts $N_{11}$ and $N_{21}$, we have
$$\left(\frac{T_n-\E[T_n]}{\sqrt{\Var[T_n]}}\right) \stackrel{\mathcal L}{\longrightarrow} N(0,1),$$
where
$\stackrel{\mathcal L}{\longrightarrow}$ stands for convergence in law,
$$\Var[T_n]=\frac{\Var[N_{11}]}{n^2_1}+\frac{\Var[N_{21}]}{n^2_2}-
\frac{2\,\Cov[N_{11},N_{21}]}{n_1\,n_2},$$
with $\Var[N_{ij}]$ are as in Equation \eqref{eqn:VarNij} and
$$\Cov[N_{11},N_{21}]=(n-R+Q)\,p_{112}+\left( n^2-3\,n-Q+R \right)\,p_{1112}-n^2\,p_{11}\,p_{12}$$
(see \cite{dixon:NNCTEco2002} for the derivation).

We propose two tests based on $T_n$:
\begin{equation}
\label{eqn:new-versions}
(i)~~Z_I=\sqrt{n}\,U_n\,\left(\frac{T_n-\E[T_n]}{\sqrt{\Var[T_n]}}\right);~~~~~
(ii)~~Z_{II}=\frac{T_n-\E[T_n]}{\sqrt{\Var[T_n]}}.
\end{equation}

The latter test statistic, $Z_{II}$ converges in distribution
to $N(0,1)$ as $n \rightarrow \infty$;
while such convergence for $Z_I$ holds conditional on $U_n$.
However, note that for large $n$, $U_n \approx 1$,
since, letting $\nu_i$ be the proportion of class $i$ in the population,
we have $\lim_{n,n_i \rightarrow \infty} n_i/n=\nu_i$;
and under CSR independence or RL,
$C_i/n\stackrel{p}{\rightarrow} \nu_i$ as $n,n_i \rightarrow \infty$ for $i=1,2$.
Furthermore, $Z_I$ and $Z_{II}$ values are positive under segregation
and negative under association alternatives.

\begin{remark}
\label{rem:ext-multi-class}
\textbf{Extension to Multi-Class Case:}
For $q$ classes, the NNCT we obtain will be of dimension $q \times q$.
The extension of the cell-specific tests to
the $q$-class case with $q > 2$ is straightforward.
But unfortunately, the asymptotic normality of the off-diagonal
cells in these NNCTs is not rigorously established yet,
although extensive Monte Carlo simulations indicate approximate
normality for large samples (\cite{dixon:NNCTEco2002}).
In the multi-class case,
a positive $z$-score, $Z_{ii}$, for the diagonal cell $(i,i)$ indicate segregation,
but it does not necessarily mean lack of association between class $i$
and class $j$ ($i\not=j$), since it could be the case that class $i$
could be associated with another class, yet not associated with another one.
Likewise for Ceyhan's cell-specific tests.
The directional tests in Section \ref{sec:new-directional} are designed for the two-class case only.
$\square$
\end{remark}

\begin{remark}
\textbf{The Status of $Q$ and $R$ under CSR Independence and RL:}
\label{rem:QandR}
$Q$ and $R$ are fixed under RL, but random under CSR independence.
The quantities given in Equations \eqref{eqn:Exp[Nij]}, \eqref{eqn:VarNij},
and all the quantities depending on these expectations also depend on $Q$ and $R$.
Hence these expressions are appropriate under the RL model.
Under the CSR independence model they are conditional variances and
covariances obtained by conditioning on $Q$ and $R$.
Hence under the CSR independence pattern,
the asymptotic distributions of the tests in Equations
\eqref{eqn:dixon-Zij}, \eqref{eqn:new-Zij}, and \eqref{eqn:new-versions} are conditional $Q$ and $R$.

The unconditional variances and covariances can be obtained
by replacing $Q$ and $R$ with their expectations (\cite{ceyhan:class2009}).
Unfortunately, given the difficulty of calculating the
expectation of $Q$ under CSR independence,
it is reasonable and convenient to use test statistics employing the
conditional variances and covariances even when assessing their
behavior under the CSR independence model.
\cite{cox:1981} calculated analytically that $\E[R|N]= 0.6215 N$ for a planar
Poisson process.
Alternatively, one can estimate the expected values of $Q$ and $R$ empirically.
For example, for homogeneous planar Poisson pattern,
we have $\E[Q|N] \approx 0.6328 N$ and $\E[R|N] \approx 0.6211 N$
(estimated empirically by 1000000 Monte Carlo simulations for various values of $N$ on unit square).
Notice that $\E[R|N]$ agrees with the analytical result of \cite{cox:1981}.
When $Q$ and $R$ are replaced by $0.63 \, n$ and $0.62 \,n$, respectively,
the so-called \emph{QR-adjusted} tests are obtained.
However, QR-adjustment does not improve on the unadjusted NNCT-tests (\cite{ECarXivQRAdjust:2008}).
$\square$
\end{remark}

\section{Empirical Significance Levels under CSR Independence}
\label{sec:empirical-size}
We only consider the two-class case with classes $X$ and $Y$.
We generate $n_1$ points from class $X$ and $n_2$ points from class $Y$
both of which are uniformly distributed on the unit square $(0,1) \times (0,1)$
for some combinations of $n_1$ and $n_2$.
Thus, we simulate the CSR independence pattern for the performance of the tests
under the null case.

We present the empirical significance
levels of the tests for the two-sided alternative in Table \ref{tab:size-two-sided},
where $\ah_{i,i}^D$ and $\ah_{i,i}^C$ are the empirical significance
levels of Dixon's and Ceyhan's cell-specific tests for cell $(i,i)$, $i=1,2$, respectively,
$\ah^P_Z$ is for the directional version of Pielou's test $Z_P$,
$\ah^{P,Z}_{mc}$ is for the empirically corrected version of $Z_P$,
$\ah_I$ and $\ah_{II}$ are for the new directional tests provided
in Equation \eqref{eqn:new-versions}.
The sizes significantly smaller (larger) than .05 are marked with $^c$ ($^{\ell}$),
which indicate that the corresponding test is conservative (liberal).
The asymptotic normal approximation to proportions is used in determining the significance of
the deviations of the empirical sizes from .05.
For these proportion tests, we also use $\alpha=.05$ as the significance level.
With $N_{mc}=10000$, empirical sizes less than .0464 are deemed conservative,
greater than .0536 are deemed liberal at $\alpha=.05$ level.
Notice that directional version of Pielou's test $Z_P$ is extremely liberal
in rejecting the null hypothesis,
similar to the two-sided version as shown in \cite{ceyhan:overall}.
The empirically corrected version of Pielou's test has much better size performance compared to $Z_P$,
but it is still extremely conservative when relative abundances are very different
for small samples (i.e., $n_i \le 30$ for both $i=1,2$);
for similar small samples it is liberal; and when the relative abundances are very different
for large samples it is also conservative.
For Dixon's cell-specific tests, if at least one sample size is small,
the normal approximation is not appropriate,
so he recommends Monte Carlo randomization
instead of the asymptotic approximation for the corresponding cell-specific tests (\cite{dixon:1994}).
For cell $(1,1)$, when $n_1 \le 10$ or when $n_1$ and $n_2$
are very different (i.e., classes have very different relative abundances),
the cell count is more likely to be $< 5$.
Hence in such cases Dixon's cell-specific test is
conservative when $n_1$ is small, and is liberal when $n_1$ is large.
The empirical size for Dixon's test for cell $(2,2)$ at sample size combination $(n_1,n_2)$
is similar to the one for cell $(1,1)$ at $(n_2,n_1)$,
so Dixon's cell-specific tests are
symmetric in the sample sizes in terms of size performance.
When cell counts are $\ge 5$ (which happens for large samples
with relative abundances not being very different),
Dixon's cell-specific tests seem to be appropriate
(i.e., they have about the desired nominal level).
On the other hand, Ceyhan's cell-specific test seems
to be conservative when both sample sizes are small ($n_i \le 30$)
or the classes have very different relative abundances.
Otherwise, they have about the desired nominal level.
Furthermore, the empirical size estimates of
Ceyhan's cell-specific tests for cells $(1,1)$ and $(2,2)$
are similar at each sample size combination.
The size performance of the new directional tests
is similar to that of Ceyhan's cell-specific tests.

The differences in the relative abundance of classes seem to affect
Dixon's tests more than the other tests.
See for example cell-specific tests for cell $(1,1)$ for sample sizes
$(30,50)$ and $(50,100)$,
where Dixon's test suggests that class $X$ (i.e., class with the smaller size)
is more segregated which is only an artifact of the difference in the relative abundance.
On the other hand, Ceyhan's cell-specific tests and the new directional
tests are more robust to differences
in the relative abundances.

\begin{table}[ht]
\centering
\begin{tabular}{|c||c|c|c|c|c|c|c|c|}
\hline
      & \multicolumn{8}{|c|}{Empirical significance levels of the tests} \\
sizes & \multicolumn{8}{|c|}{for the two-sided alternatives} \\
\hline
$(n_1,n_2)$  & $\ah_{1,1}^D$ & $\ah_{2,2}^D$ & $\ah_{1,1}^C$ & $\ah_{2,2}^C$ &
$\ah^P_Z$ & $\ah^{P,Z}_{mc}$ & $\ah_I$ & $\ah_{II}$ \\
\hline
(10,10) & .0454$^c$ & .0465 & .0452$^c$ & .0459$^c$ & .1280$^{\ell}$ & .0608$^{\ell}$ & .0503 & .0439$^c$\\
\hline
(10,30) & .0306$^c$ & .0485 & .0413$^c$ & .0420$^c$ & .1429$^{\ell}$ & .0320$^c$ & .0390$^c$ & .0410$^c$\\
\hline
(10,50) & .0270$^c$ & .0464 & .0390$^c$ & .0396$^c$ & .0664$^{\ell}$ & .0292$^c$ & .0423$^c$ & .0397$^c$\\
\hline
(30,10) & .0479 & .0275$^c$ & .0399$^c$ & .0395$^c$ & .1383$^{\ell}$ & .0282$^c$ & .0372$^c$ & .0389$^c$\\
\hline
(30,30) & .0507 & .0505 & .0443$^c$ & .0442$^c$ & .1339$^{\ell}$ & .0552$^{\ell}$ & .0465 & .0427$^c$\\
\hline
(30,50) & .0590$^{\ell}$ & .0522 & .0505 & .0510 & .1267$^{\ell}$ & .0531 & .0502 & .0505\\
\hline
(50,10) & .0524 & .0263$^c$ & .0378$^c$ & .0367$^c$ & .0654$^{\ell}$ & .0287$^c$ & .0406$^c$ & .0379$^c$\\
\hline
(50,30) & .0535 & .0597$^{\ell}$ & .0462$^c$ & .0476 & .1275$^{\ell}$ & .0534 & .0474 & .0464\\
\hline
(50,50) & .0465 & .0469 & .0500 & .0502 & .1397$^{\ell}$ & .0494 & .0520 & .0499\\
\hline
(50,100) & .0601$^{\ell}$ & .0533 & .0514 & .0515 & .1223$^{\ell}$ & .0508 & .0506 & .0519\\
\hline
(100,50) & .0490 & .0571$^{\ell}$ & .0480 & .0477 & .1190$^{\ell}$ & .0463$^c$ & .0470 & .0483\\
\hline
(100,100) & .0493 & .0463$^c$ & .0485 & .0486 & .1324$^{\ell}$ & .0524 & .0490 & .0489\\
\hline
\end{tabular}
\caption{
\label{tab:size-two-sided}
The empirical significance levels of the tests for the two-sided alternatives
under $H_o:CSR~~independence$ with $N_{mc}=10000$, $n_1,n_2$ in $\{10,30,50,100\}$ at $\alpha=.05$.
($^c$: the empirical size is significantly smaller than 0.05; i.e., the test is conservative.
$^{\ell}$: the empirical size is significantly larger than 0.05; i.e., the test is liberal.}
\end{table}

We present the empirical significance
levels of the tests for the right-sided alternative
(i.e., with respect to the segregation alternative)
in Table \ref{tab:size-seg}.
The size labeling and superscripting for conservativeness and liberalness are
as in Table \ref{tab:size-two-sided}.
Notice that directional version of Pielou's test for segregation alternative
is liberal in rejecting the null hypothesis.
The empirically corrected version of Pielou's test as in Section \ref{sec:MC-correction}
for the segregation alternative is about the desired level for larger samples,
but is conservative or liberal for smaller samples.
The size performance of Dixon's cell-specific test for cell $(1,1)$ at $(n_1,n_2)$
is similar to that for cell $(2,2)$ at $(n_2,n_1)$.
On the other hand,
at each $(n_1,n_2)$ the size performance of Ceyhan's cell-specific test
for cell $(1,1)$ is similar to that for cell $(2,2)$.
Dixon's cell-specific tests are usually liberal,
in particular for the smaller sample for different relative abundance cases.
Ceyhan's cell-specific tests are liberal when $n_i \le 30$ for both $i=1,2$
or when the relative abundances are very different.
The size performance of the new directional tests is similar
to that of Ceyhan's cell-specific tests.
As in the two-sided version, Ceyhan's cell-specific and the new
directional tests are more robust to differences in relative abundance of the classes.

\begin{table}[ht]
\centering
\begin{tabular}{|c||c|c|c|c|c|c|c|c|}
\hline
      & \multicolumn{8}{|c|}{Empirical significance levels of the tests} \\
sizes & \multicolumn{8}{|c|}{for the segregation (i.e., right-sided) alternatives} \\
\hline
$(n_1,n_2)$  & $\ah_{1,1}^D$ & $\ah_{2,2}^D$ & $\ah_{1,1}^C$ & $\ah_{2,2}^C$ &
$\ah^P_Z$ & $\ah^{P,Z}_{mc}$ & $\ah_I$ & $\ah_{II}$ \\
\hline
(10,10) & .0515 & .0489 & .0491 & .0491 & .0844$^{\ell}$ & .0422$^c$ & .0526 & .0499\\
\hline
(10,30) & .0960$^{\ell}$ & .0468 & .0631$^{\ell}$ & .0643$^{\ell}$ & .0846$^{\ell}$ & .0576$^{\ell}$ & .0613$^{\ell}$ & .0651$^{\ell}$\\
\hline
(10,50) & .0936$^{\ell}$ & .0435$^c$ & .0684$^{\ell}$ & .0677$^{\ell}$ & .0947$^{\ell}$ & .0548$^{\ell}$ & .0693$^{\ell}$ & .0678$^{\ell}$\\
\hline
(30,10) & .0430$^c$ & .0900$^{\ell}$ & .0571$^{\ell}$ & .0567$^{\ell}$ & .0760$^{\ell}$ & .0511 & .0545$^{\ell}$ & .0575$^{\ell}$\\
\hline
(30,30) & .0490 & .0530 & .0556$^{\ell}$ & .0555$^{\ell}$ & .0803$^{\ell}$ & .0557$^{\ell}$ & .0557$^{\ell}$ & .0555$^{\ell}$\\
\hline
(30,50) & .0652$^{\ell}$ & .0479 & .0482 & .0484 & .0792$^{\ell}$ & .0445$^c$ & .0479 & .0480\\
\hline
(50,10) & .0441$^c$ & .0915$^{\ell}$ & .0655$^{\ell}$ & .0665$^{\ell}$ & .0955$^{\ell}$ & .0531 & .0682$^{\ell}$ & .0655$^{\ell}$\\
\hline
(50,30) & .0492 & .0664$^{\ell}$ & .0515 & .0509 & .0829$^{\ell}$ & .0468 & .0511 & .0511\\
\hline
(50,50) & .0577$^{\ell}$ & .0546$^{\ell}$ & .0514 & .0509 & .0804$^{\ell}$ & .0421$^c$ & .0526 & .0522\\
\hline
(50,100) & .0571$^{\ell}$ & .0464 & .0509 & .0508 & .0921$^{\ell}$ & .0495 & .0524 & .0508\\
\hline
(100,50) & .0434$^c$ & .0584$^{\ell}$ & .0499 & .0500 & .0909$^{\ell}$ & .0483 & .0512 & .0498\\
\hline
(100,100) & .0515 & .0500 & .0485 & .0485 & .0927$^{\ell}$ & .0484 & .0485 & .0485\\
\hline
\end{tabular}
\caption{
\label{tab:size-seg}
The empirical significance levels of the tests for the segregation (right-sided) alternatives
under $H_o:CSR~~independence$ with $N_{mc}=10000$, $n_1,n_2$ in $\{10,30,50,100\}$ at $\alpha=.05$.
($^c$: the empirical size is significantly smaller than 0.05; i.e., the test is conservative.
$^{\ell}$: the empirical size is significantly larger than 0.05; i.e., the test is liberal.}
\end{table}

We present the empirical significance
levels of the tests for the left-sided alternative
(i.e., with respect to the association alternative)
in Table \ref{tab:size-assoc}.
The size labeling and superscripting are as in Table \ref{tab:size-two-sided}.
Notice that directional version of Pielou's test is extremely
liberal for all sample size combinations.
The empirically corrected version is still liberal
for most sample sizes and extremely conservative for $(10,50)$ and $(50,10)$ cases.
When the relative abundances of classes are very different (see $(10,50)$ and $(50,10)$ cases),
both tests are severely affected, but the corrected version is extremely conservative.
For small samples ($n_i \le 30$) Dixon's cell-specific tests are
extremely conservative for the cell associated with the
smaller sample when the relative abundances are very different.
For large samples Dixon's cell-specific tests are
liberal for the cell associated with the
smaller sample when the relative abundances are very different.
Ceyhan's cell-specific tests are extremely conservative when both samples are small
and are about the desired level for large samples.
New tests' size performance is similar to that of Ceyhan's tests.
Furthermore, the effect of the differences in relative abundances is
most severe on Dixon's cell-specific tests.

\begin{table}[ht]
\centering
\begin{tabular}{|c||c|c|c|c|c|c|c|c|}
\hline
      & \multicolumn{8}{|c|}{Empirical significance levels of the tests} \\
sizes & \multicolumn{8}{|c|}{for the association (i.e., left-sided) alternatives} \\
\hline
$(n_1,n_2)$  & $\ah_{1,1}^D$ & $\ah_{2,2}^D$ & $\ah_{1,1}^C$ & $\ah_{2,2}^C$ &
$\ah^P_Z$ & $\ah^{P,Z}_{mc}$ & $\ah_I$ & $\ah_{II}$ \\
\hline
(10,10) & .0412$^c$ & .0455$^c$ & .0467 & .0454$^c$ & .1574$^{\ell}$ & .0858$^{\ell}$ & .0484 & .0425$^c$\\
\hline
(10,30) & .0000$^c$ & .0490 & .0342$^c$ & .0362$^c$ & .1399$^{\ell}$ & .0600$^{\ell}$ & .0296$^c$ & .0362$^c$\\
\hline
(10,50) & .0000$^c$ & .0484 & .0057$^c$ & .0087$^c$ & .0574$^{\ell}$ & .0006$^c$ & .0006$^c$ & .0086$^c$\\
\hline
(30,10) & .0494 & .0000$^c$ & .0333$^c$ & .0319$^c$ & .1406$^{\ell}$ & .0556$^{\ell}$ & .0274$^c$ & .0332$^c$\\
\hline
(30,30) & .0450$^c$ & .0430$^c$ & .0504 & .0505 & .1115$^{\ell}$ & .0537$^{\ell}$ & .0505 & .0505\\
\hline
(30,50) & .0611$^{\ell}$ & .0564 & .0494 & .0494 & .1172$^{\ell}$ & .0600$^{\ell}$ & .0493 & .0495\\
\hline
(50,10) & .0545 & .0000$^c$ & .0080$^c$ & .0058$^c$ & .0544$^{\ell}$ & .0004$^c$ & .0004$^c$ & .0079$^c$\\
\hline
(50,30) & .0520 & .0594$^{\ell}$ & .0475 & .0479 & .1173$^{\ell}$ & .0572$^{\ell}$ & .0467 & .0477\\
\hline
(50,50) & .0486 & .0494 & .0503 & .0500 & .1041$^{\ell}$ & .0580$^{\ell}$ & .0522 & .0517\\
\hline
(50,100) & .0548$^{\ell}$ & .0491 & .0487 & .0491 & .1090$^{\ell}$ & .0534 & .0475 & .0486\\
\hline
(100,50) & .0485 & .0515 & .0465 & .0464 & .1063$^{\ell}$ & .0515 & .0453$^c$ & .0464\\
\hline
(100,100) & .0478 & .0493 & .0475 & .0475 & .1092$^{\ell}$ & .0592$^{\ell}$ & .0476 & .0475\\
\hline
\end{tabular}
\caption{
\label{tab:size-assoc}
The empirical significance levels of the tests for the association (left-sided) alternatives
under $H_o:CSR~~independence$ with $N_{mc}=10000$, $n_1,n_2$ in $\{10,30,50,100\}$ at $\alpha=.05$.
($^c$: the empirical size is significantly smaller than 0.05; i.e., the test is conservative.
$^{\ell}$: the empirical size is significantly larger than 0.05; i.e., the test is liberal.}
\end{table}

\section{Empirical Power Analysis}
\label{sec:emp-power}
We consider three cases for each of segregation and association alternatives.
Based on the empirical size estimates provided in Section \ref{sec:empirical-size},
we omit the directional versions of Pielou's test $Z_P$ and the empirically corrected version of it
(see  Section \ref{sec:MC-correction}) from further consideration.
$Z_P$ is extremely liberal in rejecting the null hypothesis,
so it is likely to give more false alarms than we can tolerate.
On the other hand, the empirically corrected version is only valid for rectangular regions
for similar sample sizes, hence might miss the correct pattern due to these restrictions.

\subsection{Empirical Power Analysis under Segregation Alternatives}
\label{sec:power-comp-seg}

For the segregation alternatives, we generate
$X_i \stackrel{iid}{\sim} \U((0,1-s)\times(0,1-s))$ and
$Y_j \stackrel{iid}{\sim} \U((s,1)\times(s,1))$
for $i=1,\ldots,n_1$ and $j=1,\ldots,n_2$.
Notice the level of segregation is determined by the magnitude of $s \in (0,1)$.
We consider the following three segregation alternatives:
\begin{equation}
\label{eqn:seg-alt}
H_S^I: s=1/6, \;\;\; H_S^{II}: s=1/4, \text{ and } H_S^{III}: s=1/3.
\end{equation}

Observe that, from $H_S^I$ to $H_S^{III}$, the segregation gets stronger
in the sense that $X$ and $Y$ points tend to form one-class clumps or clusters.

\begin{figure}[hbp]
\centering
\rotatebox{-90}{ \resizebox{2. in}{!}{\includegraphics{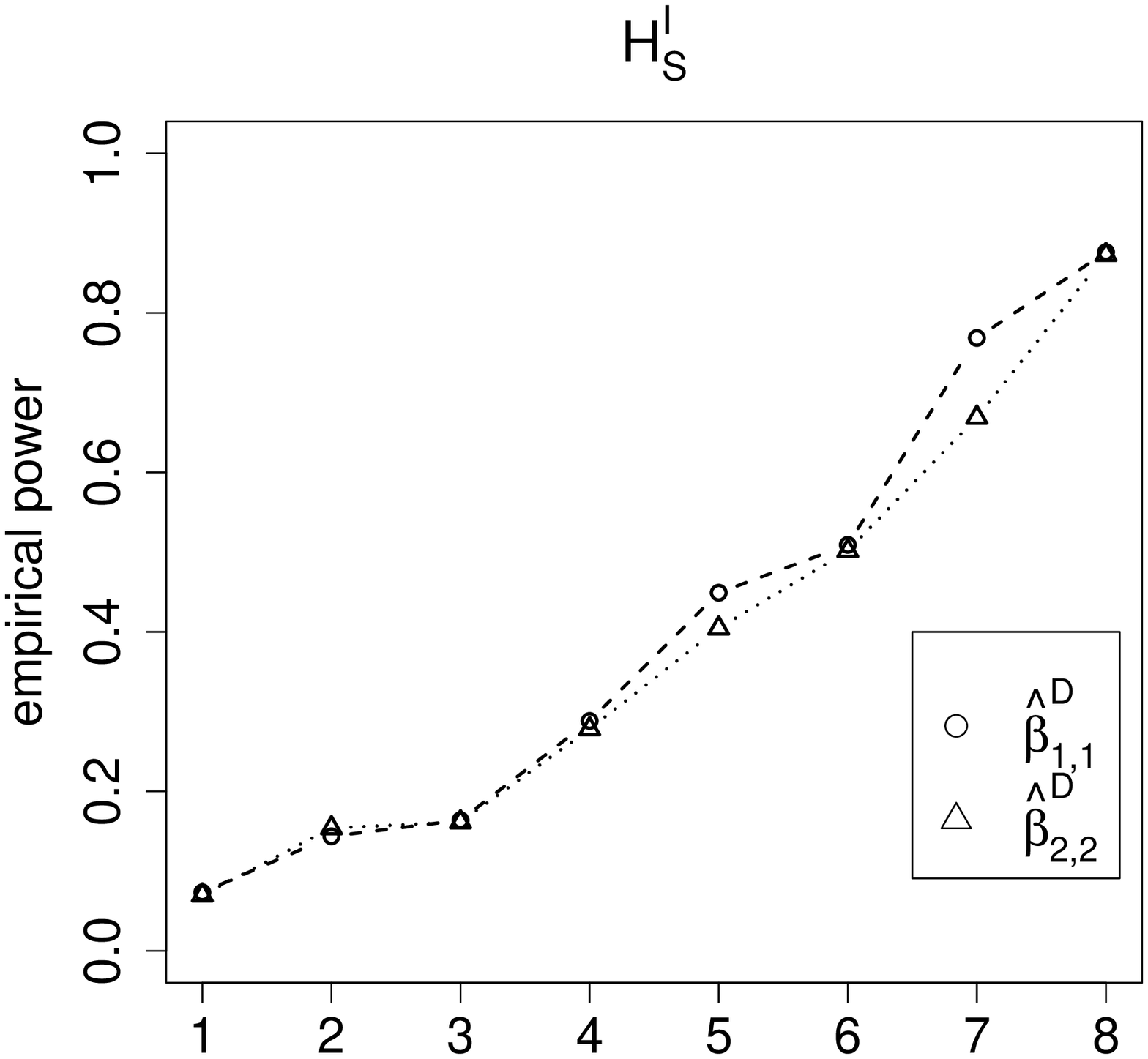} }}
\rotatebox{-90}{ \resizebox{2. in}{!}{\includegraphics{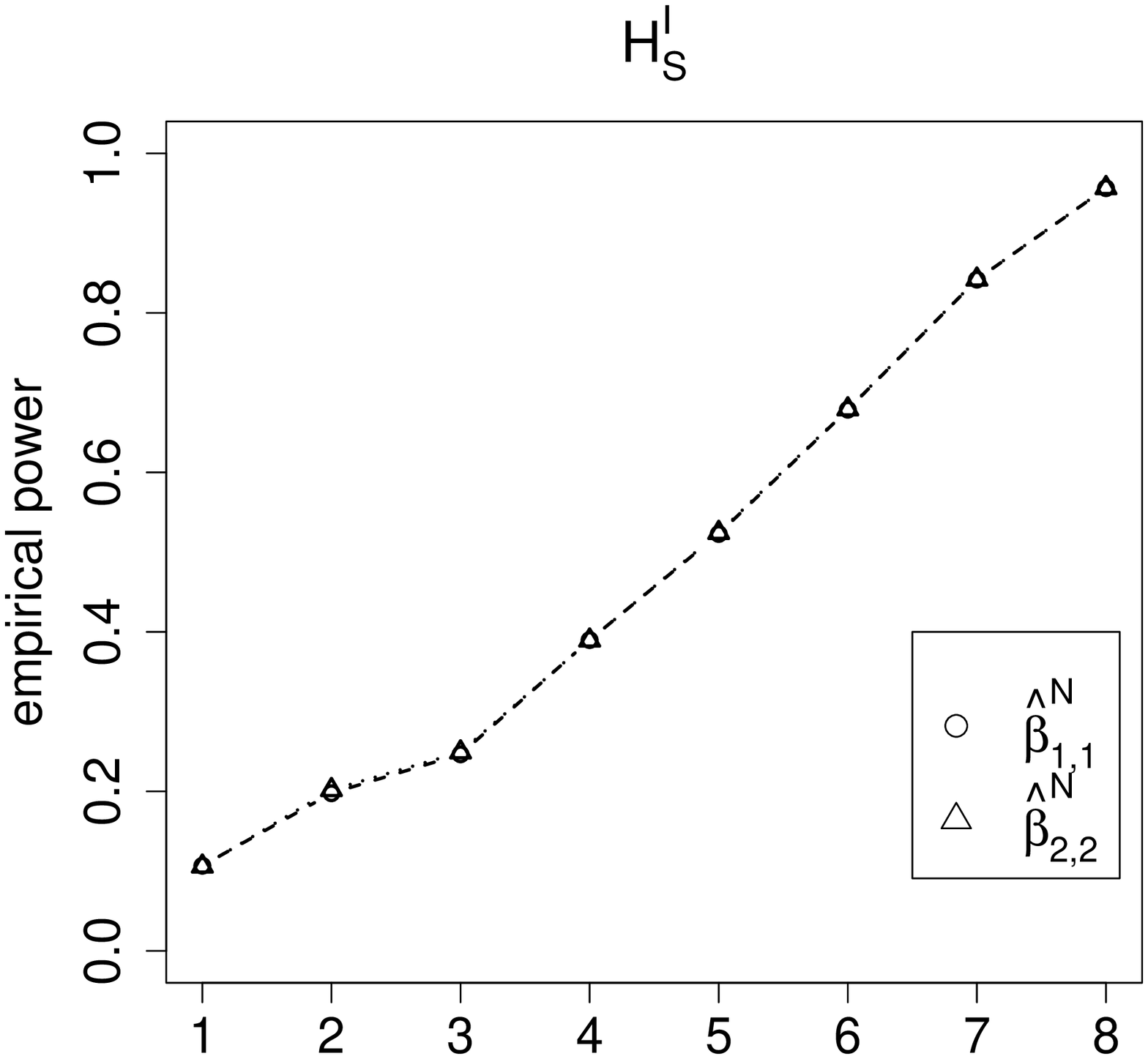} }}
\rotatebox{-90}{ \resizebox{2. in}{!}{\includegraphics{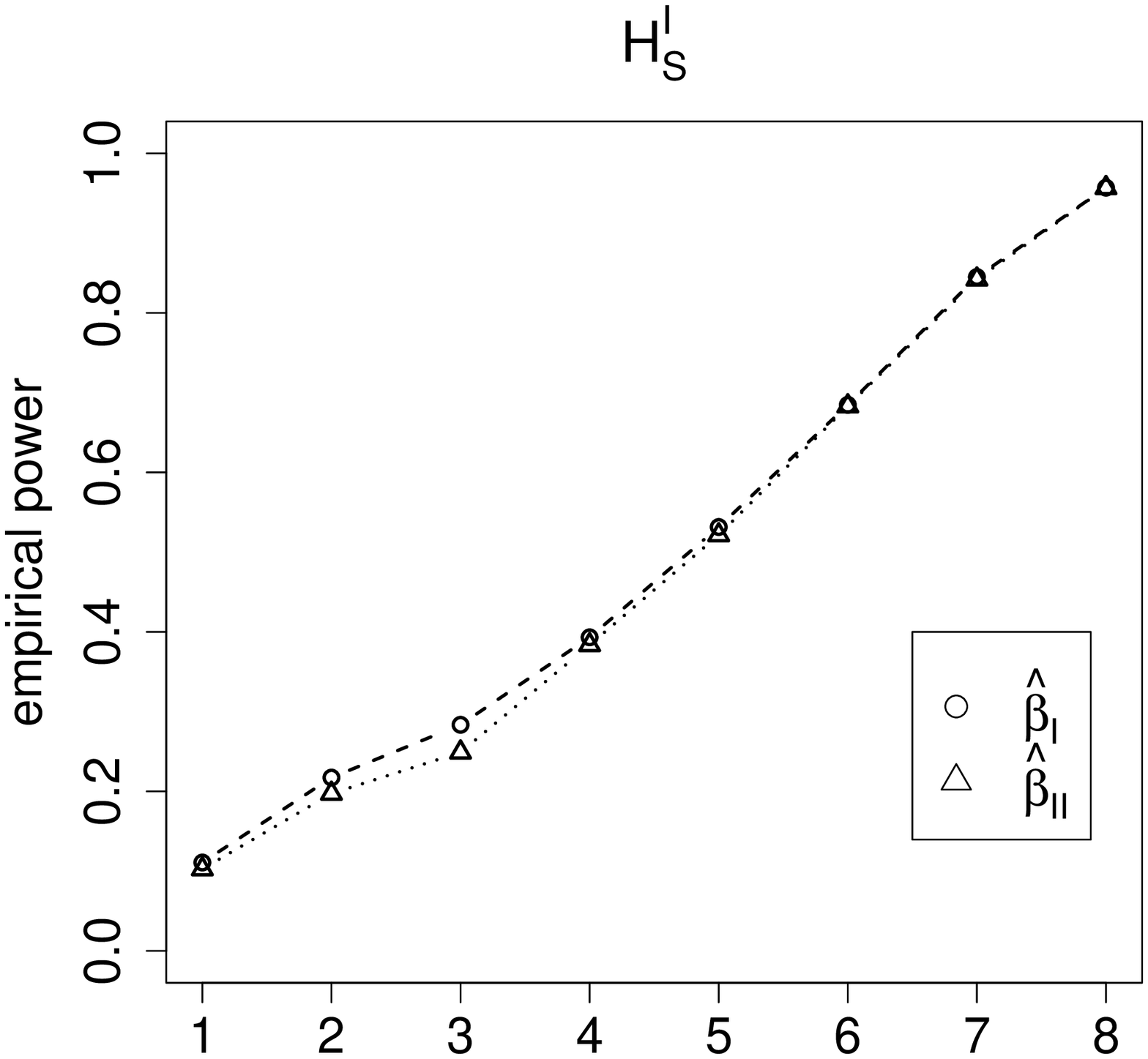} }}
\rotatebox{-90}{ \resizebox{2. in}{!}{\includegraphics{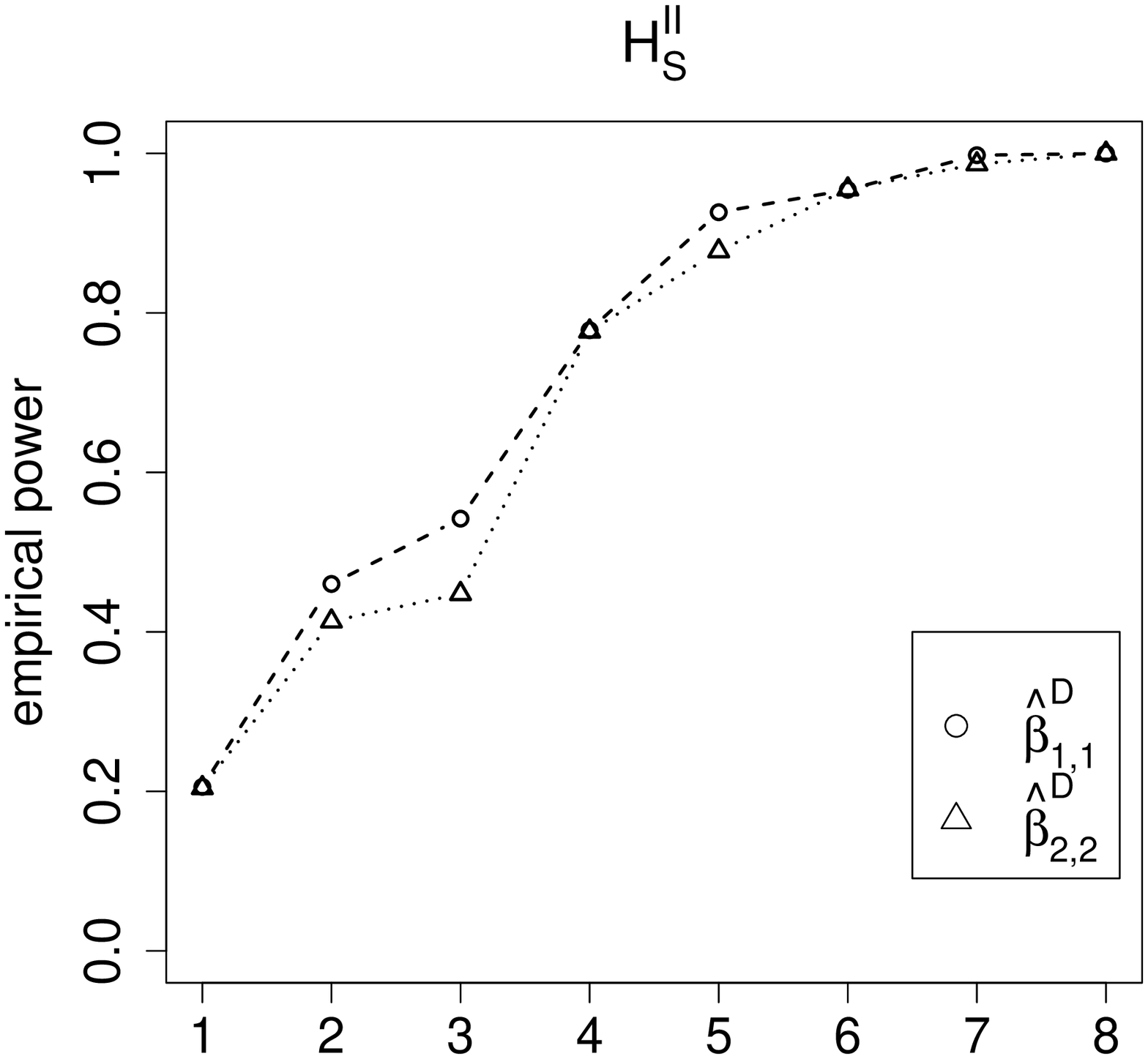} }}
\rotatebox{-90}{ \resizebox{2. in}{!}{\includegraphics{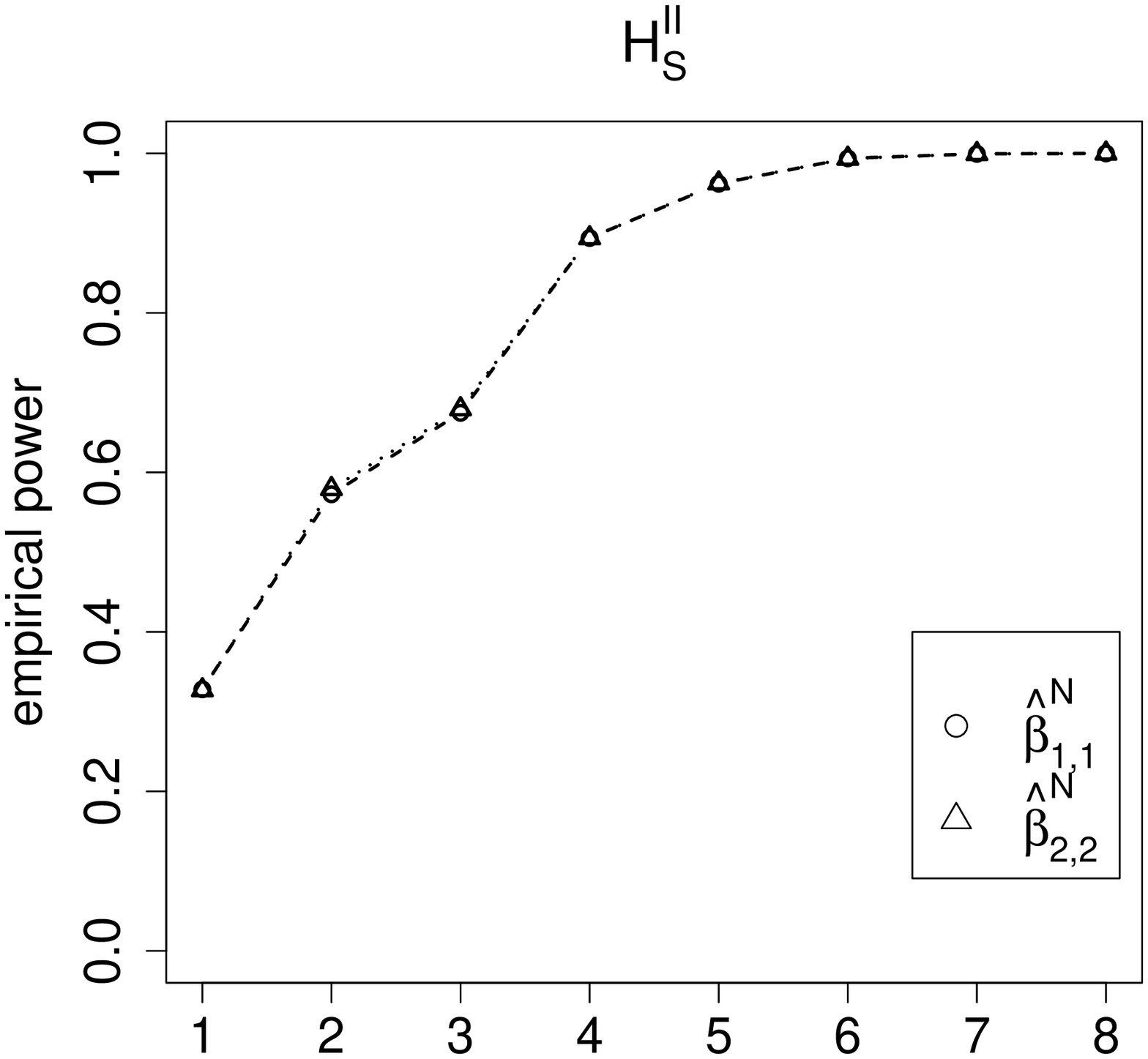} }}
\rotatebox{-90}{ \resizebox{2. in}{!}{\includegraphics{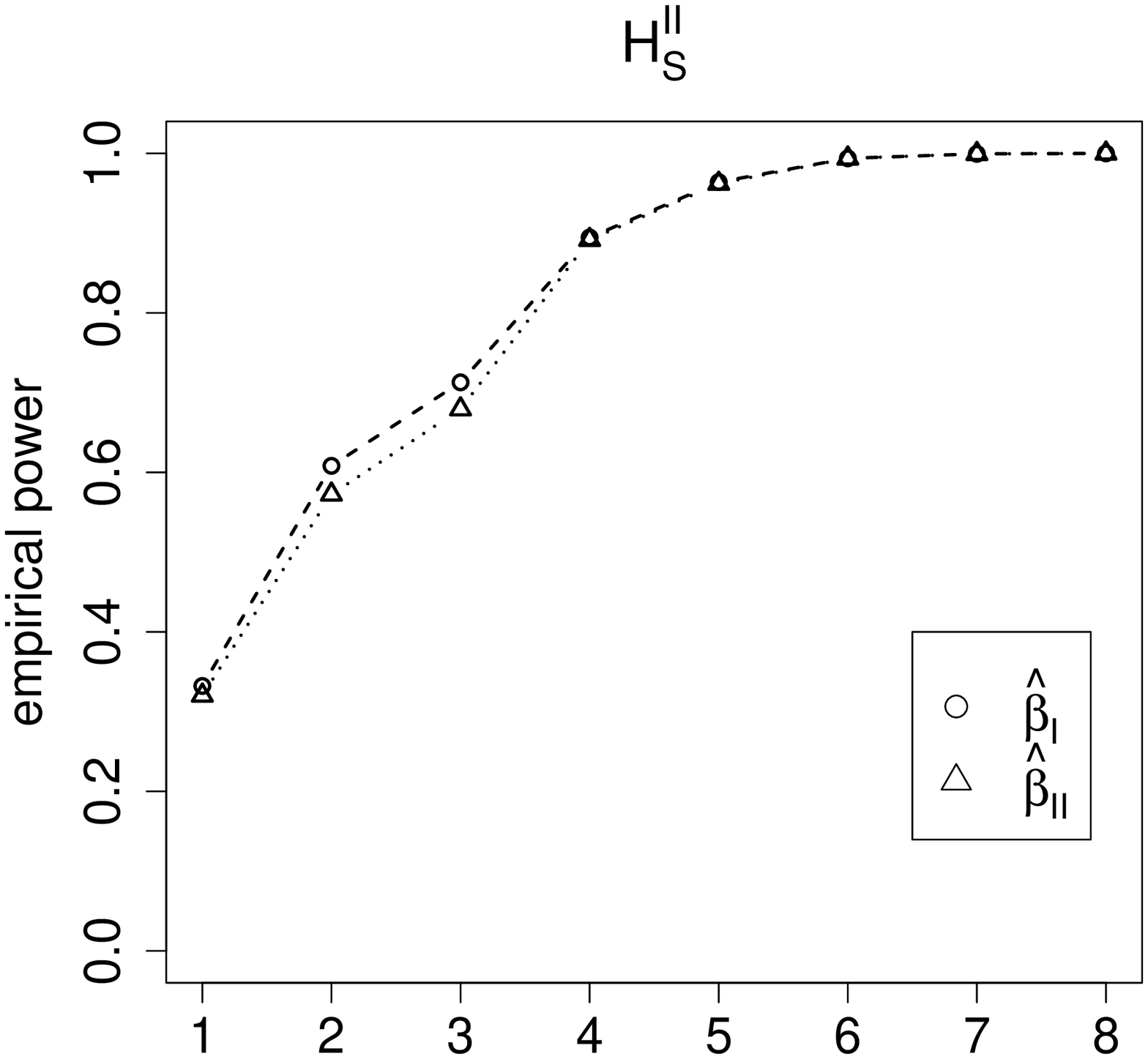} }}
\rotatebox{-90}{ \resizebox{2. in}{!}{\includegraphics{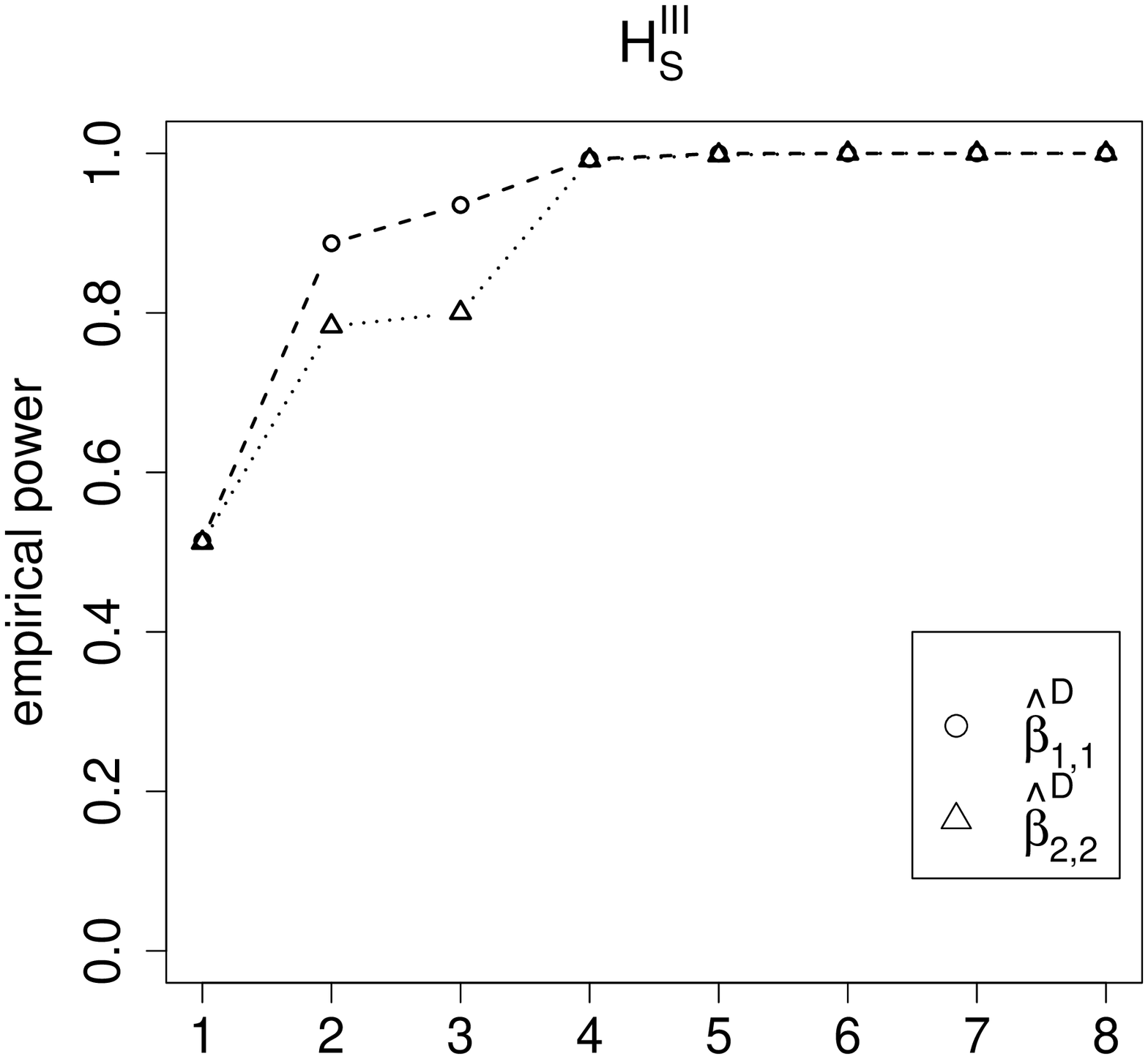} }}
\rotatebox{-90}{ \resizebox{2. in}{!}{\includegraphics{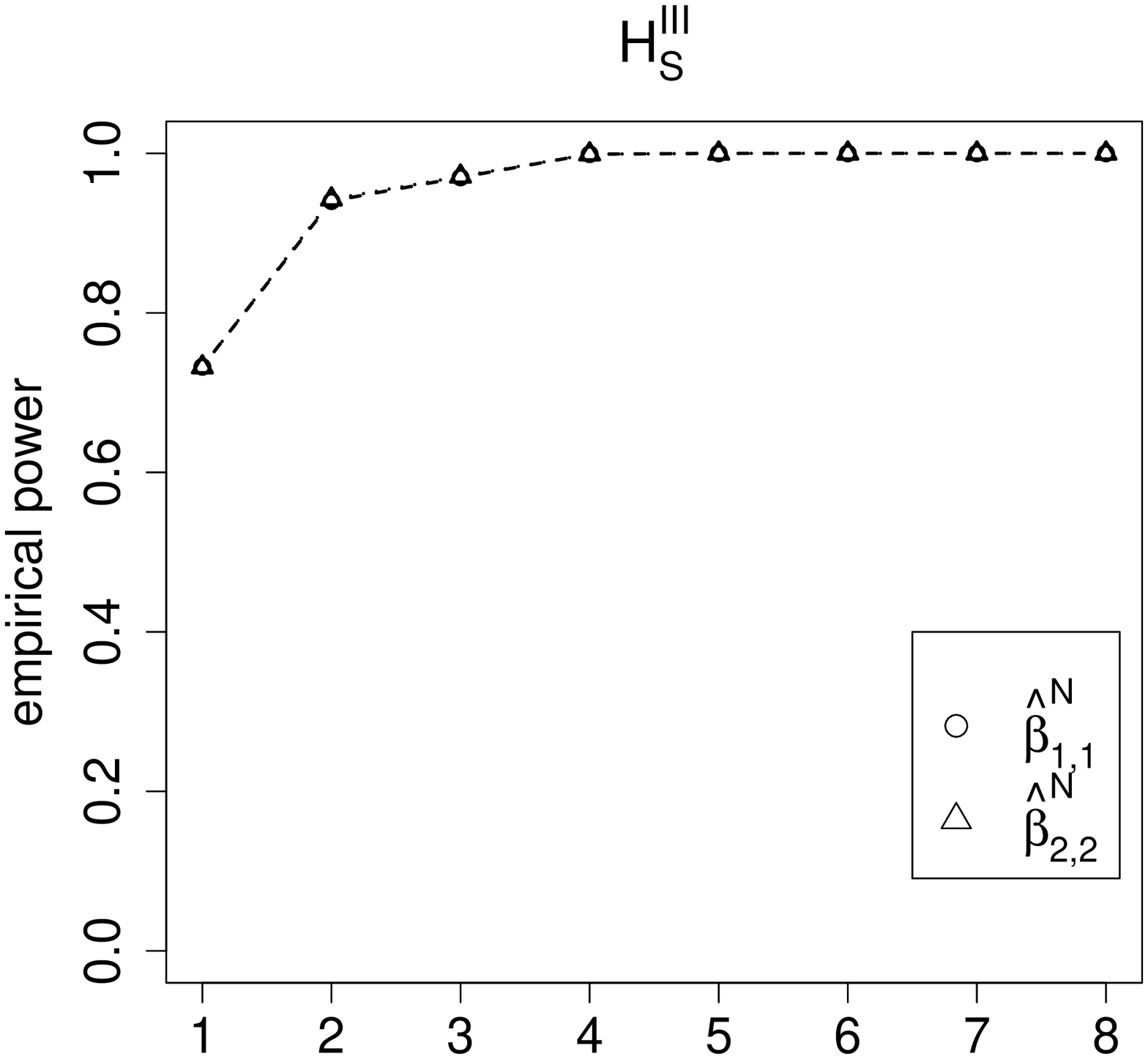} }}
\rotatebox{-90}{ \resizebox{2. in}{!}{\includegraphics{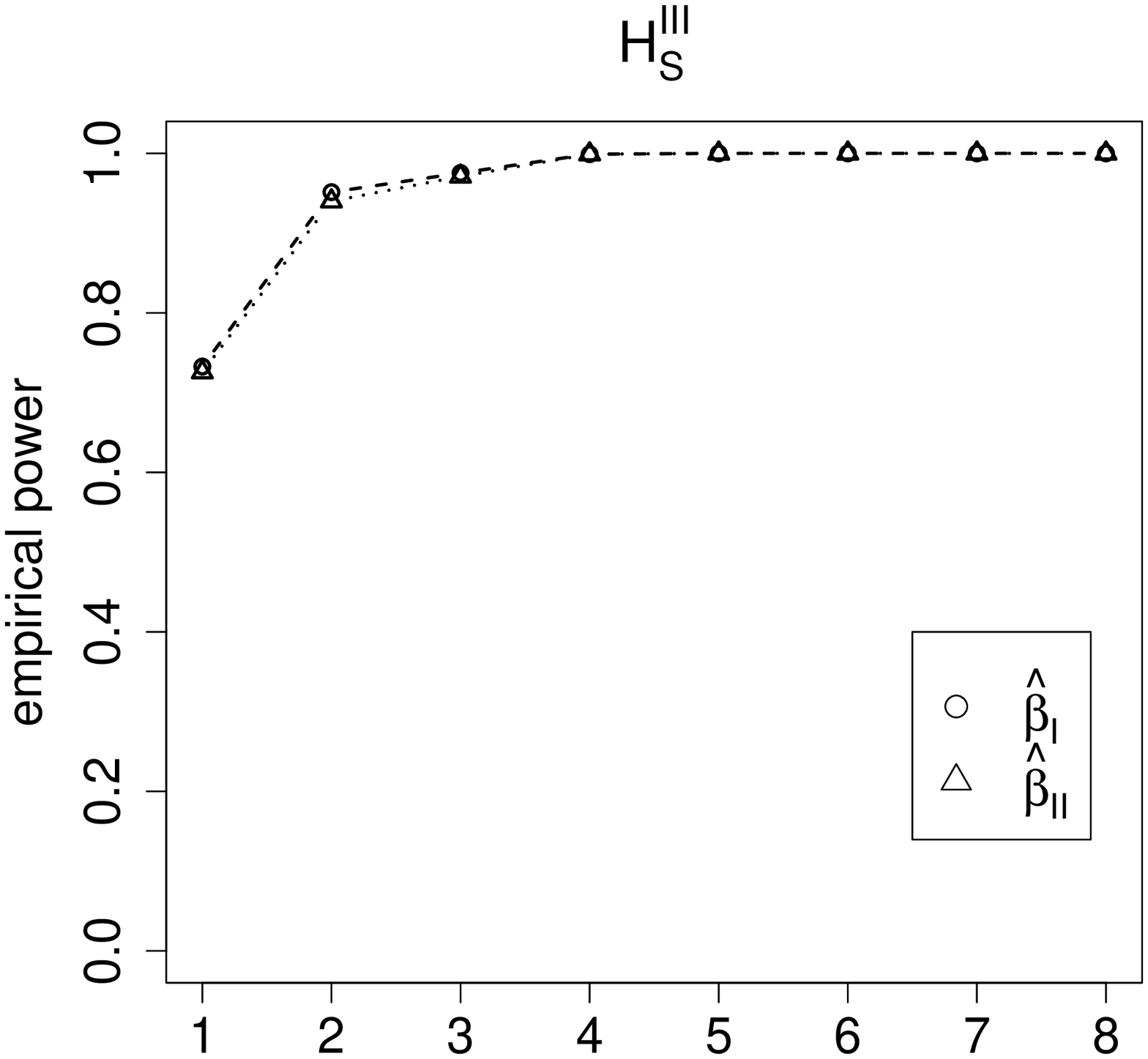} }}
\caption{
\label{fig:power-seg-TS}
The empirical power estimates for Dixon's and Ceyhan's cell-specific tests
for cell $(i,i)$, $i=1,2$ and the new directional tests
under the segregation alternatives in the two-class case for the two-sided alternatives.
$\bh^D_{i,i}$ and $\bh^N_{i,i}$ stand for Dixon's and Ceyhan's cell-specific tests for cell $(i,i)$ $i=1,2$,
respectively,
and $\bh_I$ and $\bh_{II}$ stand for the versions I and II of the new directional tests of segregation.
The horizontal axis labels are
1=(10,10), 2=(10,30), 3=(10,50), 4=(30,30), 5=(30,50),
6=(50,50), 7=(50,100), 8=(100,100).
}
\end{figure}

\begin{figure}[hbp]
\centering
\rotatebox{-90}{ \resizebox{2. in}{!}{\includegraphics{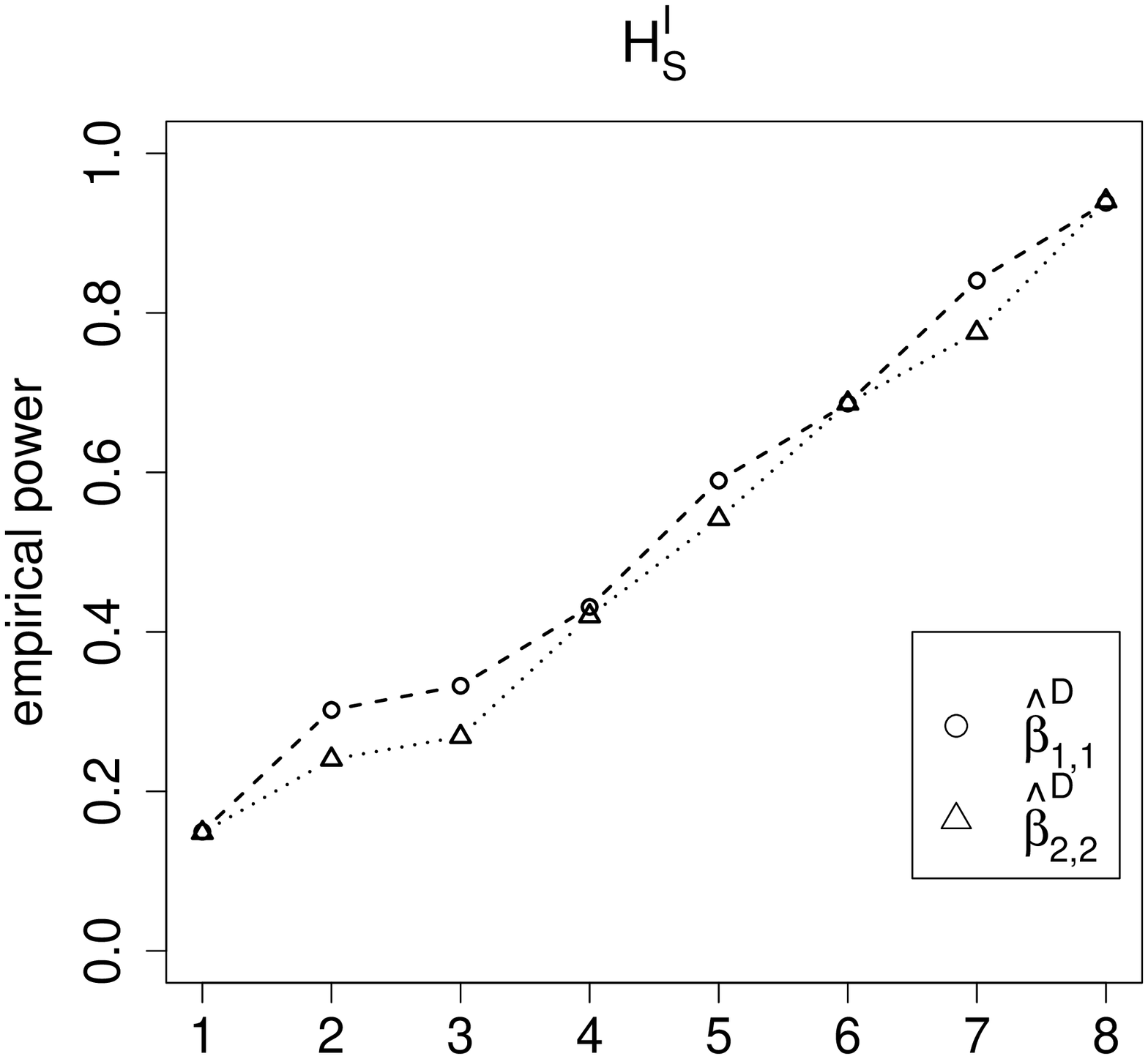} }}
\rotatebox{-90}{ \resizebox{2. in}{!}{\includegraphics{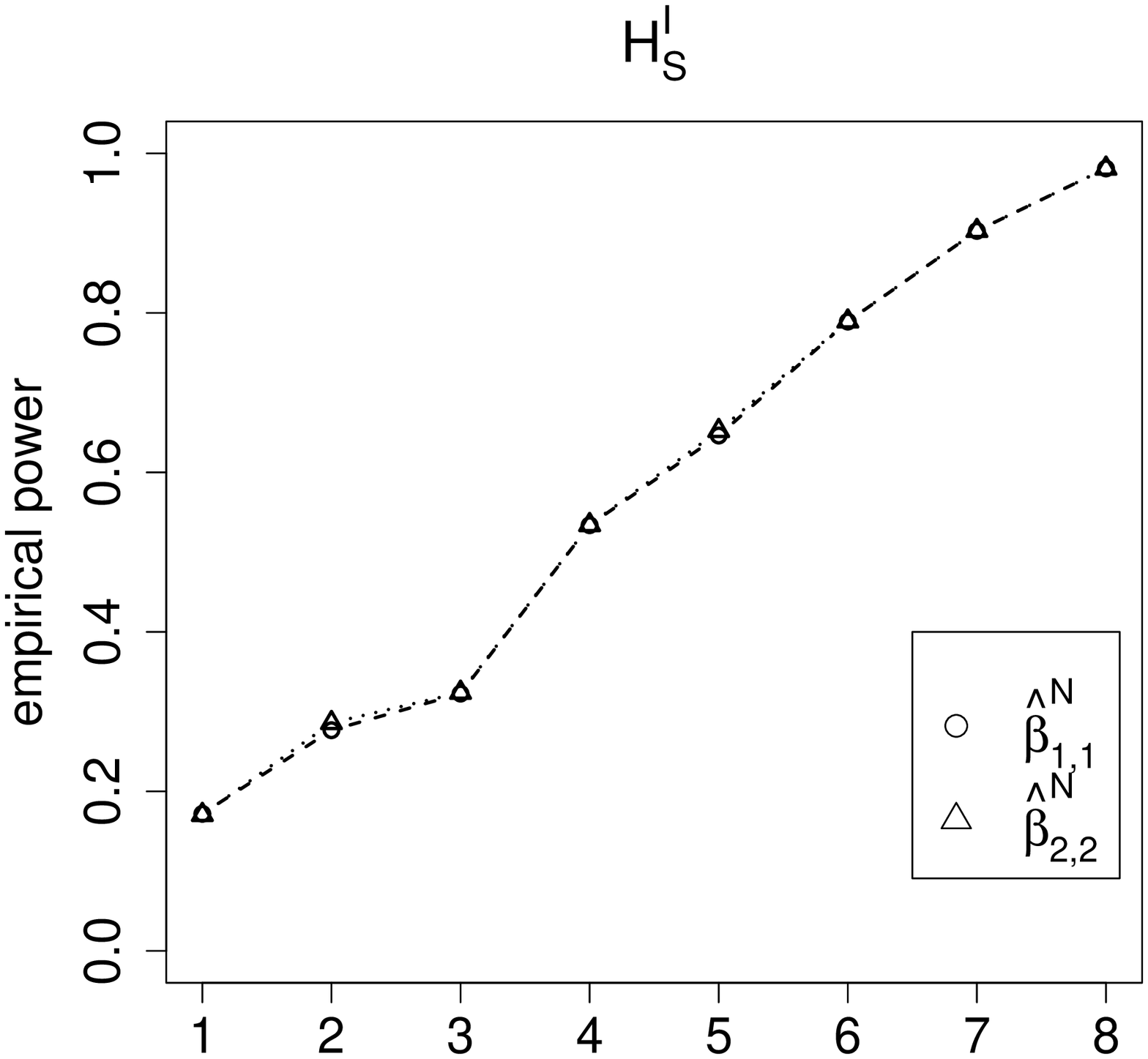} }}
\rotatebox{-90}{ \resizebox{2. in}{!}{\includegraphics{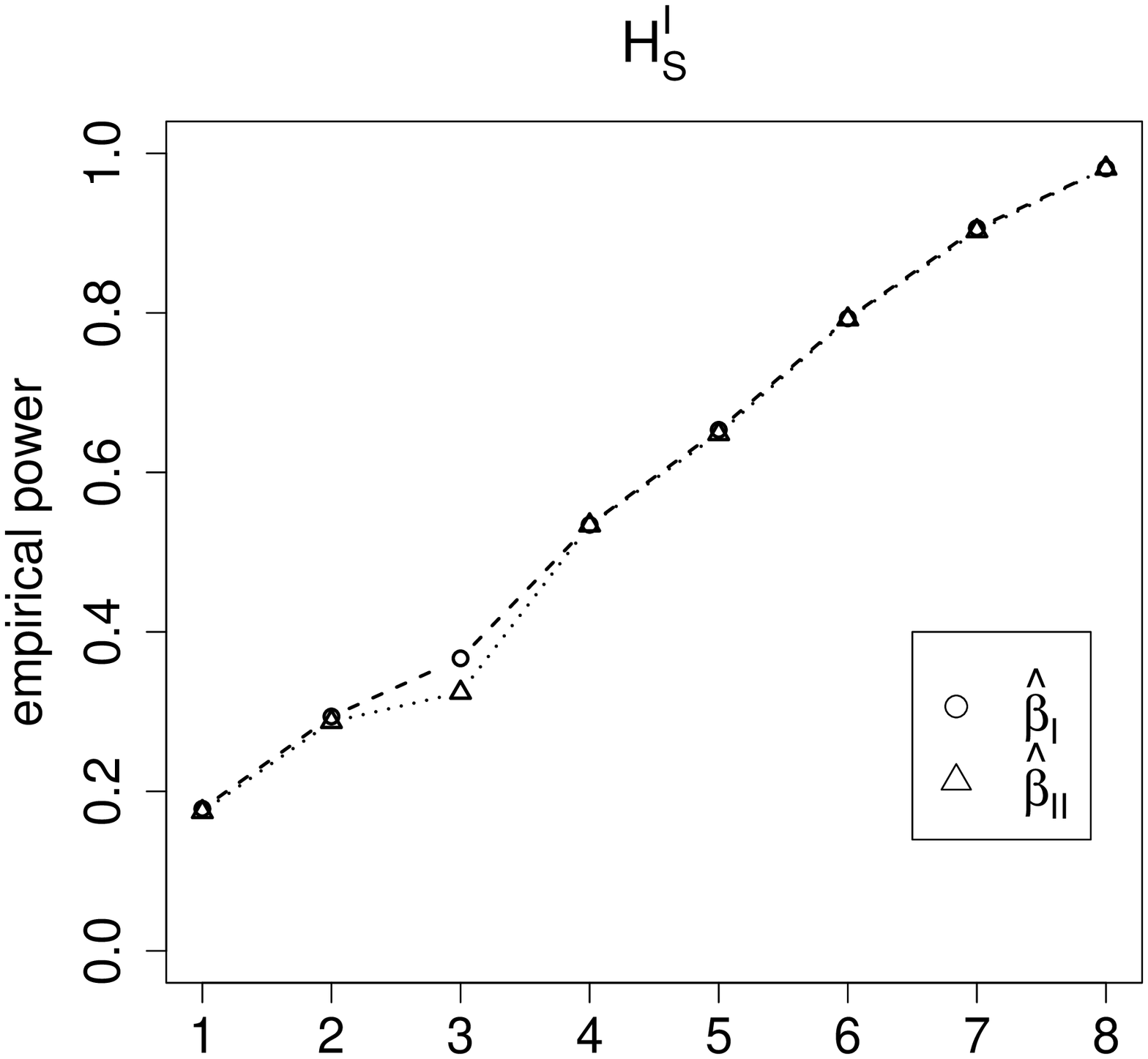} }}
\rotatebox{-90}{ \resizebox{2. in}{!}{\includegraphics{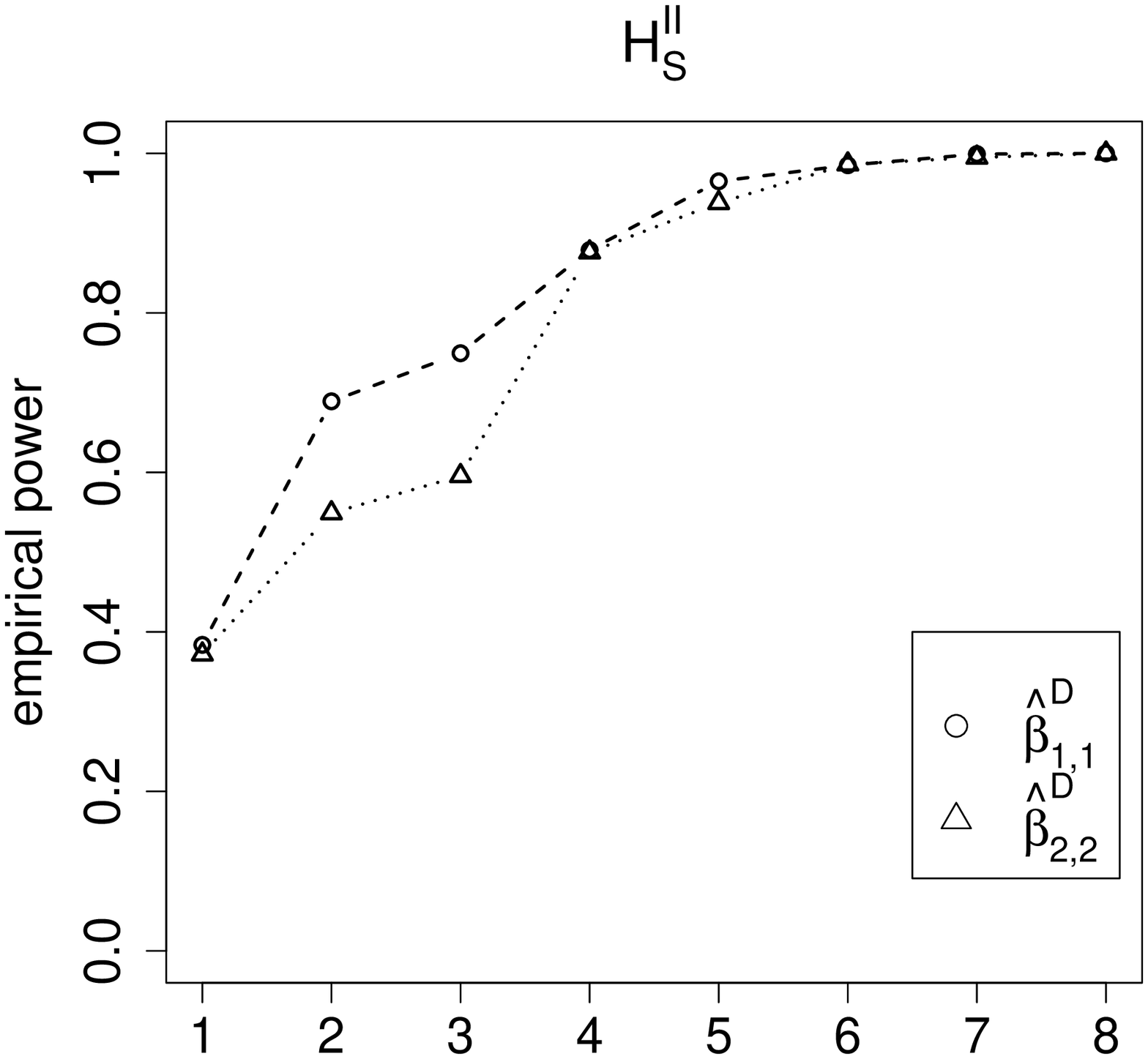} }}
\rotatebox{-90}{ \resizebox{2. in}{!}{\includegraphics{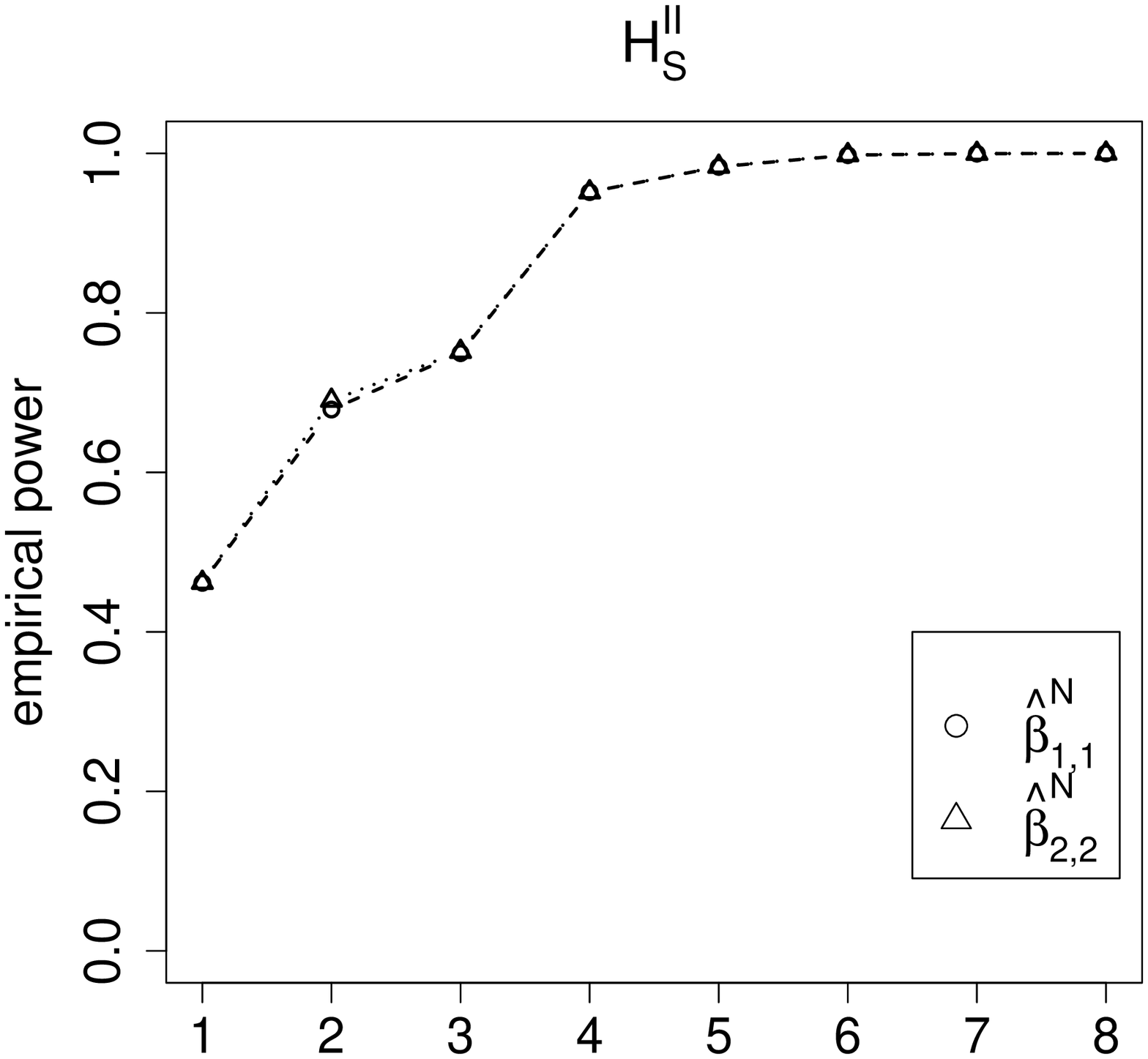} }}
\rotatebox{-90}{ \resizebox{2. in}{!}{\includegraphics{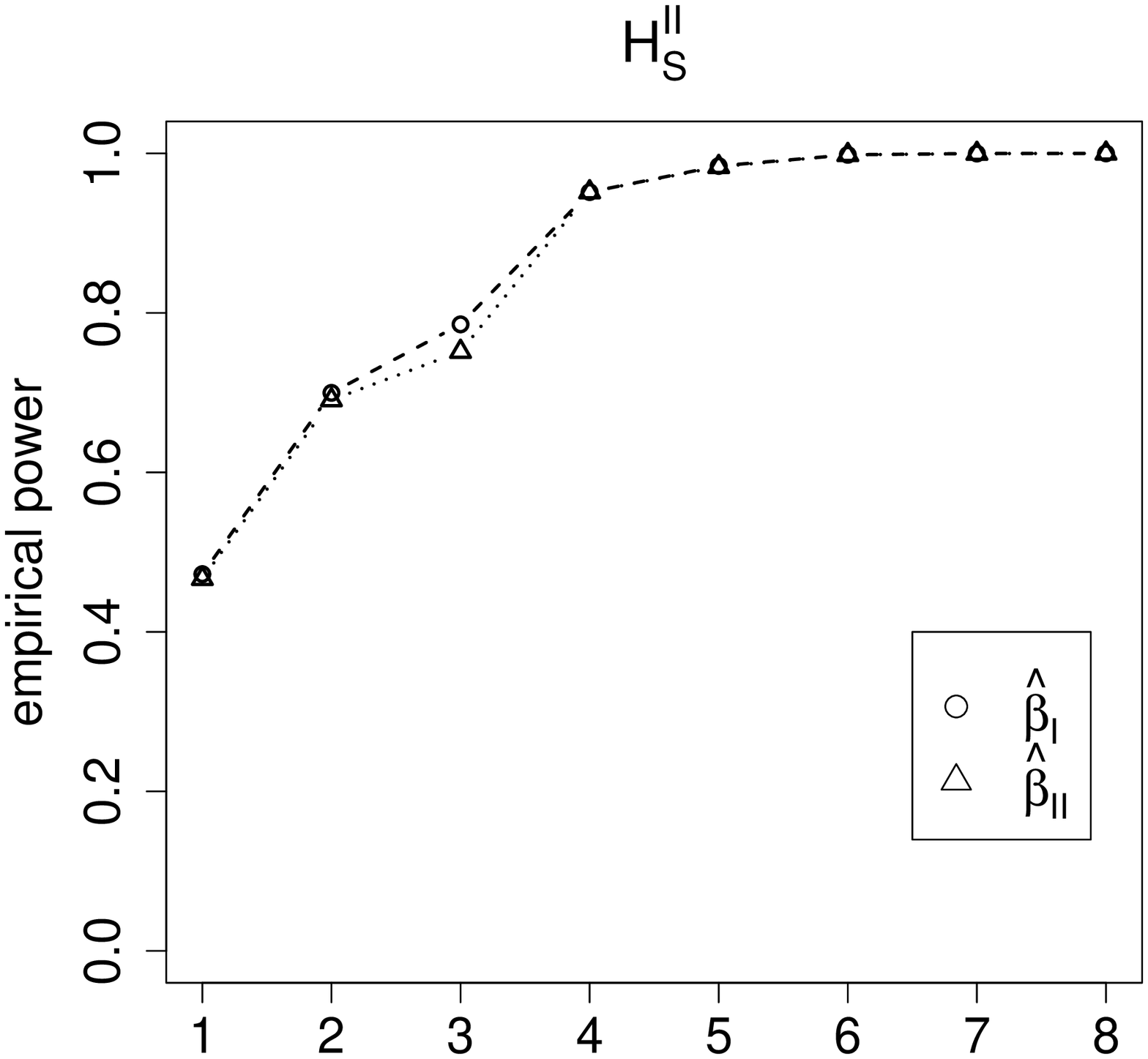} }}
\rotatebox{-90}{ \resizebox{2. in}{!}{\includegraphics{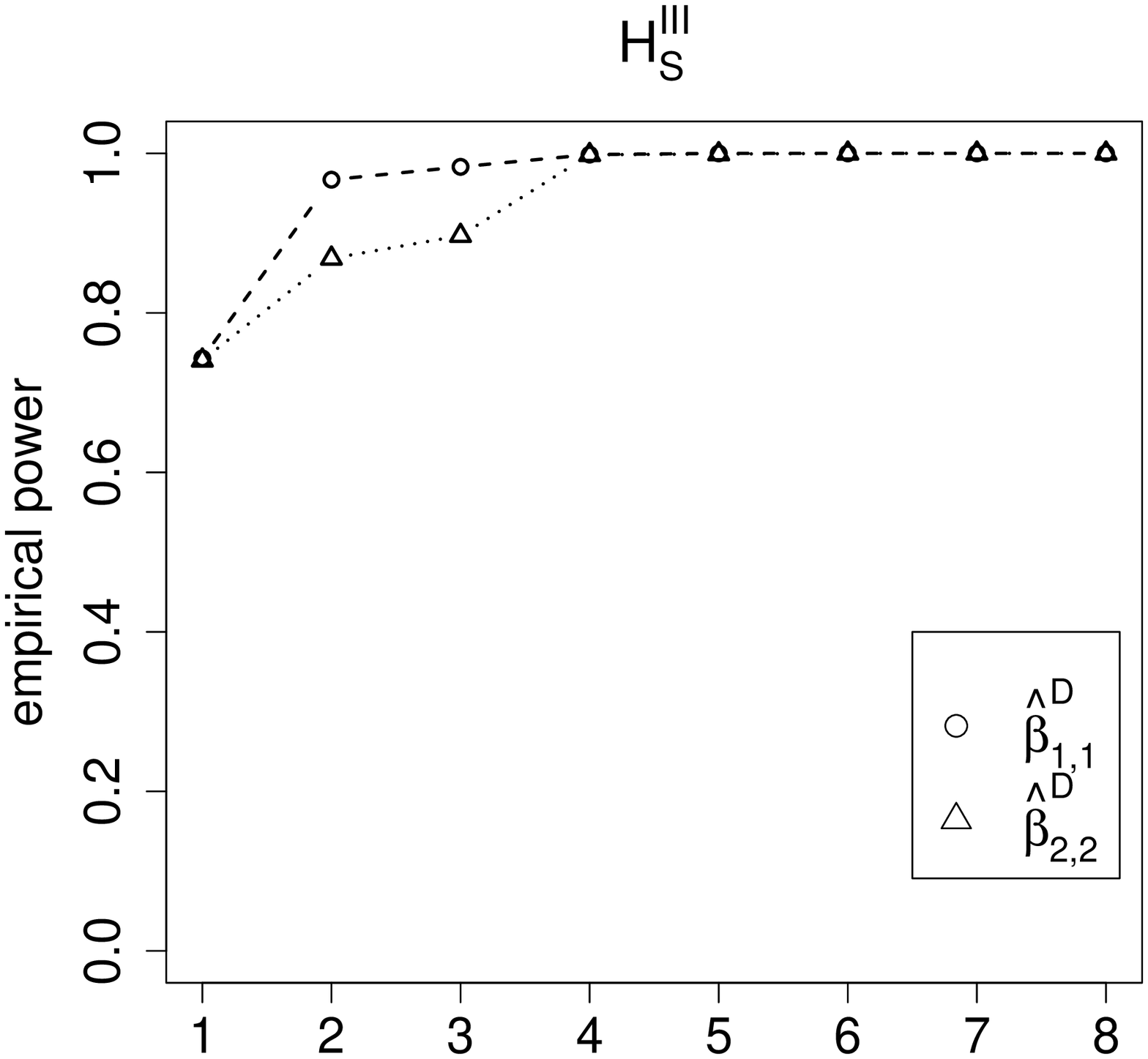} }}
\rotatebox{-90}{ \resizebox{2. in}{!}{\includegraphics{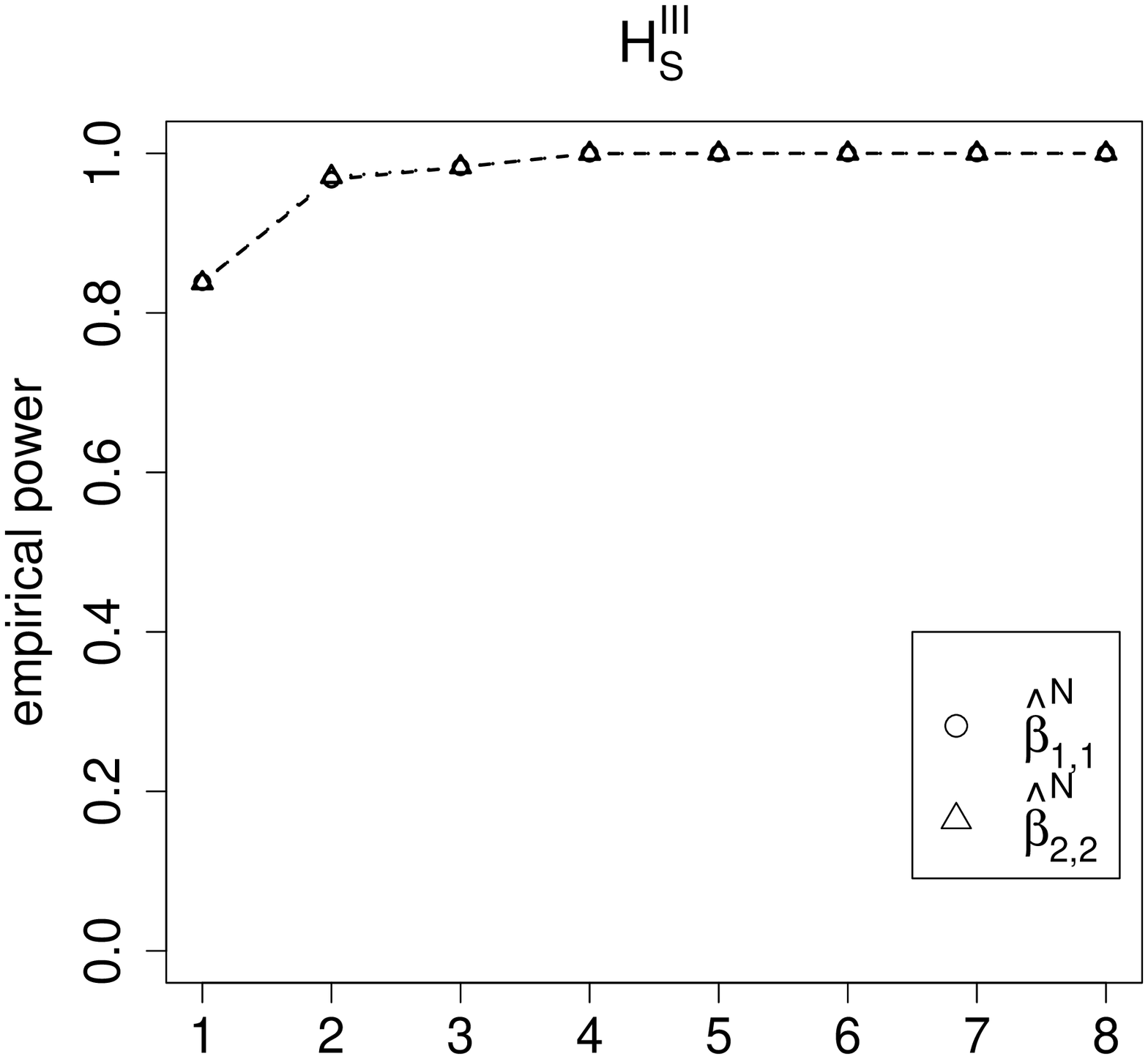} }}
\rotatebox{-90}{ \resizebox{2. in}{!}{\includegraphics{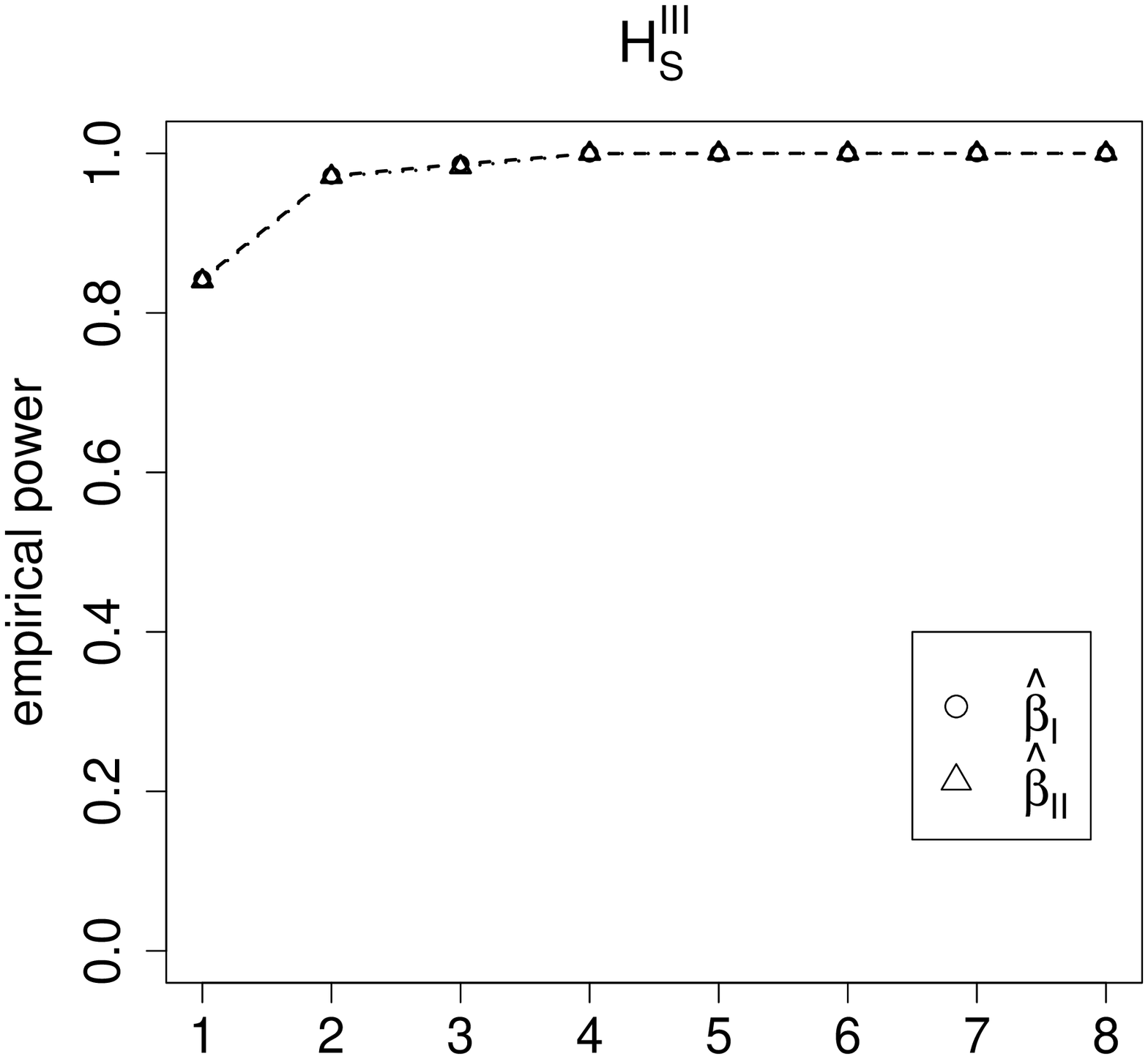} }}
\caption{
\label{fig:power-seg-RS}
The empirical power estimates for Dixon's and Ceyhan's cell-specific tests
for cell $(i,i)$, $i=1,2$ and the new directional tests
under the segregation alternatives in the two-class case for the right-sided alternatives
(which is sensitive for the segregation pattern).
The power and horizontal axis labeling is as in Figure \ref{fig:power-seg-TS}.
}
\end{figure}

The power estimates for the two-sided versions and right-sided versions under segregation
alternatives are presented in Tables \ref{tab:emp-power-seg-2s} and  \ref{tab:emp-power-seg-rt},
and plotted in Figures \ref{fig:power-seg-TS} and  \ref{fig:power-seg-RS},
respectively, where $\bh_{i,i}^D$ and $\bh_{i,i}^C$ are the
empirical power estimates for Dixon's and Ceyhan's cell-specific tests for cell $(i,i)$, for $i=1,2$
and $\bh_I$ and $\bh_{II}$ are for the new directional tests.
We omit the power estimates of the tests for the left-sided alternative
under the segregation alternatives as they are virtually zero.
Observe that, for all directional tests,
as $n=(n_1+n_2)$ gets larger, the power estimates get
larger under each segregation alternative;
for the same $n=(n_1+n_2)$ values, the power estimate is
larger for classes with similar sample sizes;
and as the segregation gets stronger, the power estimates
get larger at each $(n_1,n_2)$ combination.
The power estimates for the right-sided tests are
all higher than their corresponding two-sided estimates (as expected).
The power estimates for Ceyhan's cell-specific tests and the
new versions of the directional tests are similar and are higher than
those for Dixon's cell-specific tests.
Furthermore, version I of the new directional tests seems to have
the highest power estimates.

Considering the empirical significance levels and power estimates,
we recommend the version I of the new directional tests ($Z_I$) in the right-sided form
when testing against the segregation alternatives,
as it is at the desired level for similar sample sizes,
slightly conservative for very different sample sizes,
but have higher power for each sample size combination.
On the other hand, $Z_I$ is a conditional test (conditional on column sums),
while $Z_{II}$ is unconditional and the
empirical size and power estimates are about the same as $Z_I$.
Hence $Z_{II}$ can also be used instead.

\begin{table}[ht]
\centering
\begin{tabular}{|c|c||c|c|c|c|c|c|}
\hline
\multicolumn{8}{|c|}{Empirical power estimates for the two-sided tests} \\
\multicolumn{8}{|c|}{under the segregation alternatives} \\
\hline
& $(n_1,n_2)$  & $\bh_{1,1}^D$ & $\bh_{2,2}^D$ & $\bh_{1,1}^C$ & $\bh_{2,2}^C$ & $\bh_I$ & $\bh_{II}$ \\
\hline
\hline
 & $(10,10)$ & .0734 & .0698 & .1068 & .1060  & .1108  & .1026 \\
\cline{2-8}
 & $(10,30)$ & .1436 & .1540  & .1977  & 2019  & .2173  & .1973 \\
\cline{2-8}
 & $(10,50)$ & .1639 & .1615  & .2465  & .2491  & .2835  & .2490 \\
\cline{2-8}
 & $(30,30)$ & .2883 & .2783  & .3898  & .3894  & .3932  & .3837 \\
\cline{2-8} \raisebox{4.ex}[0pt]{$H_S^I$}
 & $(50,10)$ & .1636 & .1520  & .2395  & .2367  & .2741  & .2395 \\
\cline{2-8}
 & $(50,50)$ & .5091 & .5016  & .6786  & .6793  & .6847  & .6831 \\
\hline
\hline
 & $(10,10)$ & .2057 & .2044  & .3280  & .3270  & .3322  & .3203 \\
\cline{2-8}
 & $(10,30)$ & .4601 & .4133  & .5725  & .5793  & .6082  & .5724 \\
\cline{2-8}
 & $(10,50)$ & .5420 & .4477  & .6747  & .6794  & .7129  & .6793 \\
\cline{2-8}
 & $(30,30)$ & .7783 & .7769  & .8939  & .8938  & .8946  & .8917 \\
\cline{2-8}\raisebox{4.ex}[0pt]{$H_S^{II}$}
 & $(50,10)$ & .4453 & .5383  & .6756  & .6721  & .7116  & .6754 \\
\cline{2-8}
 & $(50,50)$ & .9543 & .9551  & .9938  & .9936  & .9938  & .9938 \\
\hline
\hline
 & $(10,10)$ & .5144 & .5121  & .7324  & .7320  & .7327  & .7257 \\
\cline{2-8}
 & $(10,30)$ & .8873 & .7833  & .9402  & .9425  & .9514  & .9400 \\
\cline{2-8}
 & $(10,50)$ & .9353 & .8002  & .9699  & .9711  & .9754  & .9711 \\
\cline{2-8}
 & $(30,30)$ & .9929 & .9915  & .9990  & .9990  & .9990  & .9990 \\
\cline{2-8} \raisebox{4.ex}[0pt]{$H_S^{III}$}
 & $(50,10)$ & .7989 & .9393  & .9720  & .9712  & .9772  & .9720 \\
\cline{2-8}
 & $(50,50)$ & .9999 & 1.000 & 1.000 & 1.000 & 1.000 & 1.000  \\
\hline
\end{tabular}
\caption{
\label{tab:emp-power-seg-2s}
The empirical power estimates for the two-sided tests under the segregation alternatives,
$H_S^I$, $H_S^{II}$, and $H_S^{III}$ for the two-class case
with $N_{mc}=10000$, for some combinations of $n_1,n_2 \in
\{10,30,50\}$ at $\alpha=.05$.}
\end{table}

\begin{table}[ht]
\centering
\begin{tabular}{|c||c|c|c|c|c|c|c|}
\hline
\multicolumn{8}{|c|}{Empirical power estimates for the right-sided tests} \\
\multicolumn{8}{|c|}{under the segregation alternatives} \\
\hline
& $(n_1,n_2)$  & $\bh_{1,1}^D$ & $\bh_{2,2}^D$ & $\bh_{1,1}^C$ & $\bh_{2,2}^C$ & $\bh_I$ & $\bh_{II}$ \\
\hline
\hline
 & $(10,10)$ & .1495 & .1482  & .1719  & .1707  & .1782  & .1743 \\
\cline{2-8}
 & $(10,30)$ & .3021 & .2401  & .2767  & .2859  & .2939  & .2873 \\
\cline{2-8}
 & $(10,50)$ & .3324 & .2686  & .3226  & .3239  & .3668  & .3239 \\
\cline{2-8}
 & $(30,30)$ & .4313 & .4199  & .5338  & .5342  & .5343  & .5338 \\
\cline{2-8}\raisebox{4.ex}[0pt]{$H_S^I$}
 & $(50,10)$ & .2755 & .3332  & .3226  & .3220  & .3647  & .3226 \\
\cline{2-8}
 & $(50,50)$ & .6863 & .6866  & .7891  & .7895  & .7931  & .7920 \\
\hline
\hline
 & $(10,10)$ & .3837 & .3719  & .4615  & .4618  & .4724  & .4666 \\
\cline{2-8}
 & $(10,30)$ & .6891 & .5490  & .6788  & .6901  & .6997  & .6908 \\
\cline{2-8}
 & $(10,50)$ & .7494 & .5956  & .7495  & .7514  & .7856  & .7514 \\
\cline{2-8}
 & $(30,30)$ & .8788 & .8761  & .9515  & .9515  & .9515  & .9515 \\
\cline{2-8}\raisebox{4.ex}[0pt]{$H_S^{II}$}
 & $(50,10)$ & .5995 & .7448  & .7480  & .7454  & .7880  & .7480 \\
\cline{2-8}
 & $(50,50)$ & .9851 & .9864  & .9978  & .9981  & .9982  & .9982 \\
\hline
\hline
 & $(10,10)$ & .7429  & .7405  & .8386  & .8375  & .8424  & .8399 \\
\cline{2-8}
 & $(10,30)$ & .9670 & .8683  & .9672  & .9706   & .9719  & .9708 \\
\cline{2-8}
 & $(10,50)$ & .9831 & .8968  & .9827  & .9829   & .9866  & .9829 \\
\cline{2-8}
 & $(30,30)$ & .9985 & .9982  & .9997  & .9997   & .9997  & .9997 \\
\cline{2-8}\raisebox{4.ex}[0pt]{$H_S^{III}$}
 & $(50,10)$ & .8918 & .9836  & .9850  & .9842   & .9893  & .9850 \\
\cline{2-8}
 & $(50,50)$ & 1.000 & 1.000 & 1.000 & 1.000 & 1.000 & 1.000 \\
\hline
\end{tabular}
\caption{
\label{tab:emp-power-seg-rt}
The empirical power estimates for the right-sided tests under the segregation alternatives,
$H_S^I$, $H_S^{II}$, and $H_S^{III}$ for the two-class case
with $N_{mc}=10000$, for some combinations of $n_1,n_2 \in
\{10,30,50\}$ at $\alpha=.05$.}
\end{table}

\subsubsection{Empirical Power Analysis under Association Alternatives}
 \label{sec:power-comp-assoc}
For the association alternatives, we consider three cases.
First, we generate $X_i \stackrel{iid}{\sim} \U((0,1)\times(0,1))$ for $i=1,2,\ldots,n_1$.
Then we generate $Y_j$ for $j=1,2,\ldots,n_2$ as follows.
For each $j$, we pick an $i$ randomly, then generate $Y_j$ as
$X_i+R_j\,(\cos T_j, \sin T_j)'$ where
$R_j \stackrel{iid}{\sim} \U(0,r)$ with $r \in (0,1)$ and
$T_j \stackrel{iid}{\sim} \U(0,2\,\pi)$.
In the pattern generated, appropriate choices of
$r$ will imply association between classes $X$ and $Y$.
That is, it will be  more likely to have $(X,Y)$ or $(Y,X)$
NN pairs than same-class NN pairs (i.e., $(X,X)$ or $(Y,Y)$).
The three values of $r$ we consider constitute
the following three association alternatives;
\begin{equation}
\label{eqn:assoc-alt}
H_A^{I}: r=1/4,\;\;\; H_A^{II}: r=1/7, \text{ and } H_A^{III}: r=1/10.
\end{equation}
Observe that, from $H_A^I$ to $H_A^{III}$,
the association gets stronger
in the sense that $X$ and $Y$ points tend to occur
together more and more frequently.
By construction, for similar sample sizes the association between $X$ and $Y$ are
at about the same degree as association between $Y$ and $X$.
For very different sample sizes,
smaller sample is associated with the larger
but the abundance of the larger sample
confounds its association with the smaller.

\begin{figure}[hbp]
\centering
\rotatebox{-90}{ \resizebox{2. in}{!}{\includegraphics{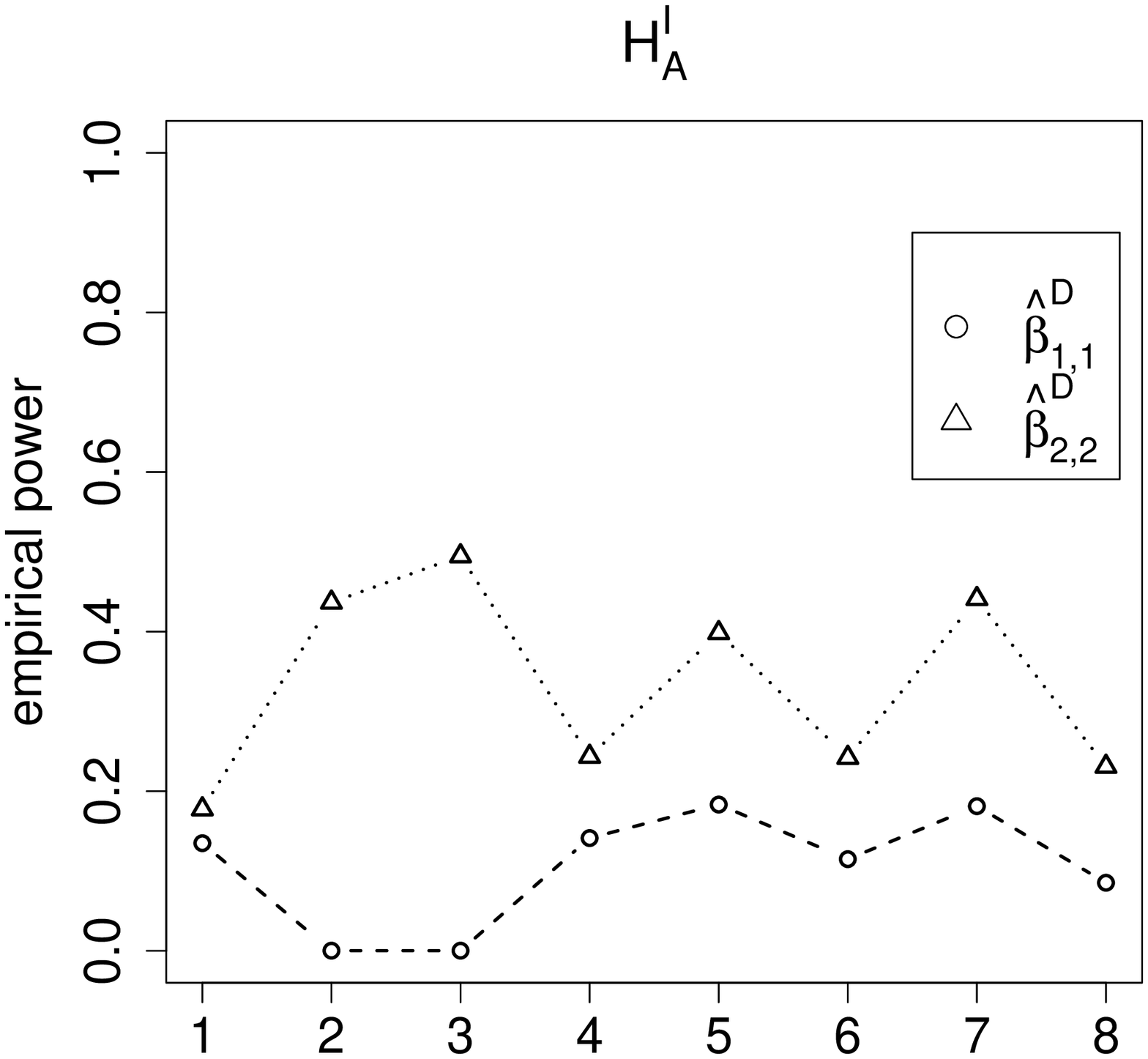} }}
\rotatebox{-90}{ \resizebox{2. in}{!}{\includegraphics{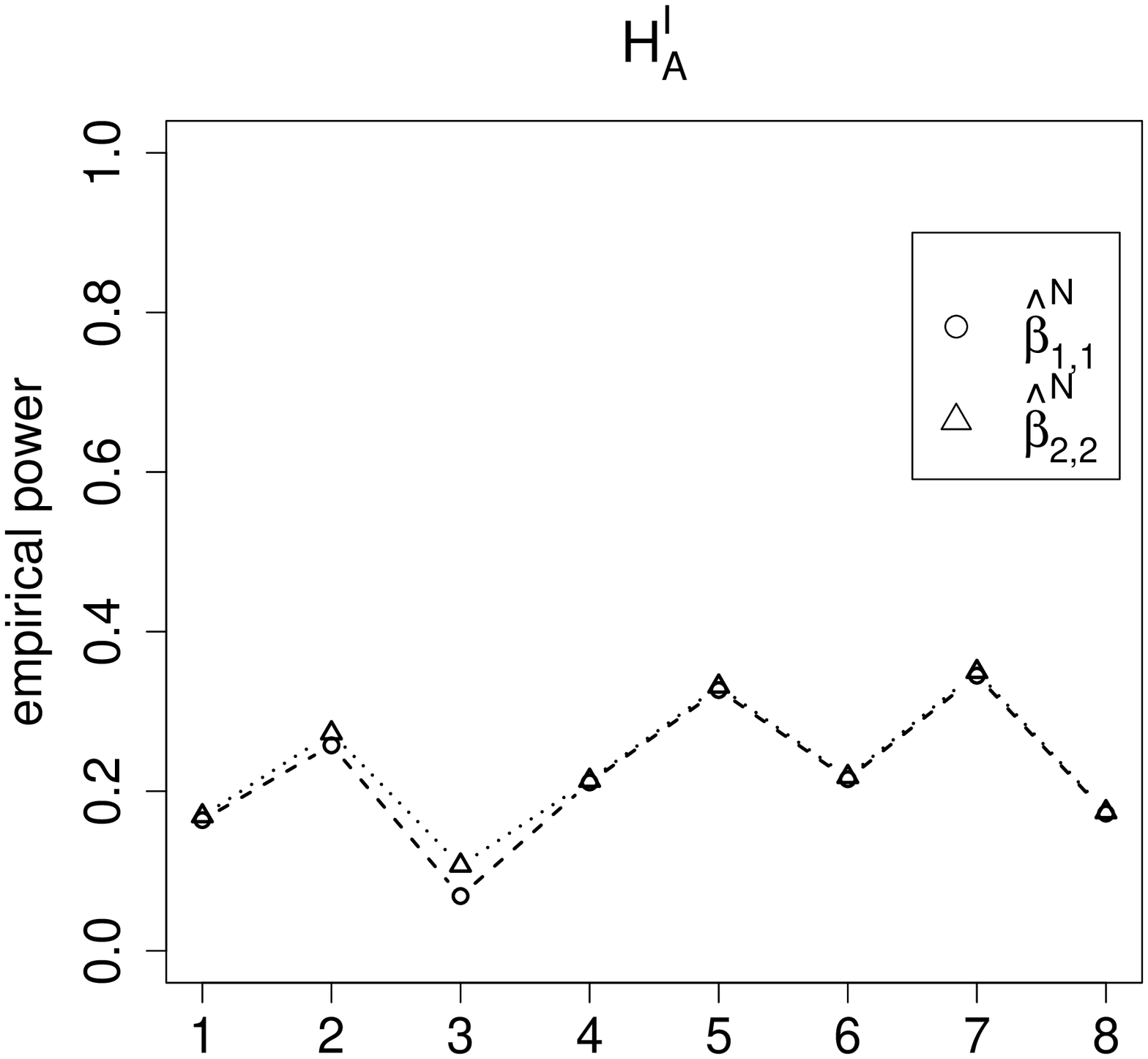} }}
\rotatebox{-90}{ \resizebox{2. in}{!}{\includegraphics{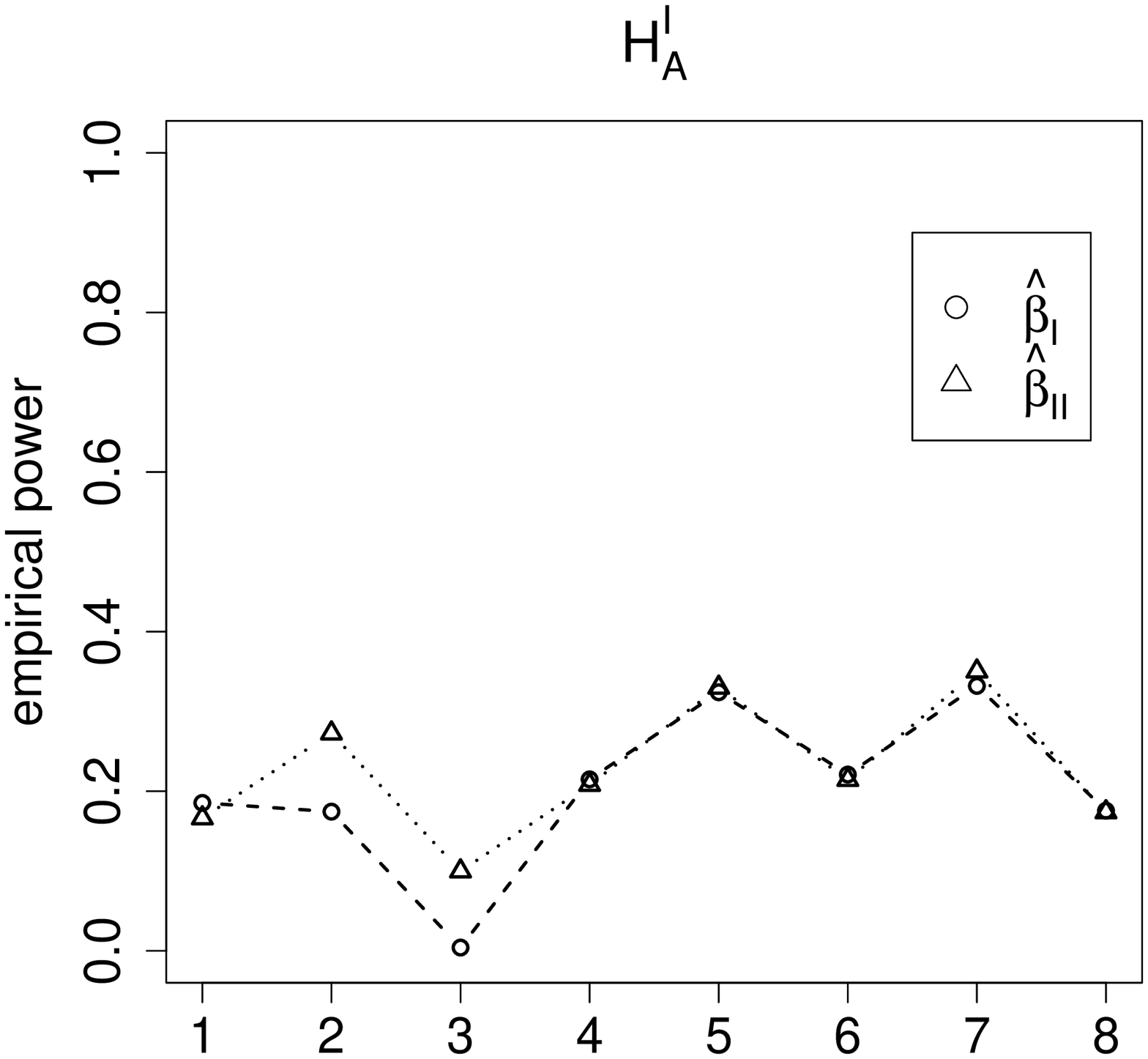} }}
\rotatebox{-90}{ \resizebox{2. in}{!}{\includegraphics{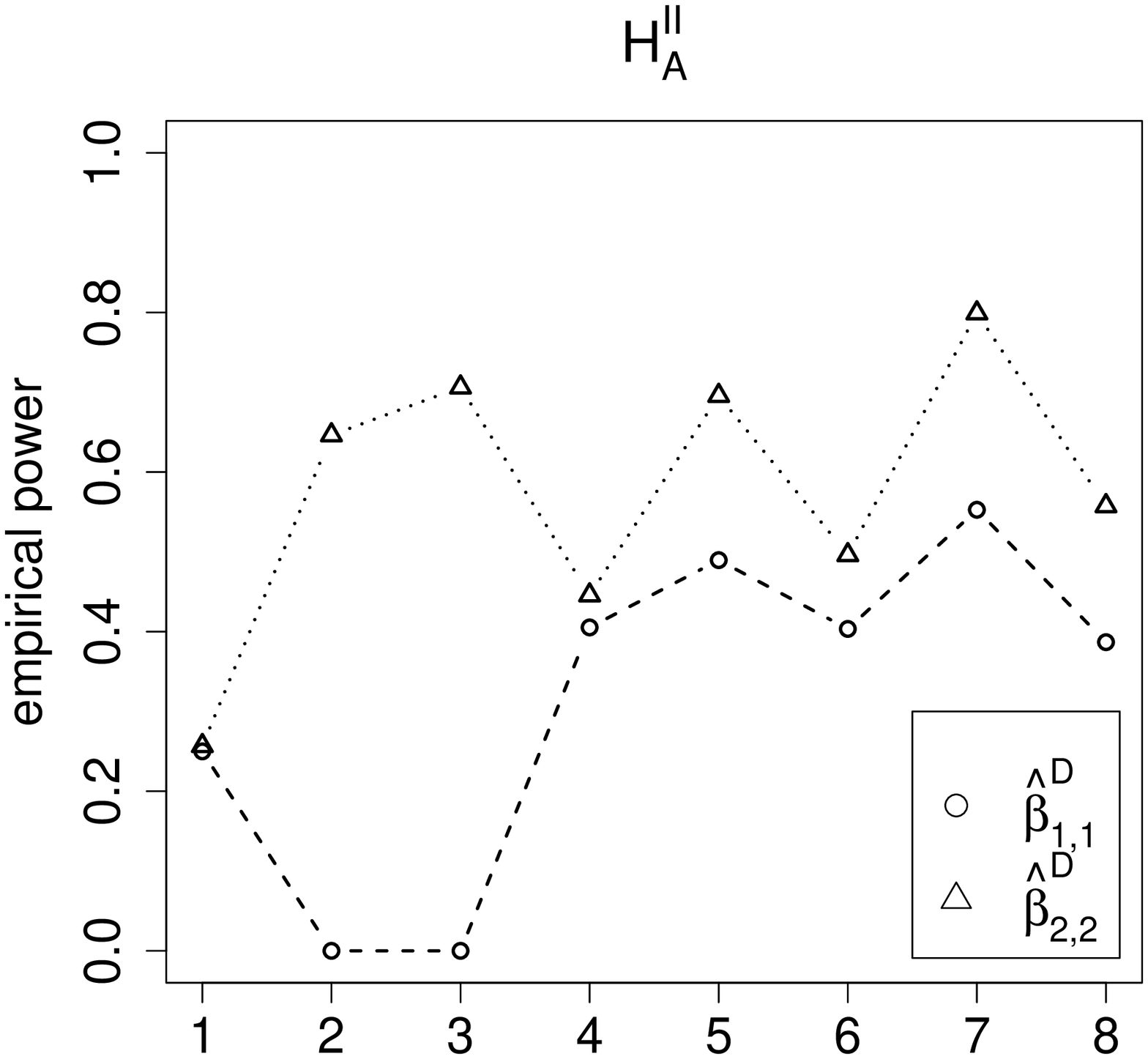} }}
\rotatebox{-90}{ \resizebox{2. in}{!}{\includegraphics{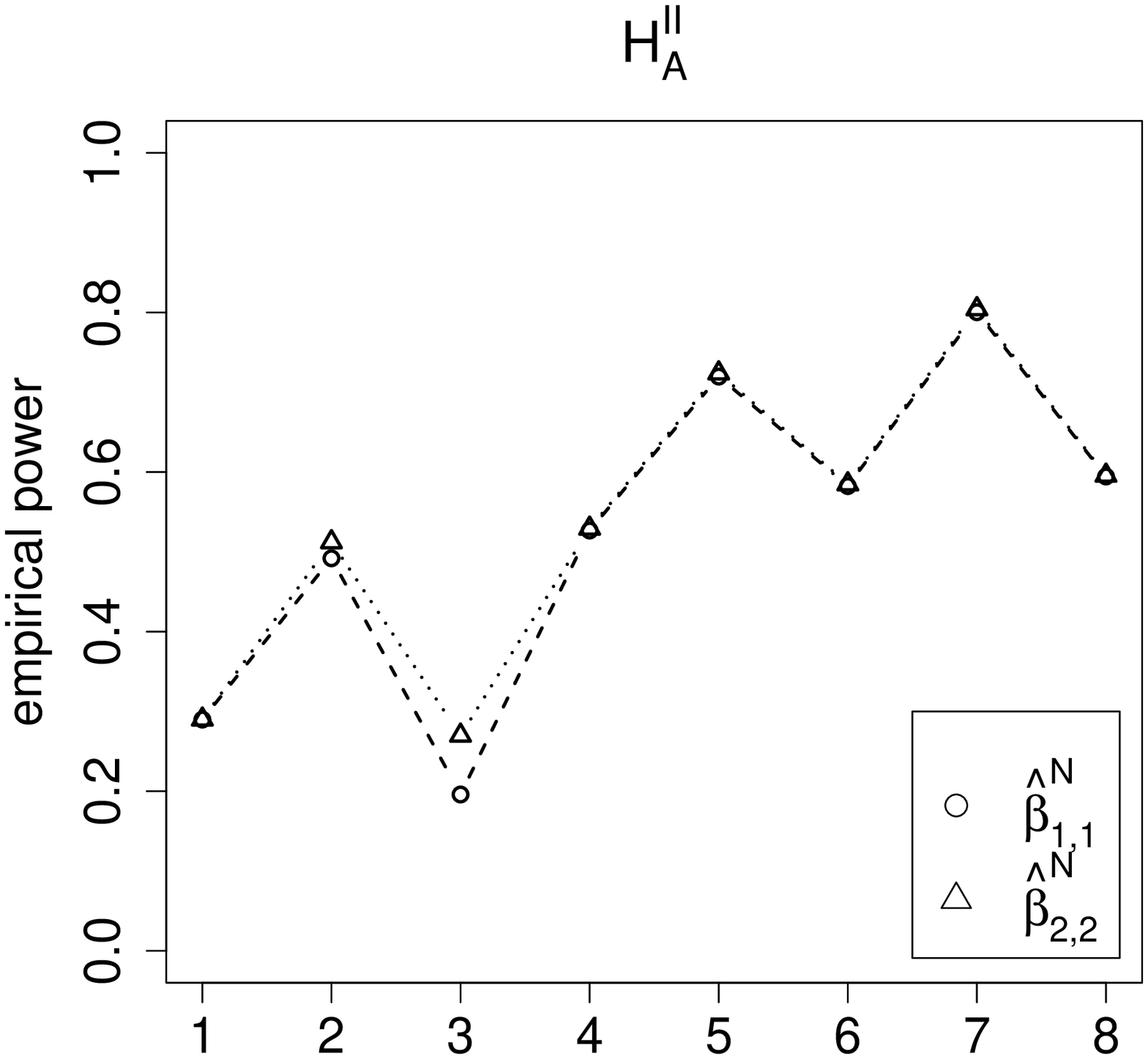} }}
\rotatebox{-90}{ \resizebox{2. in}{!}{\includegraphics{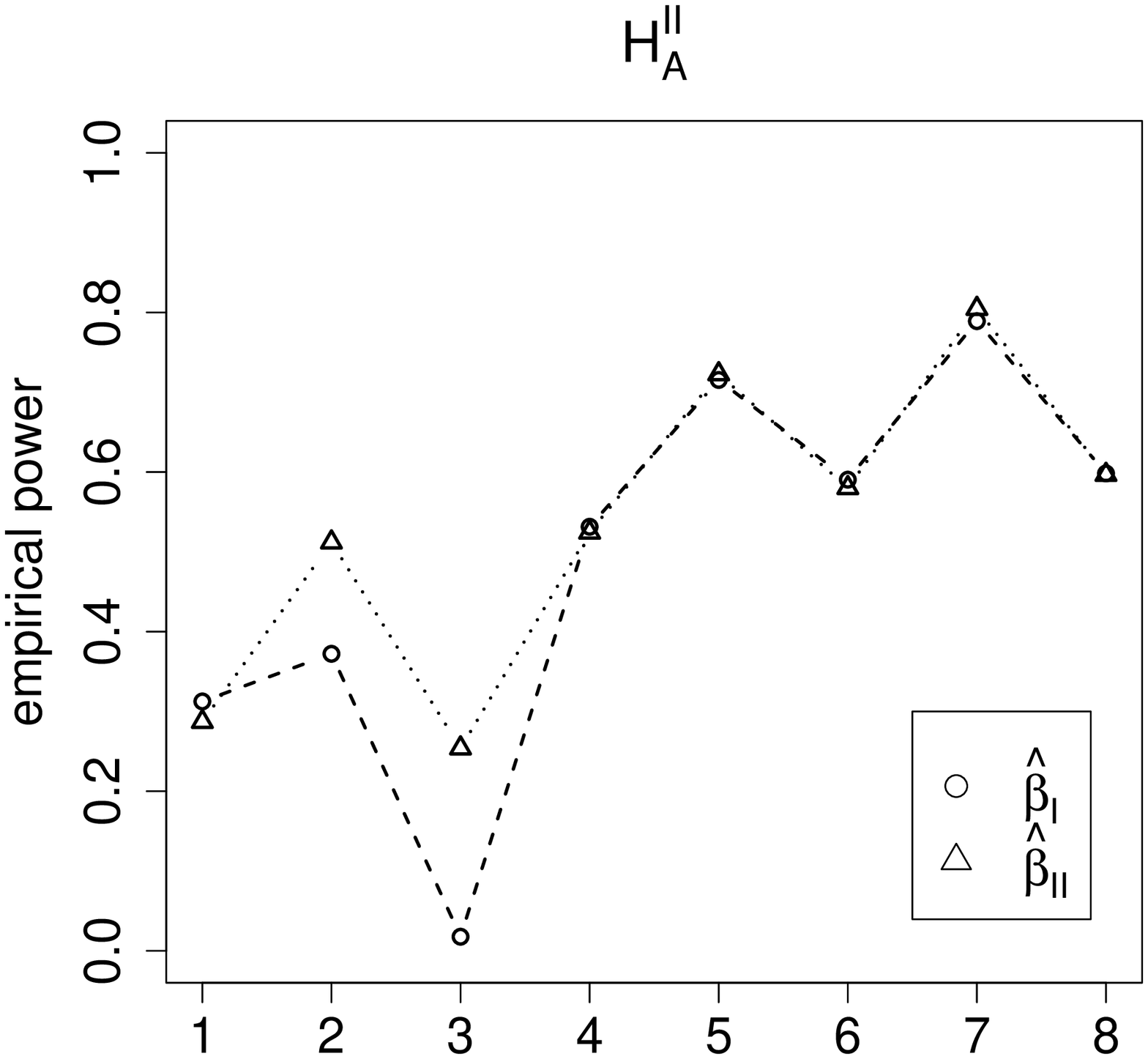} }}
\rotatebox{-90}{ \resizebox{2. in}{!}{\includegraphics{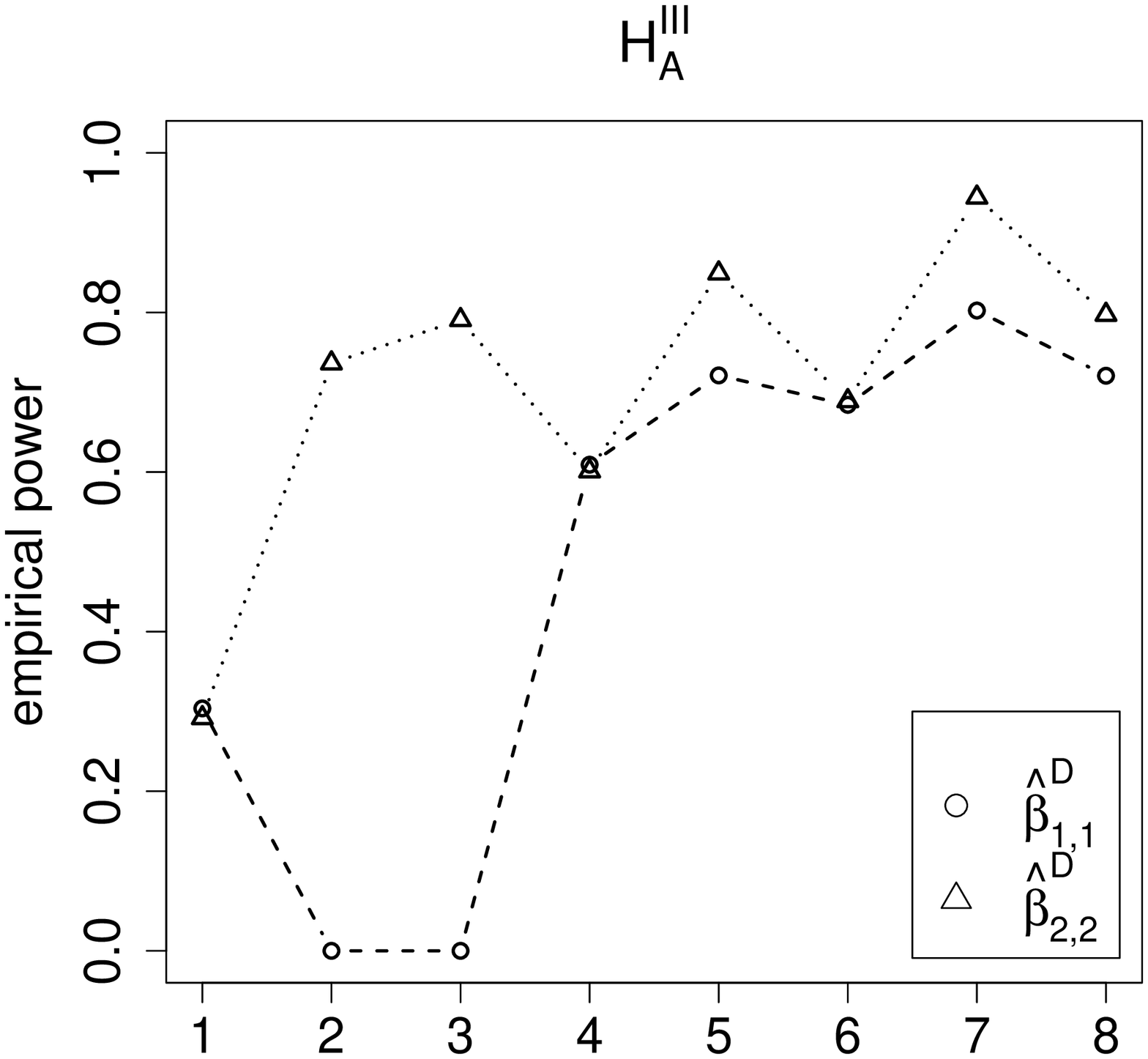} }}
\rotatebox{-90}{ \resizebox{2. in}{!}{\includegraphics{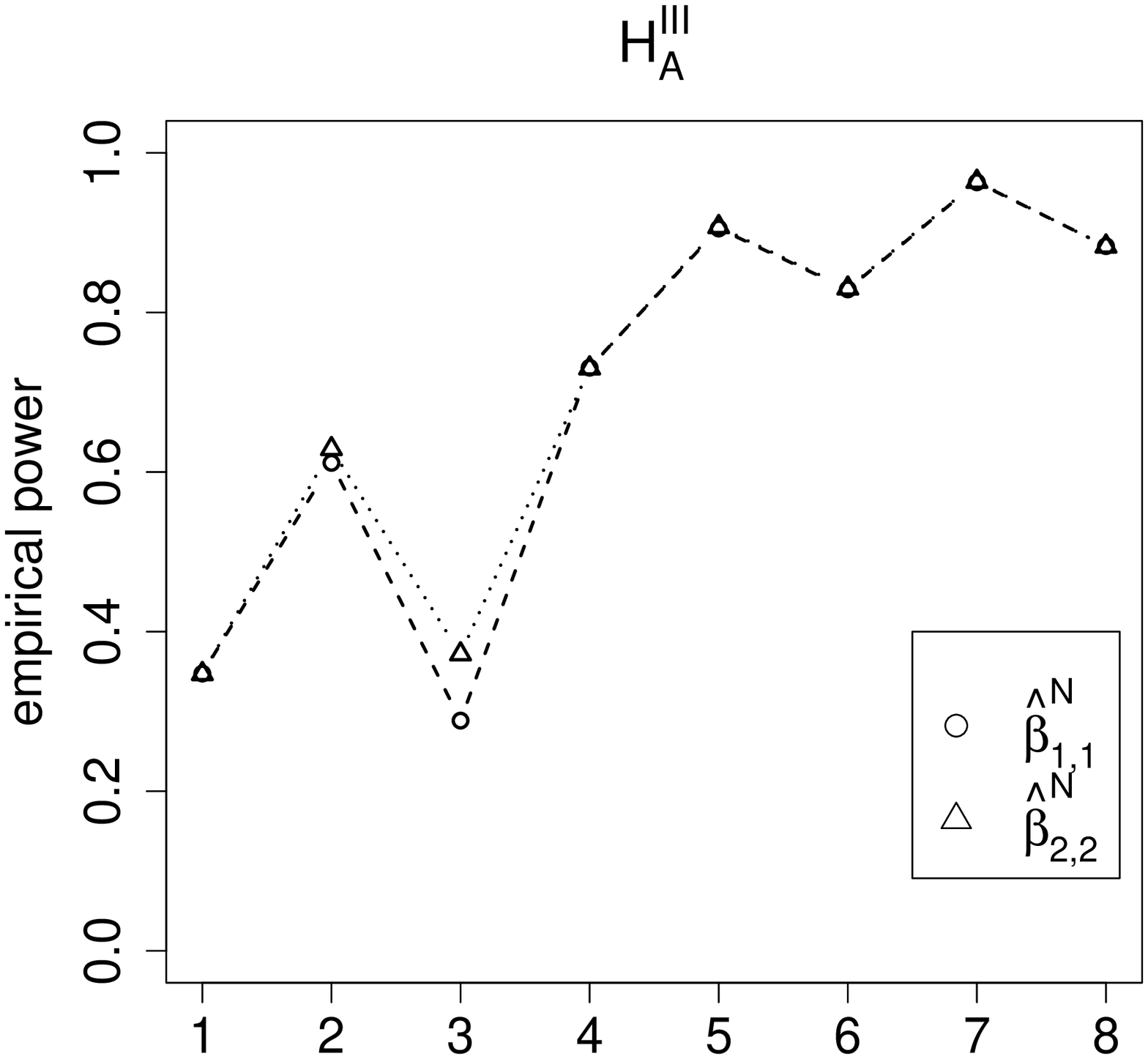} }}
\rotatebox{-90}{ \resizebox{2. in}{!}{\includegraphics{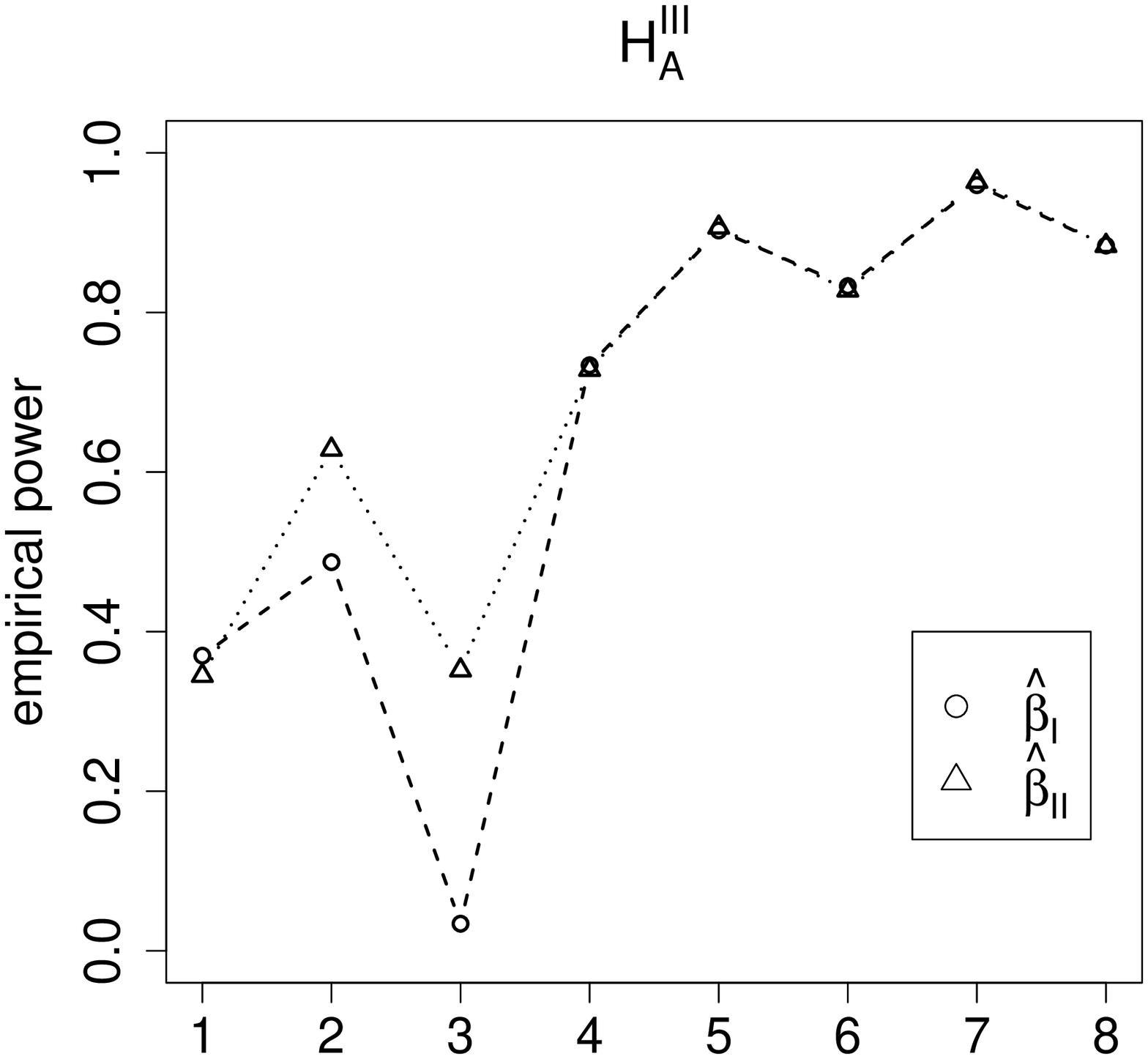} }}
\caption{
\label{fig:power-assoc-TS}
The empirical power estimates for Dixon's and Ceyhan's cell-specific tests
for cell $(i,i)$, $i=1,2$ and the new directional tests
under the association alternatives in the two-class case for the two-sided alternatives.
The power and horizontal axis labeling is as in Figure \ref{fig:power-seg-TS}.
}
\end{figure}

\begin{figure}[hbp]
\centering
\rotatebox{-90}{ \resizebox{2. in}{!}{\includegraphics{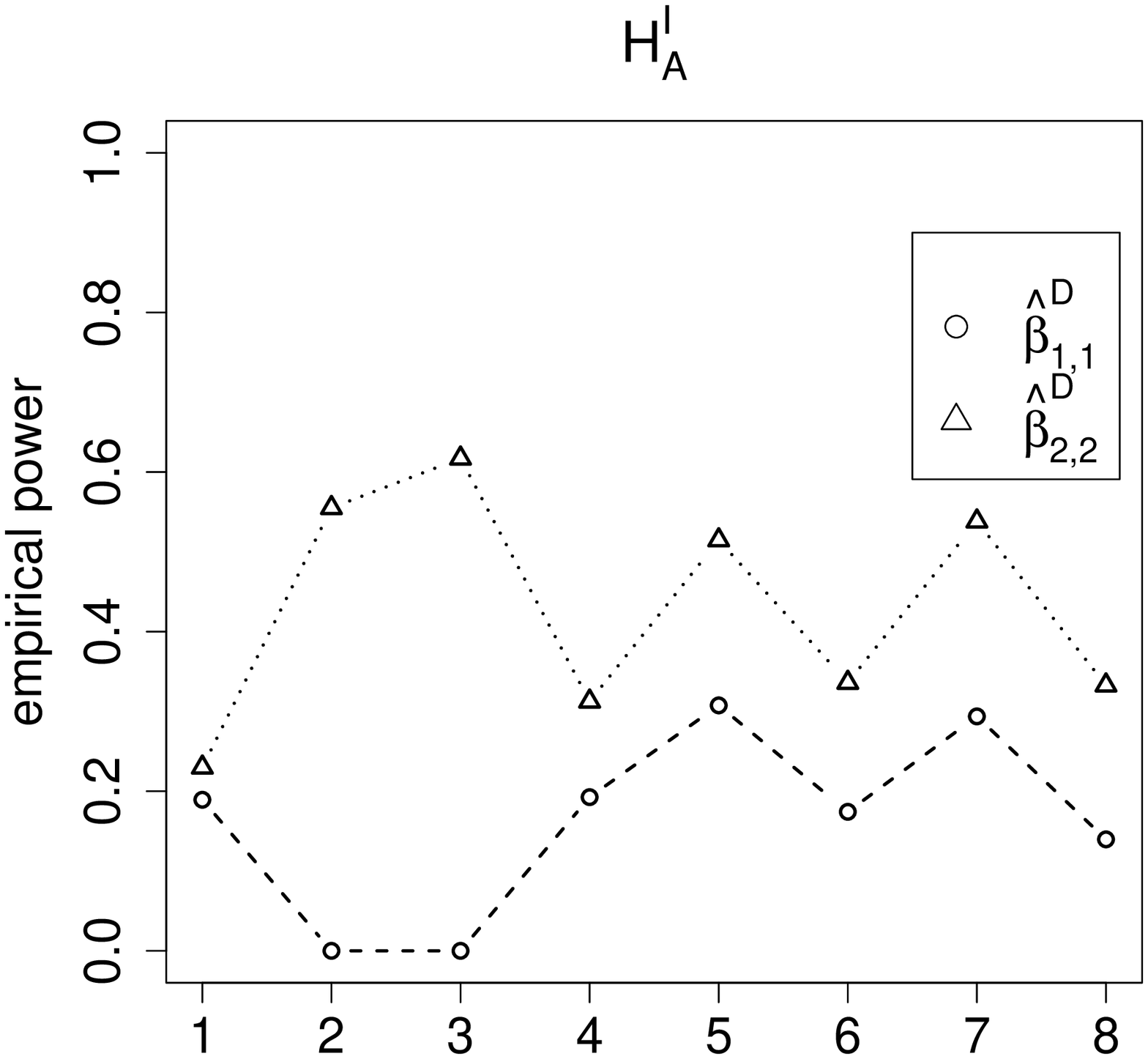} }}
\rotatebox{-90}{ \resizebox{2. in}{!}{\includegraphics{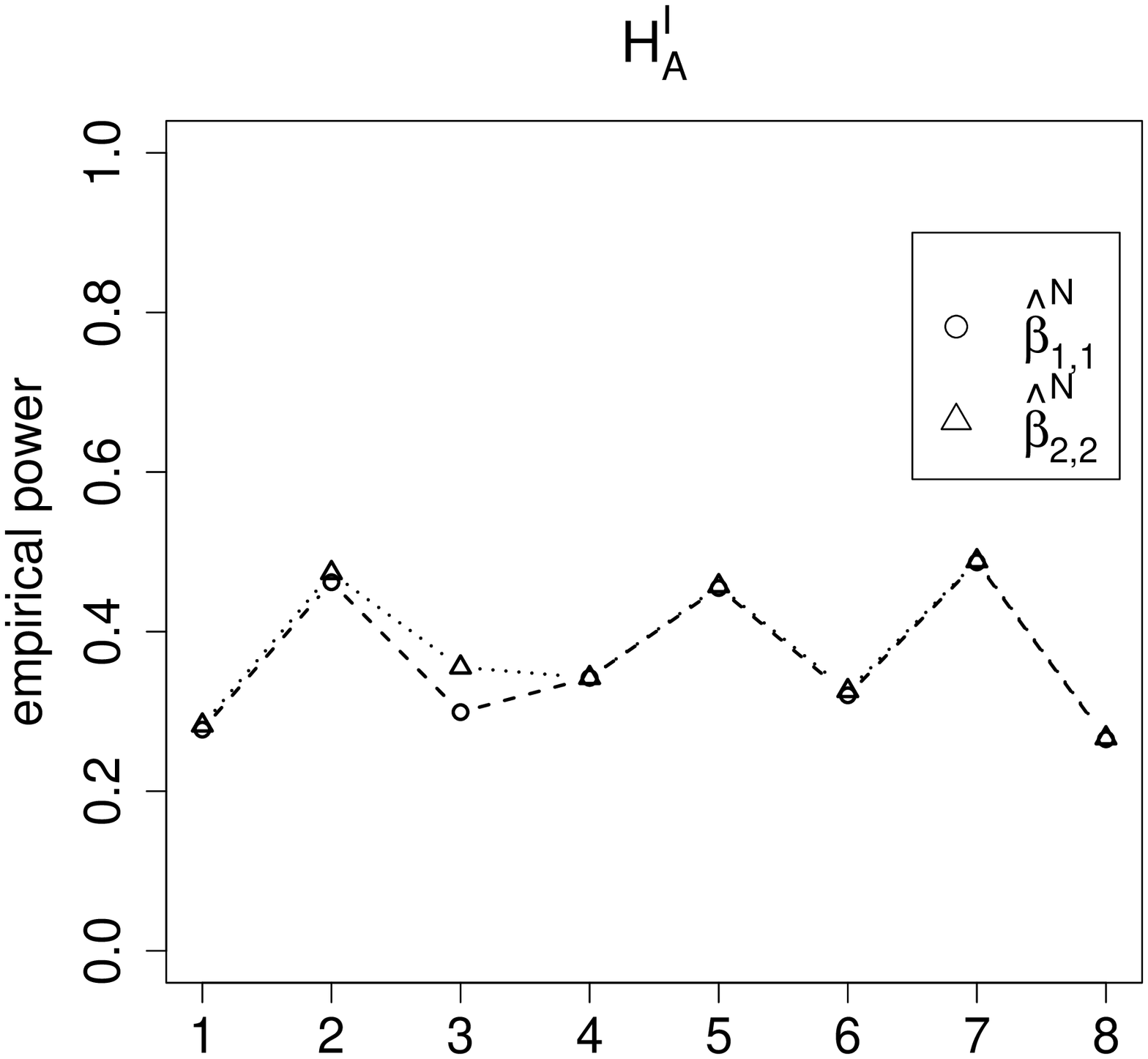} }}
\rotatebox{-90}{ \resizebox{2. in}{!}{\includegraphics{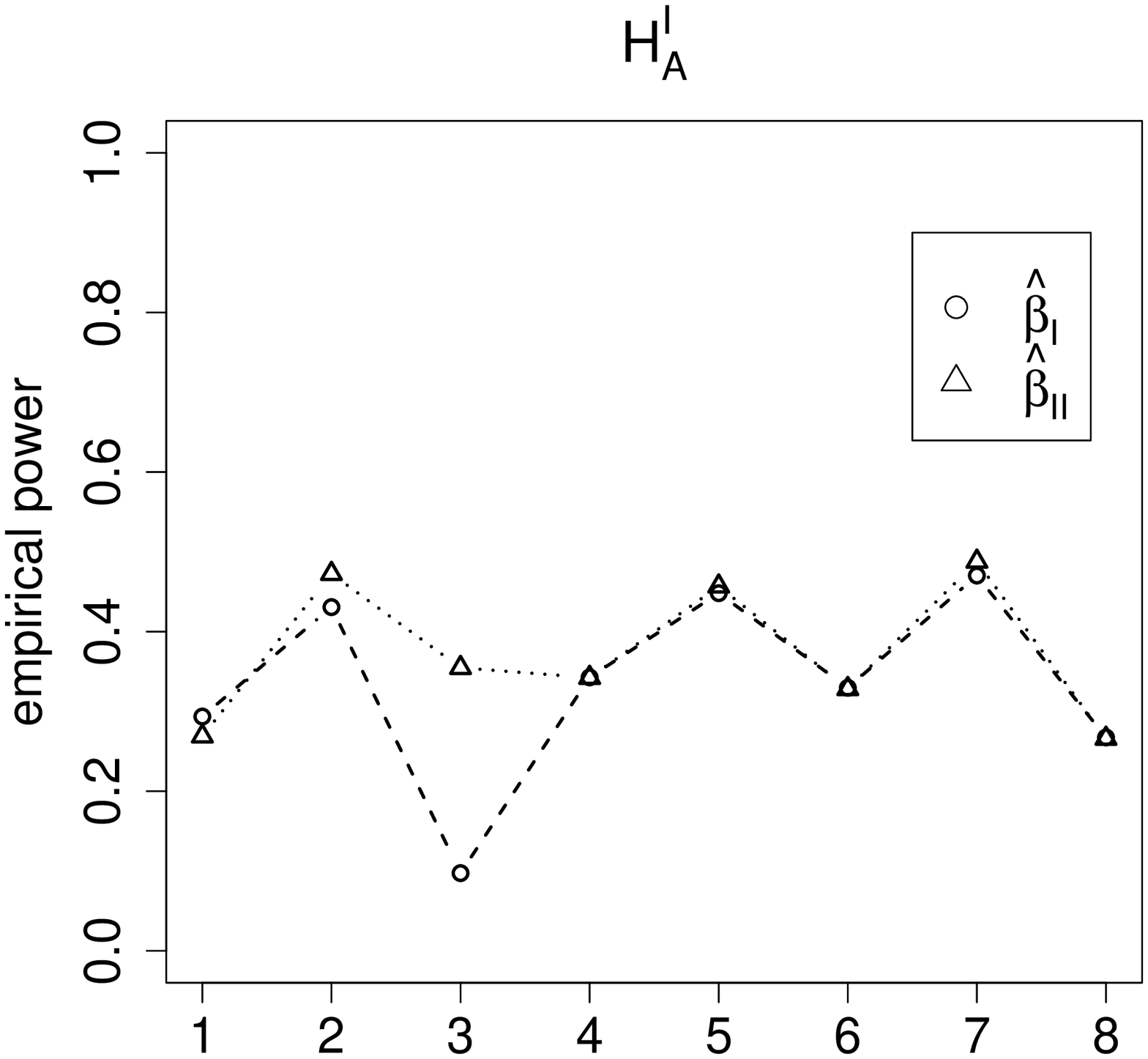} }}
\rotatebox{-90}{ \resizebox{2. in}{!}{\includegraphics{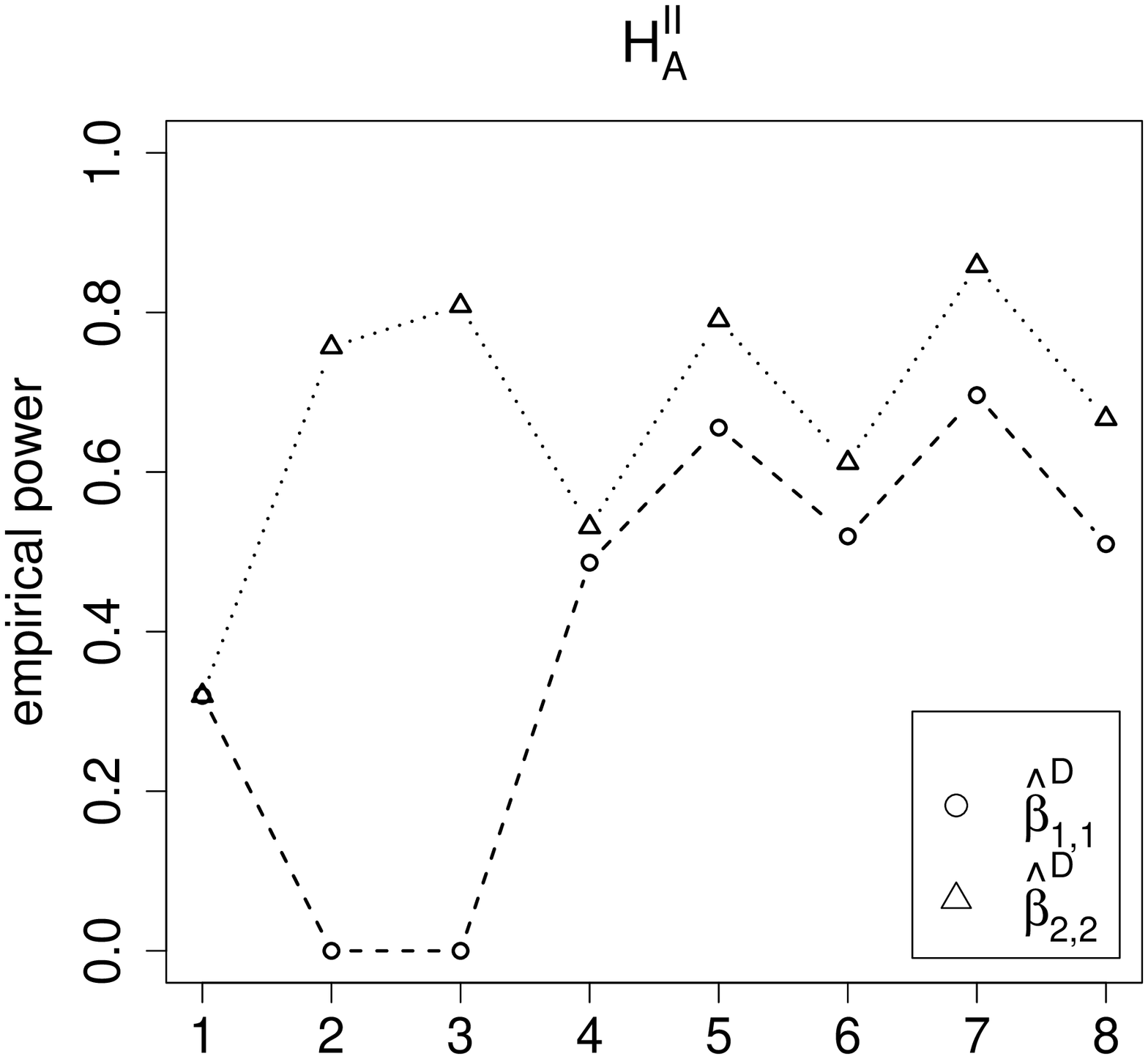} }}
\rotatebox{-90}{ \resizebox{2. in}{!}{\includegraphics{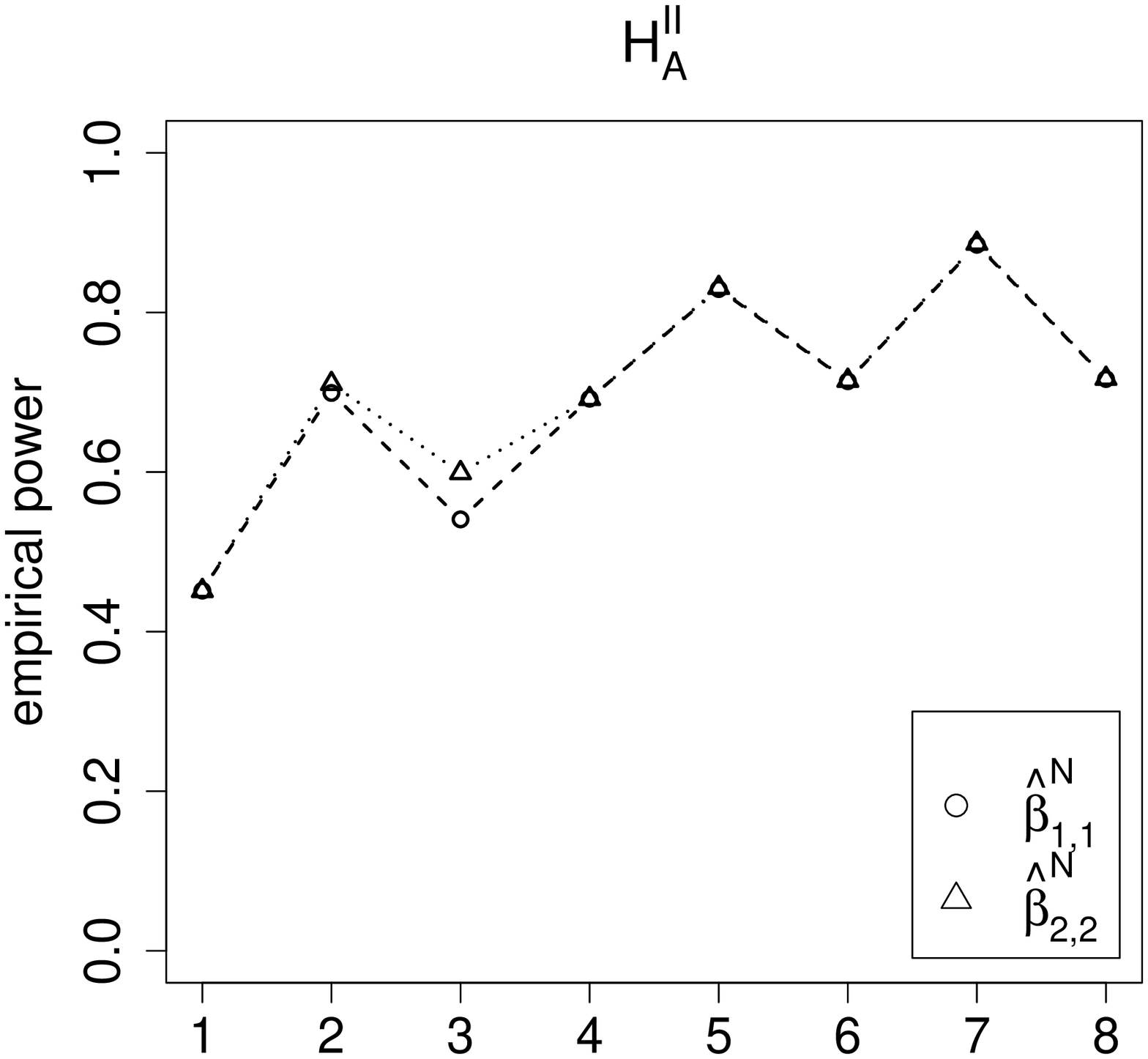} }}
\rotatebox{-90}{ \resizebox{2. in}{!}{\includegraphics{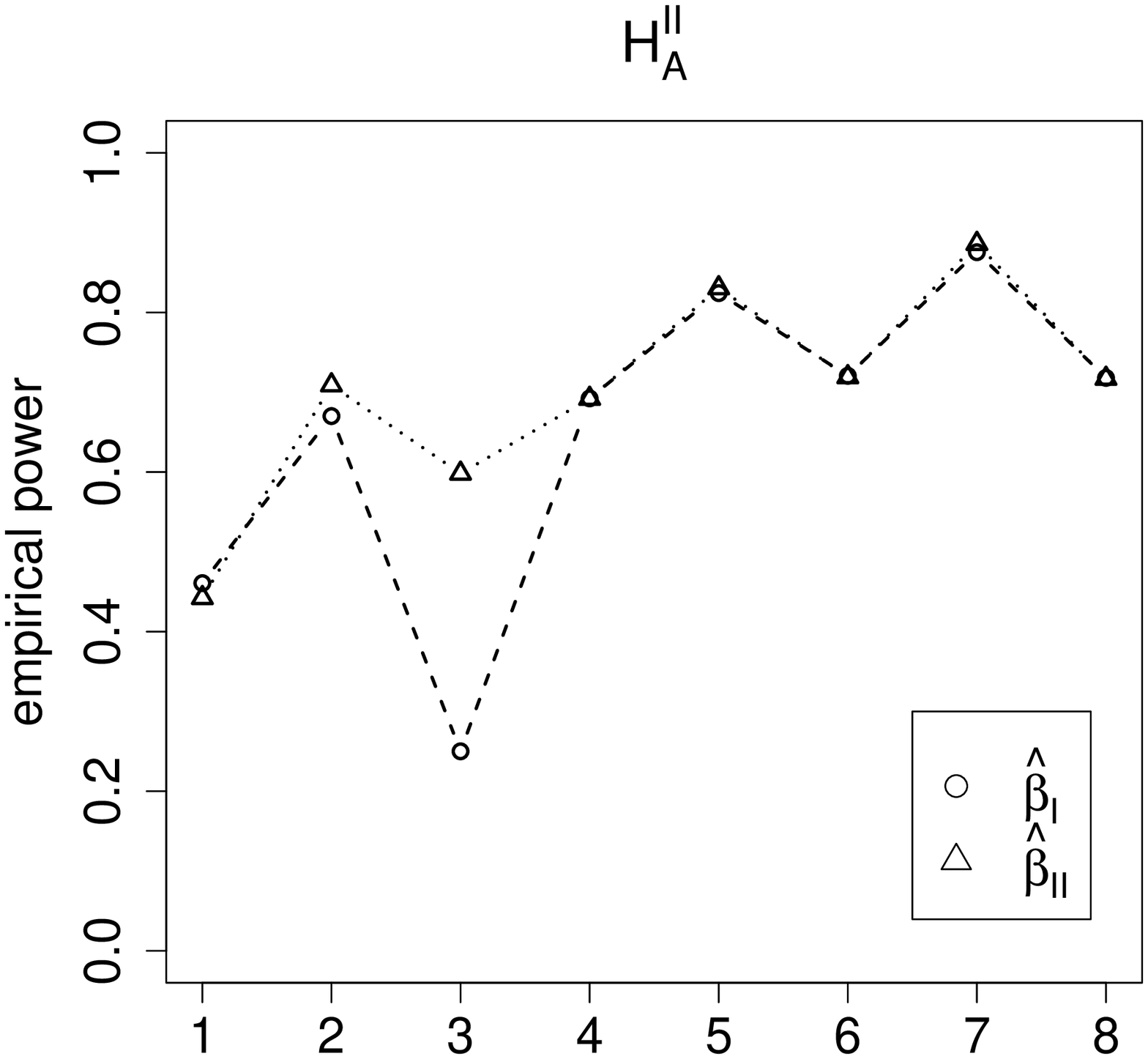} }}
\rotatebox{-90}{ \resizebox{2. in}{!}{\includegraphics{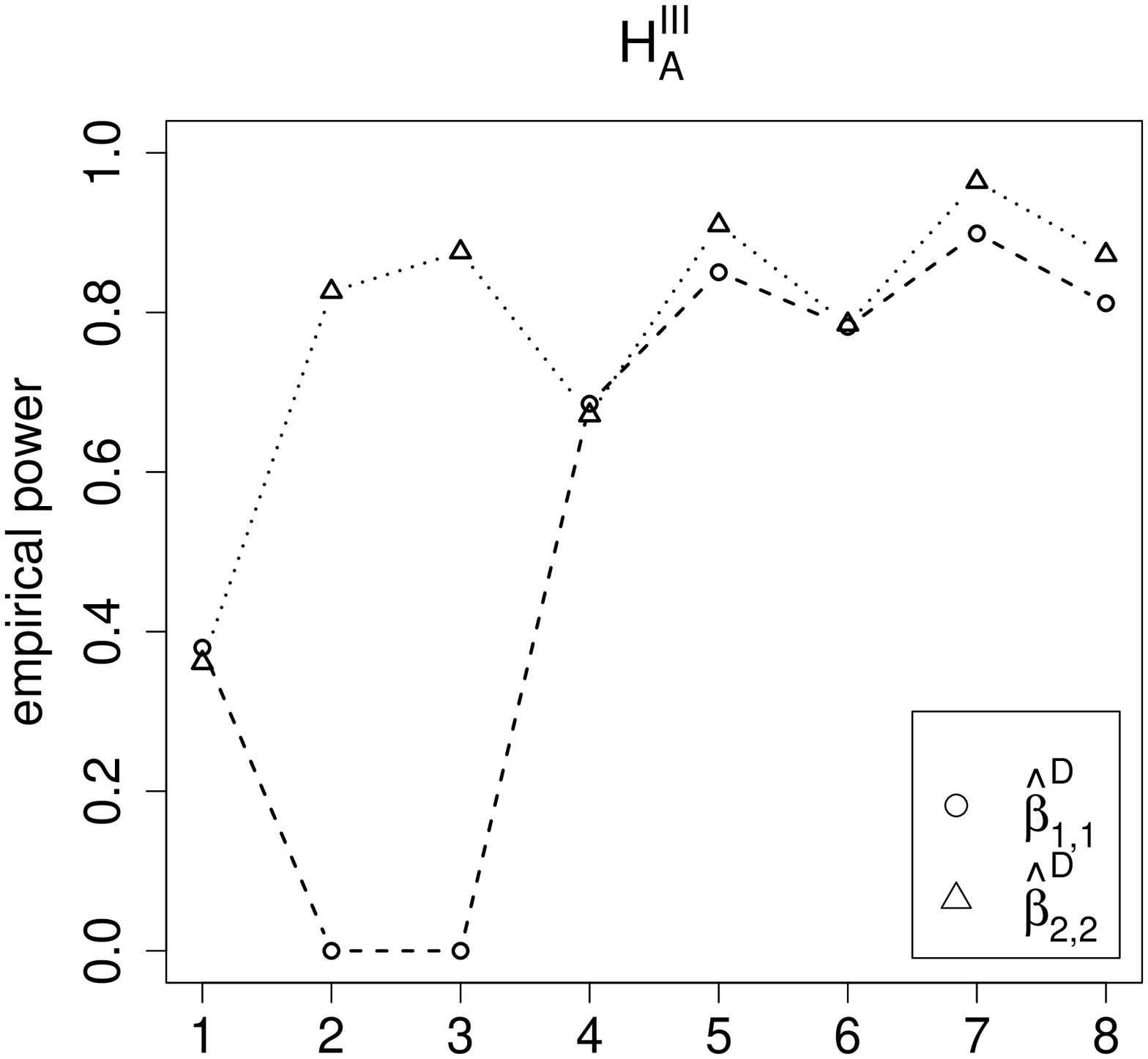} }}
\rotatebox{-90}{ \resizebox{2. in}{!}{\includegraphics{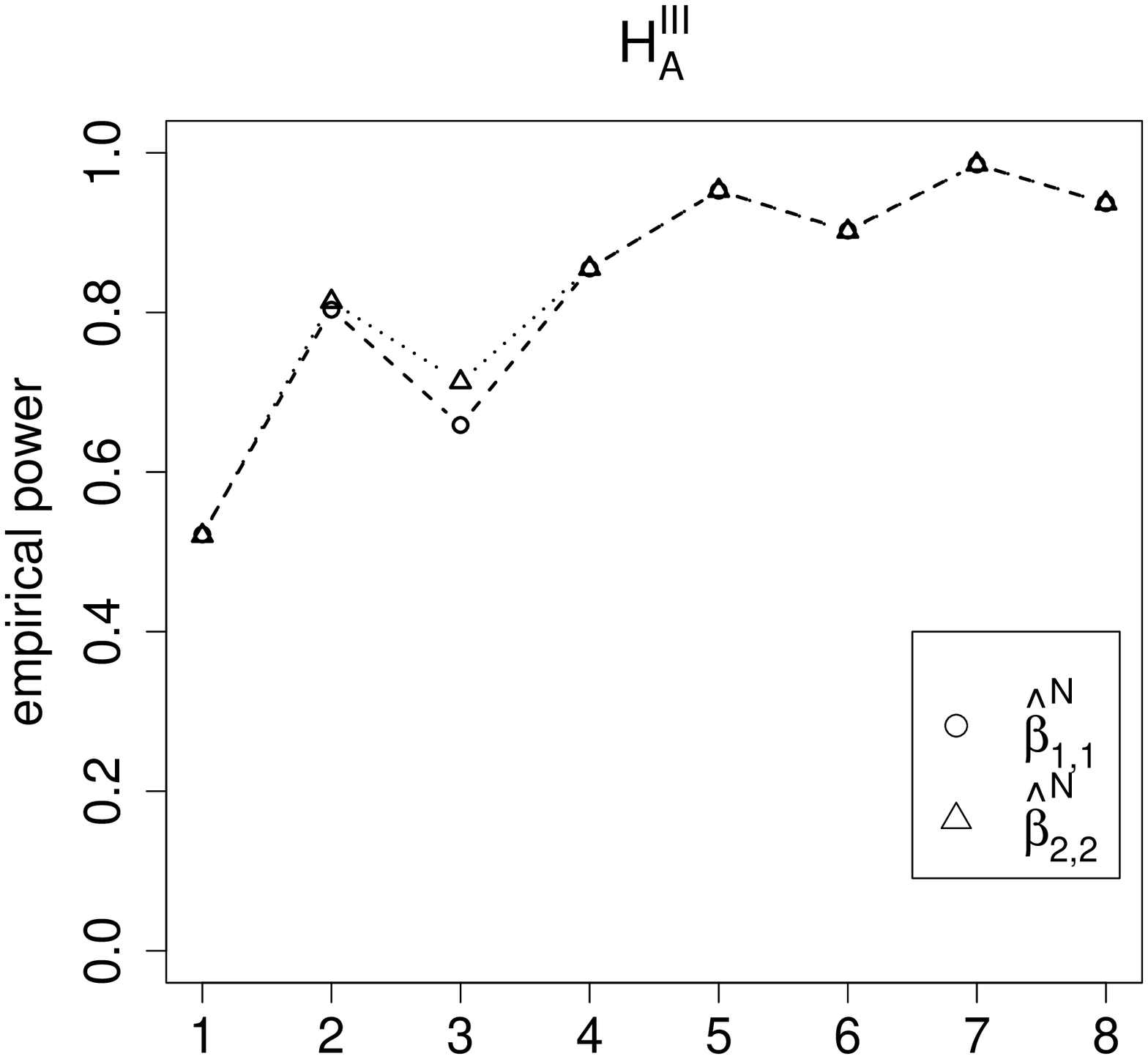} }}
\rotatebox{-90}{ \resizebox{2. in}{!}{\includegraphics{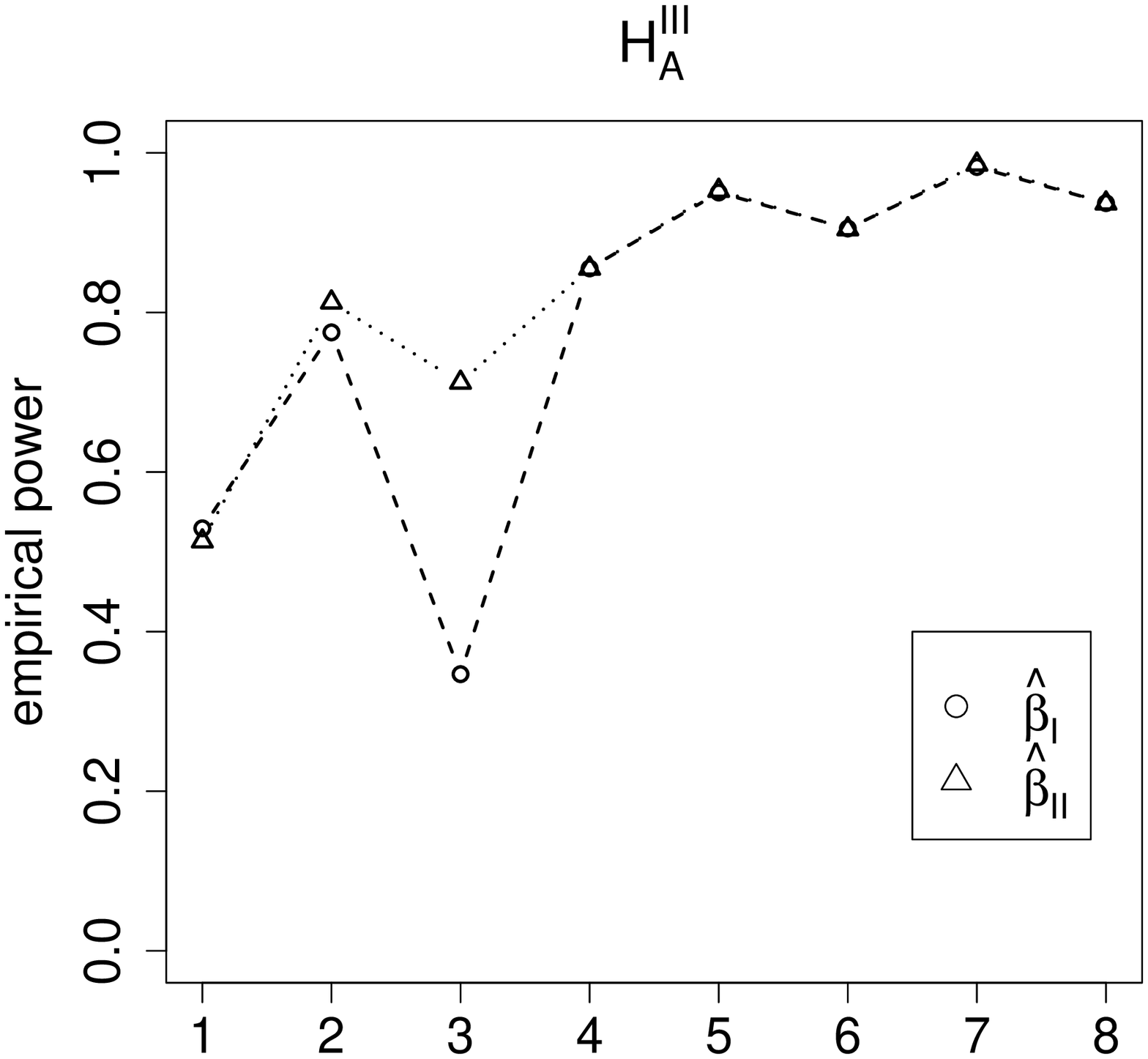} }}
\caption{
\label{fig:power-assoc-LS}
The empirical power estimates for Dixon's and Ceyhan's cell-specific tests
for cell $(i,i)$, $i=1,2$ and the new directional tests
under the association alternatives in the two-class case for the left-sided alternatives.
(which is sensitive for the association pattern).
The power and horizontal axis labeling is as in Figure \ref{fig:power-seg-TS}.
}
\end{figure}

The power estimates for the two-sided versions and left-sided versions under association
alternatives are presented in Tables \ref{tab:emp-power-assoc-2s} and  \ref{tab:emp-power-assoc-lt},
and plotted in Figures \ref{fig:power-assoc-TS} and  \ref{fig:power-assoc-LS},
respectively, where power labeling is as
in Section \ref{sec:power-comp-seg}.
We omit the power estimates of the tests for the right-sided alternative
under the association alternatives as they are virtually zero.
Observe that when sample sizes are similar (see $n_1=n_2$ cases),
for all tests, as $n=(n_1+n_2)$ gets larger, the power estimates get larger;
and as the association gets stronger, the power estimates get larger.
The power estimates for the left-sided tests are
all higher than their corresponding two-sided estimates (as expected).
For such sample sizes, version I has the highest power estimates.

For smaller samples, Dixon's test has the highest power,
while for larger samples new versions and
Ceyhan's test have similar power performance,
but they have higher power compared to Dixon's tests.
The power performance is highly dependent on the
level of relative abundances of the classes.
This might be due to the fact that by construction,
when class $X$ is much larger than class $Y$,
the two classes are not strongly associated,
since NN of $X$ points could also be from
the same class with a high probability.
The lack of association when class $Y$ is
larger occurs for the same reason.

Considering the empirical significance levels and power estimates,
we recommend the version II of the new directional
tests (i.e., $Z_{II}$) in the left-sided form
when testing against the association alternatives,
as it is at the desired level for most sample sizes,
and has considerably higher power for each sample size combination.

\begin{table}[ht]
\centering
\begin{tabular}{|c|c||c|c|c|c|c|c|}
\hline
\multicolumn{8}{|c|}{Empirical power estimates for the two-sided tests} \\
\multicolumn{8}{|c|}{under the association alternatives} \\
\hline
& $(n_1,n_2)$  & $\bh_{1,1}^D$ & $\bh_{2,2}^D$ & $\bh_{1,1}^C$ & $\bh_{2,2}^C$ & $\bh_I$ & $\bh_{II}$ \\
\hline
\hline
 & $(10,10)$ & .1349 & .1776 & .1638 & .1689 & .1854 & .1663 \\
\cline{2-8}
 & $(10,30)$ & .0002 & .4366 & .2575 & .2728 & .1744 & .2727 \\
\cline{2-8}
 & $(10,50)$ & .0002 & .4947 & .0686 & .1071 & .0041 & .0998 \\
\cline{2-8}
 & $(30,10)$ & .0725 & .0047  & .0294  & .0275 & .0155 & .0292 \\
\cline{2-8}
 & $(30,30)$ & .1413 & .2434 & .2110 & .2134 & .2149 & .2082 \\
\cline{2-8}
 & $(30,50)$ & .1833 & .3984 & .3268 & .3314 & .3240 & .3300 \\
\cline{2-8} \raisebox{4.ex}[0pt]{$H_A^I$}
 & $(50,10)$ & .0547 & .0067 & .0122 & .0114 & .0113 & .0120 \\
\cline{2-8}
 & $(50,30)$ & .0882 & .1484 & .1218 & .1224 & .1281 & .1218 \\
\cline{2-8}
 & $(50,50)$ & .1149 & .2421 & .2151 & .2181 & .2209 & .2144 \\
\hline
\hline
 & $(10,10)$ & .2499 & .2569 & .2898 & .2900 & .3125 & .2871 \\
\cline{2-8}
 & $(10,30)$ & .0000 & .6463 & .4919 & .5123 & .3722 & .5120 \\
\cline{2-8}
 & $(10,50)$ & .0000 & .7062 & .1959 & .2699 & .0177 & .2539 \\
\cline{2-8}
 & $(30,10)$ & .1818 & .0019 & .0770 & .0703 & .0341 & .0769 \\
\cline{2-8}
 & $(30,30)$ & .4053 & .4457 & .5267 & .5293 & .5314 & .5243 \\
\cline{2-8}
 & $(30,50)$ & .4896 & .6957 & .7196 & .7239 & .7154 & .7228 \\
\cline{2-8} \raisebox{4.ex}[0pt]{$H_A^{II}$}
 & $(50,10)$ & .1268 & .0032 & .0109 & .0071 & .0036 & .0102 \\
\cline{2-8}
 & $(50,30)$ & .2957 & .2735 & .3387 & .3363 & .3392 & .3383 \\
\cline{2-8}
 & $(50,50)$ & .4034 & .4961 & .5824 & .5848 & .5904 & .5802 \\
\hline
\hline
 & $(10,10)$ & .3038 & .2918 & .3475 & .3471 & .3699 & .3448 \\
\cline{2-8}
 & $(10,30)$ & .0000 & .7364 & .6115 & .6290 & .4871 & .6283 \\
\cline{2-8}
 & $(10,50)$ & .0000 & .7907 & .2885 & .3718 & .0340 & .3517 \\
\cline{2-8}
 & $(30,10)$ & .2957 & .0018 & .1371 & .1258 & .0669 & .1371 \\
\cline{2-8}
 & $(30,30)$ & .6092 & .6011 & .7308 & .7301 & .7338 & .7283 \\
\cline{2-8}
 & $(30,50)$ & .7211 & .8491 & .9052 & .9072 & .9027 & .9066 \\
\cline{2-8} \raisebox{4.ex}[0pt]{$H_A^{III}$}
 & $(50,10)$ & .2277 & .0016 & .0162 & .0089 & .0015 & .0144 \\
\cline{2-8}
 & $(50,30)$ & .5414 & .3919 & .5632 & .5588 & .5573 & .5617 \\
\cline{2-8}
 & $(50,50)$ & .6842 & .6891 & .8289 & .8301 & .8330 & .8278 \\
\hline
\end{tabular}
\caption{
\label{tab:emp-power-assoc-2s}
The empirical power estimates for the two-sided tests under the association alternatives,
$H_A^I$, $H_A^{II}$, and $H_A^{III}$ for the two-class case
with $N_{mc}=10000$, for some combinations of $n_1,n_2 \in
\{10,30,50\}$ at $\alpha=.05$.}
\end{table}

\begin{table}[ht]
\centering
\begin{tabular}{|c||c|c|c|c|c|c|c|}
\hline
\multicolumn{8}{|c|}{Empirical power estimates for the left-sided tests} \\
\multicolumn{8}{|c|}{under the association alternatives} \\
\hline
& $(n_1,n_2)$  & $\bh_{1,1}^D$ & $\bh_{2,2}^D$ & $\bh_{1,1}^C$ & $\bh_{2,2}^C$ & $\bh_I$ & $\bh_{II}$ \\
\hline
\hline
 & $(10,10)$ & .1893 & .2299 & .2772 & .2826 & .2936 & .2689 \\
\cline{2-8}
 & $(10,30)$ & .0000 & .5551 & .4619 & .4736 & .4307 & .4725 \\
\cline{2-8}
 & $(10,50)$ & .0000 & .6171 & .2991 & .3553 & .0974 & .3547 \\
\cline{2-8}
 & $(30,10)$ & .1027 & .0000 & .0884 & .0849 & .0763 & .0883 \\
\cline{2-8}
 & $(30,30)$ & .1926 & .3126 & .3421 & .3422 & .3427 & .3423 \\
\cline{2-8}
 & $(30,50)$ & .3076 & .5147 & .4547 & .4572 & .4482 & .4568 \\
 \cline{2-8} \raisebox{4.ex}[0pt]{$H_A^I$}
 & $(50,10)$ & .0748 & .0000 & .0172 & .0118 & .0013 & .0172 \\
 \cline{2-8}
 & $(50,30)$ & .1373 & .2559 & .2040 & .2070 & .2080 & .2061 \\
 \cline{2-8}
 & $(50,50)$ & .1742 & .3361 & .3202 & .3258 & .3296 & .3283 \\
\hline
\hline
 & $(10,10)$ & .3195 & .3199 & .4514 & .4512 & .4608 & .4423 \\
\cline{2-8}
 & $(10,30)$ & .0000 & .7566 & .6991 & .7106 & .6701 & .7089 \\
\cline{2-8}
 & $(10,50)$ & .0000 & .8083 & .5407 & .5989 & .2500 & .5982 \\
\cline{2-8}
 & $(30,10)$ & .2631 & .0000 & .2108 & .2013 & .1829 & .2106 \\
\cline{2-8}
 & $(30,30)$ & .4864 & .5312 & .6919 & .6918 & .6923 & .6920 \\
\cline{2-8}
 & $(30,50)$ & .6556 & .7906 & .8293 & .8310 & .8242 & .8306 \\
 \cline{2-8} \raisebox{4.ex}[0pt]{$H_A^{II}$}
 & $(50,10)$ & .1901 & .0000 & .0490 & .0361 & .0054 & .0487 \\
 \cline{2-8}
 & $(50,30)$ & .4129 & .4177 & .4737 & .4735 & .4703 & .4747 \\
 \cline{2-8}
 & $(50,50)$ & .5194 & .6117 & .7138 & .7146 & .7206 & .7191 \\
\hline
\hline
 & $(10,10)$ & .3799 & .3610 & .5218 & .5202 & .5295 & .5133 \\
\cline{2-8}
 & $(10,30)$ & .0000 & .8261 & .8034 & .8135 & .7750 & .8127 \\
\cline{2-8}
 & $(10,50)$ & .0000 & .8757 & .6589 & .7128 & .3467 & .7122 \\
\cline{2-8}
 & $(30,10)$ & .4010 & .0000 & .3090 & .2942 & .2670 & .3087 \\
\cline{2-8}
 & $(30,30)$ & .6856 & .6714 & .8550 & .8550 & .8552 & .8551 \\
\cline{2-8}
 & $(30,50)$ & .8504 & .9096 & .9525 & .9527 & .9503 & .9527 \\
 \cline{2-8} \raisebox{4.ex}[0pt]{$H_A^{III}$}
 & $(50,10)$ & .3215 & .0000 & .0993 & .0771 & .0107 & .0991 \\
 \cline{2-8}
 & $(50,30)$ & .6571 & .5503 & .7010 & .6994 & .6924 & .7012 \\
 \cline{2-8}
 & $(50,50)$ & .7819 & .7850 & .9026 & .9015 & .9052 & .9045 \\
\hline
\end{tabular}
\caption{
\label{tab:emp-power-assoc-lt}
The empirical power estimates for the left-sided tests under the association alternatives,
$H_A^I$, $H_A^{II}$, and $H_A^{III}$ for the two-class case
with $N_{mc}=10000$, for some combinations of $n_1,n_2 \in
\{10,30,50\}$ at $\alpha=.05$.}
\end{table}

\section{Example Data}
\label{sec:examples}
We illustrate the tests on four example data sets:
Pielou's Douglas-fir/ponderosa pine data, Dixon's swamp tree data,
pyramidal neuron data, and an artificial data set.

\subsection{Pielou's Data}
\label{sec:pielou-data}
Pielou used a completely mapped data that is
comprised of ponderosa pine (\emph{Pinus ponderosa}) and Douglas-fir
trees (\emph{Pseudotsuga menziesii} formerly \emph{P. taxifolia})
from a region in British Columbia (\cite{pielou:1961}).
Her data is also used by Dixon as an illustrative example (\cite{dixon:1994}).
The question of interest is whether the two tree species are
segregated, associated, or do not significantly deviate from CSR independence.
The corresponding NNCT and the percentages are provided in Table \ref{tab:NNCT-pielou}.
The percentages for the cells are based on
the sample sizes of each species, that is, for example, \% 86 of
Douglas-firs have NNs from Douglas firs, and remaining \% 15 NNs are from ponderosa pines.
The row and column percentages are marginal
percentages with respect to the total sample size.
The percentage values are suggestive of segregation for both species.

\begin{table}[ht]
\centering
\begin{tabular}{cc}

\begin{tabular}{cc|cc|c}
\multicolumn{2}{c}{}& \multicolumn{2}{c}{NN}& \\
\multicolumn{2}{c}{}&    D.F. &  P.P.   &   sum  \\
\hline
&D.F.&    137  &   23    &   160  \\
\raisebox{1.5ex}[0pt]{base}
&P.P. &    38 &  30    &   68  \\
\hline
&sum     &    175   & 53             &  228  \\
\end{tabular}
&
\begin{tabular}{cc|cc|c}
\multicolumn{2}{c}{}& \multicolumn{2}{c}{NN}& \\
\multicolumn{2}{c}{}&    D.F. &  P.P.   &    \\
\hline
&D.F.&    86 \% &    15 \%    &  70 \%  \\
&P.P. &   56 \% &    44 \%    &  30 \%  \\
\hline
&     &   77 \% &    23 \%    &  100 \%  \\
\end{tabular}

\end{tabular}
\caption{ \label{tab:NNCT-pielou}
The NNCT for Pielou's data (left) and the corresponding percentages (right).
D.F. = Douglas-fir, P.P.= ponderosa pine.}
\end{table}

The raw data is not available, but fortunately,
\cite{pielou:1961} provided $Q=162$ and $R=134$.
The test statistics are provided in Table \ref{tab:Pielou-test-stat},
and the corresponding $p$-values for the two-sided, right-sided (segregation),
and left-sided (association) alternatives are provided below the test statistics.
Observe that all two-sided tests are significant,
implying significant deviation from CSR independence or RL;
and among the directional tests right-sided $p$-values are significant;
hence there is significant segregation between Douglas-firs and
ponderosa pines.

\begin{table}[ht]
\centering
\begin{tabular}{|c||c|c|c|c|c|c|c|c|c|c|}
\hline
\multicolumn{11}{|c|}{Test statistics and the associated $p$-values for Pielou's data} \\
\hline
 & $Z^D_{11}$ & $Z^D_{22}$ &  $Z^C_{11}$ &  $Z^C_{22}$ &  $Z_P$ & $Z_{mc}$ & $Z^a_{mc}$ & $Z^s_{mc}$ &  $Z_I$ &  $Z_{II}$ \\
\hline
test statistics  & 4.36 & 2.29 & 3.63 & 3.61 & 4.86 & 3.81 & 3.69 & 3.86 & 3.92 & 3.62 \\
\hline
two-sided & $<.0001$ & .0221 & .0003 &  .0003 &  $<.0001$ &  .0001 & --- & --- &  .0001 &  .0003 \\
segregation & $<.0001$ & .0110 & .0001 & .0002 & $<.0001$ & --- & --- & .0001 & $<.0001$ & .0001 \\
association & $\approx 1.0$ & .9890 & .9999 & .9998 & $\approx 1.0$ & --- & .9999 & --- & $\approx 1.0$ & .9999 \\
\hline
\end{tabular}
\caption{ \label{tab:Pielou-test-stat}
Test statistics and the associated $p$-values based on asymptotic approximation
for the two-sided and directional alternatives for Pielou's data.
}
\end{table}

\subsection{Swamp Tree Data}
\label{sec:swamp-data}
\begin{figure}[hbp]
\rotatebox{-90}{ \resizebox{3.5 in}{!}{
\includegraphics{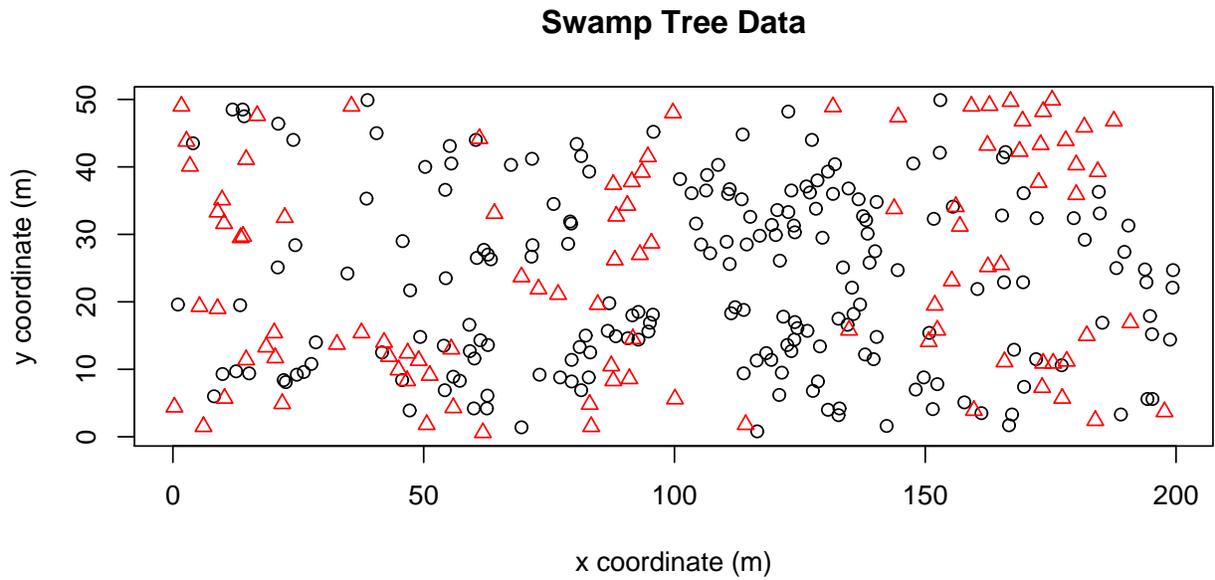} }}
\caption{
\label{fig:Swamp}
The scatter plots of the locations of black gum trees (circles $\circ$) and
bald cypress trees (triangles $\triangle$).}
\end{figure}

Dixon illustrates NN-methods on a 50m
$\times$ 200m rectangular plot of hardwood swamp in South Carolina, USA (\cite{dixon:EncycEnv2002}).
The plot contains 13 different tree species,
of which we only consider two, namely, bald cypress (\emph{Taxodium distichum})
and black gum trees (\emph{Nyssa sylvatica}).
The question of interest is whether these tree species are segregated,
associated, or satisfy CSR independence.
For more detail on the data, see (\cite{dixon:EncycEnv2002}).
The locations of these trees in the study
region are plotted in Figure \ref{fig:Swamp} and the corresponding
NNCT together with percentages are provided in Table \ref{tab:NNCT-swamp}.
Observe that the percentages are suggestive of segregation
for both species.

\begin{table}[ht]
\centering
\begin{tabular}{cc}

\begin{tabular}{cc|cc|c}
\multicolumn{2}{c}{}& \multicolumn{2}{c}{NN}& \\
\multicolumn{2}{c}{}&    B.G. &  B.C.   &   sum  \\
\hline
& B.G.&    149  &   33    &   182  \\
\raisebox{1.5ex}[0pt]{base}
& B.C. &    43 &  48    &   91  \\
\hline
&sum     &    192   & 581            &  273  \\
\end{tabular}
&
\begin{tabular}{cc|cc|c}
\multicolumn{2}{c}{}& \multicolumn{2}{c}{NN}& \\
\multicolumn{2}{c}{}&    B.G. &  B.C.   &    \\
\hline
&B.G.&   82 \%  &   18 \%    &   67 \%  \\
&B.C. &    47 \% &  53 \%    &   23 \%  \\
\hline
&     &    34 \%   & 66 \%             &  100 \%  \\
\end{tabular}

\end{tabular}
\caption{ \label{tab:NNCT-swamp}
The NNCT for swamp tree data and the corresponding percentages (in parenthesis).
B.G. = black gum trees, B.C. = bald cypress trees.}
\end{table}

\begin{table}[ht]
\centering
\begin{tabular}{|c||c|c|c|c|c|c|c|c|c|c|}
\hline
\multicolumn{11}{|c|}{Test statistics and the associated $p$-values for the swamp tree data} \\
\hline
 & $Z^D_{11}$ & $Z^D_{22}$ &  $Z^C_{11}$ &  $Z^C_{22}$ &  $Z_P$ & $Z_{mc}$ & $Z^a_{mc}$ & $Z^s_{mc}$ &  $Z_I$ &  $Z_{II}$ \\
\hline
test statistics  & 4.47   & 3.54  & 4.62  & 4.61  & 5.90  & 4.62  & 4.48  & 4.67  & 4.76  & 4.61 \\
\hline
\multicolumn{11}{|c|}{against the two-sided alternative} \\
\hline
$\pasy$ & $<.0001$ & .0004 & $<.0001$ & $<.0001$ & $<.0001$ & $<.0001$ & --- & --- & $<.0001$ & $<.0001$ \\
\hline
$\pmc$ & $<.0001$ & .0003 & $<.0001$ & $<.0001$ & --- & --- & --- & --- & $<.0001$ & $<.0001$ \\
\hline
$\prand$ & $<.0001$ & .0004 & $<.0001$ & $<.0001$ & $<.0001$ & $<.0001$ & --- & --- & $<.0001$ & $<.0001$ \\
\hline
\multicolumn{11}{|c|}{against the right-sided (i.e., segregation) alternative} \\
\hline
$\pasy$ & $<.0001$ & .0002 & $<.0001$ & $<.0001$ & $<.0001$ & --- & --- & $<.0001$ & $<.0001$ & $<.0001$ \\
\hline
$\pmc$ & $<.0001$ & .0002 & $<.0001$ & $<.0001$ & --- & --- & --- & --- & $<.0001$ & $<.0001$ \\
\hline
$\prand$ & $<.0001$ & .0003 & $<.0001$ & $<.0001$ & --- & --- & --- & --- & $<.0001$ & $<.0001$ \\
\hline
\multicolumn{11}{|c|}{against the left-sided (i.e., association) alternative} \\
\hline
$\pasy$  & $\approx 1.0$ & .9998 & $\approx 1.0$ & $\approx 1.0$ & $\approx 1.0$ & --- & $\approx 1.0$ & --- & $\approx 1.0$ & $\approx 1.0$ \\
\hline
$\pmc$  & $\approx 1.0$ & .9998 & $\approx 1.0$ & $\approx 1.0$ & --- & --- & --- & --- & $\approx 1.0$ & $\approx 1.0$ \\
\hline
$\prand$ & $\approx 1.0$ & .9998 & $\approx 1.0$ & $\approx 1.0$ & --- & --- & --- & --- & $\approx 1.0$ & $\approx 1.0$ \\
\hline
\end{tabular}
\caption{ \label{tab:swamp-test-stat}
Test statistics and the associated $p$-values for the two-sided and directional alternatives
for the swamp tree data.
$\pasy$, $\pmc$, and $\prand$ stand for the $p$-values based on the asymptotic
approximation, Monte Carlo simulation, and randomization of the tests, respectively.}
\end{table}

The locations of the tree species can be viewed a priori resulting from different processes,
so the more appropriate null hypothesis is the CSR independence pattern.
Hence our inference will be a conditional one (see Remark \ref{rem:QandR}).
We calculate $Q= 178$ and $R=156$ for this data set.
We present Dixon's and Ceyhan's cell-specific and new directional
test statistics and the associated $p$-values
the two-, right-, and left-sided alternatives in Table \ref{tab:swamp-test-stat},
where $\pasy$ stands for the $p$-value based on the asymptotic approximation,
$\pmc$ is the $p$-value based on $10000$ Monte Carlo replication
of CSR independence in the same plot and $\prand$ is based on Monte Carlo
randomization of the labels on the given locations of the trees 10000 times.
Notice that $\pasy$, $\pmc$, and $\prand$ are very similar for each test.
All tests are significant for the two-sided alternative
implying deviation from CSR independence,
and among the directional tests, right-sided tests are significant,
indicating significant segregation between black gums and bald cypresses.

\begin{figure}[hbp]
\centering
\rotatebox{-90}{ \resizebox{2 in}{!}{\includegraphics{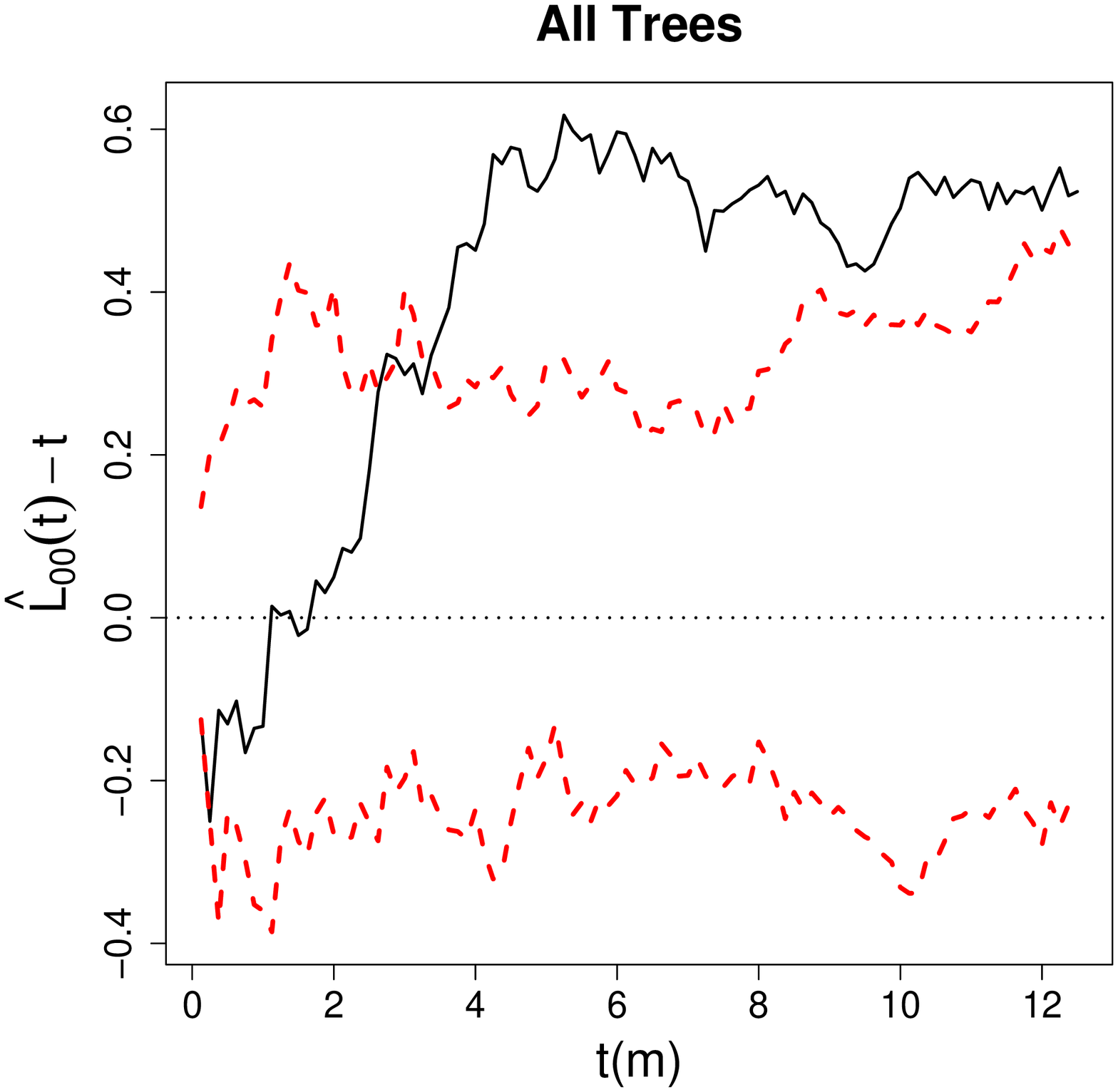} }}
\rotatebox{-90}{ \resizebox{2 in}{!}{\includegraphics{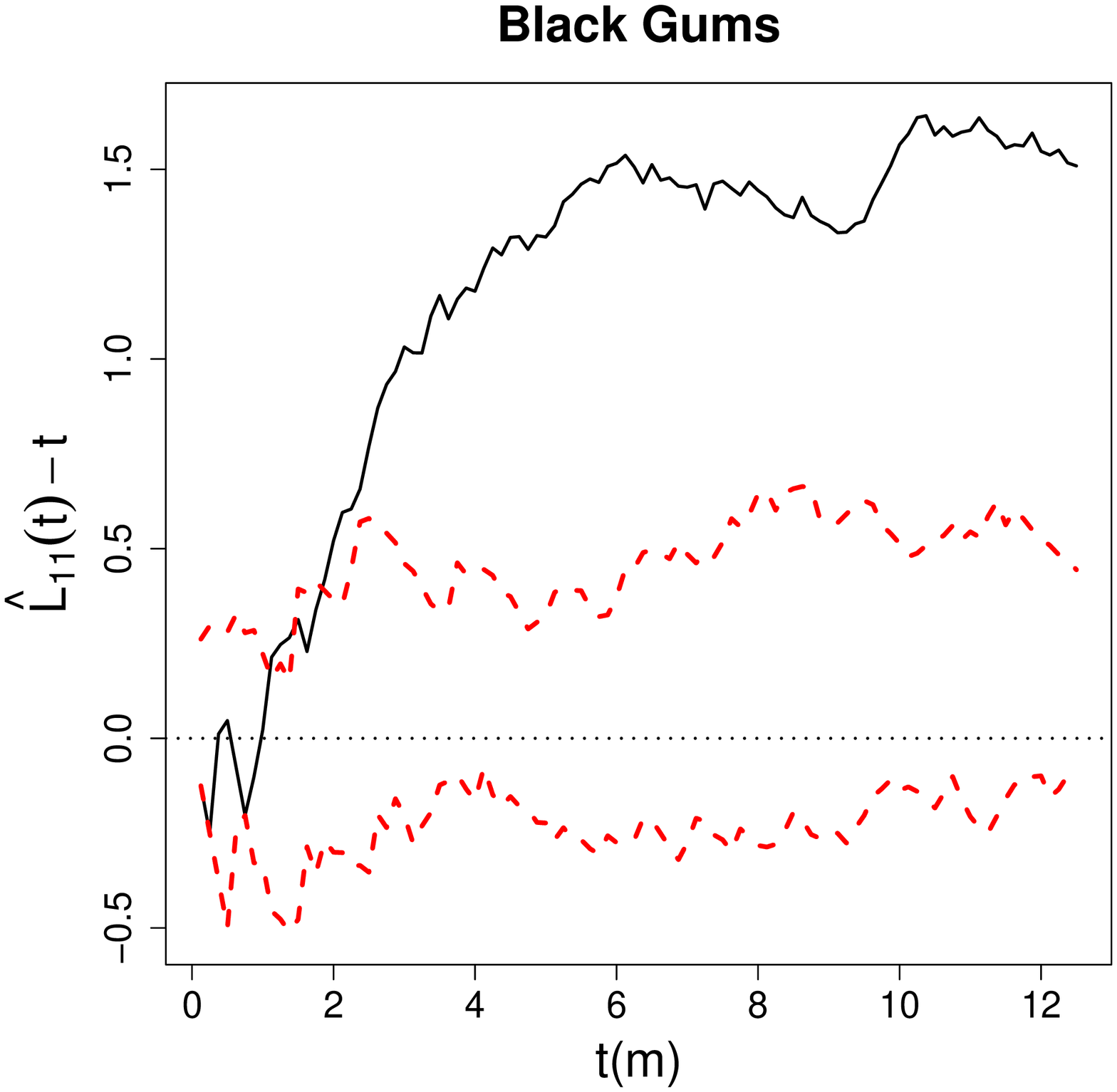} }}
\rotatebox{-90}{ \resizebox{2 in}{!}{\includegraphics{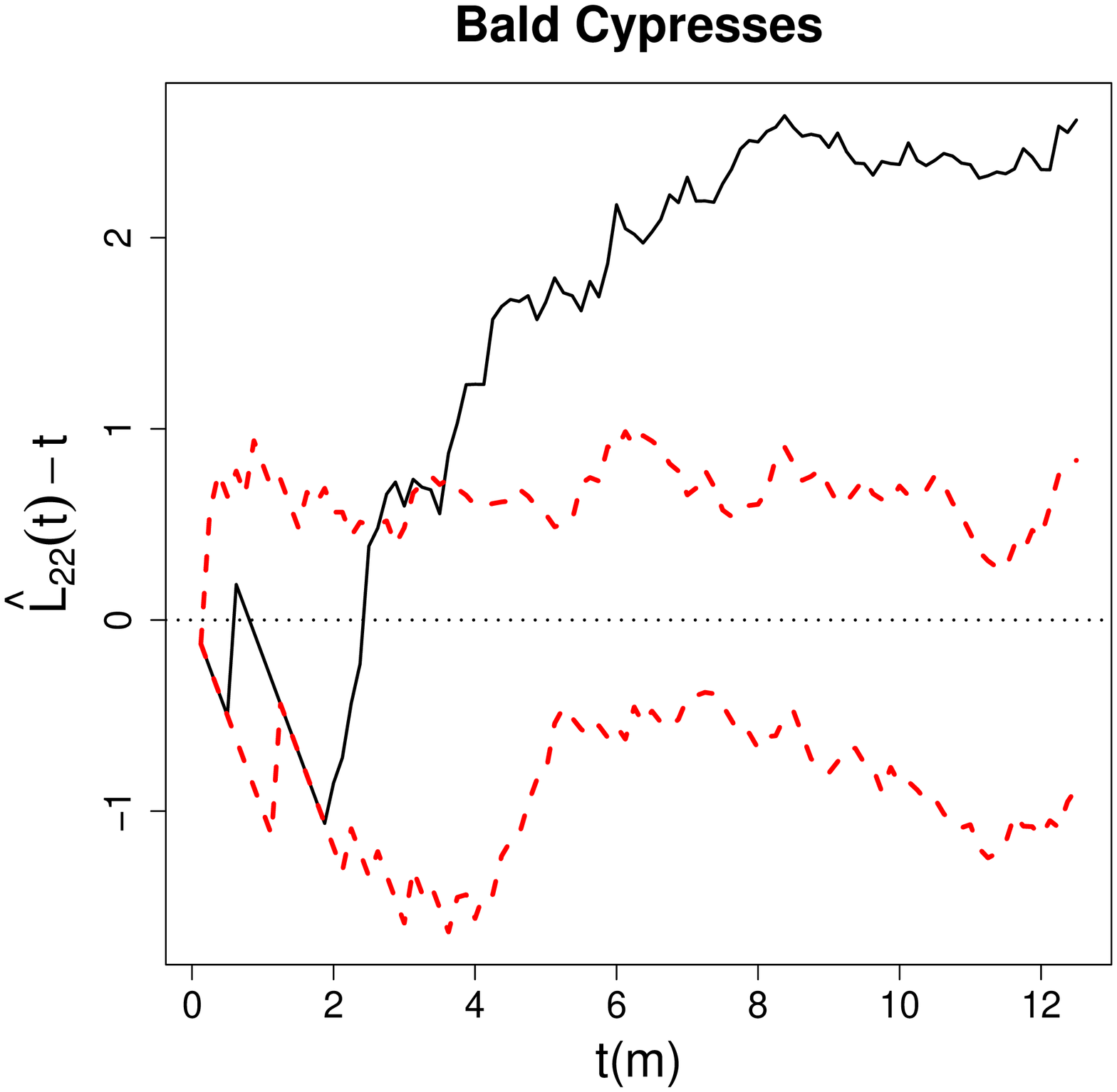} }}
\rotatebox{-90}{ \resizebox{2 in}{!}{\includegraphics{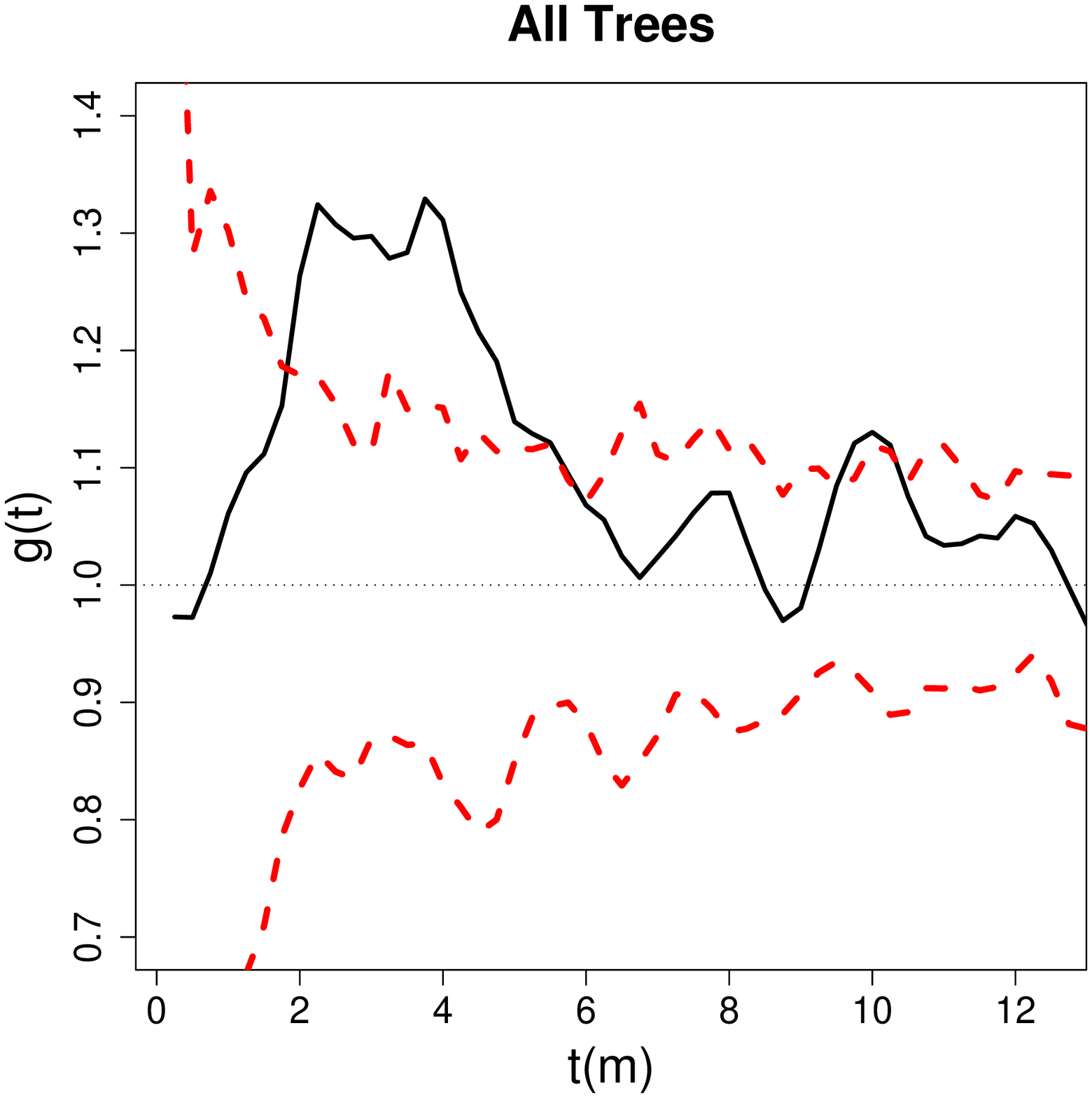} }}
\rotatebox{-90}{ \resizebox{2 in}{!}{\includegraphics{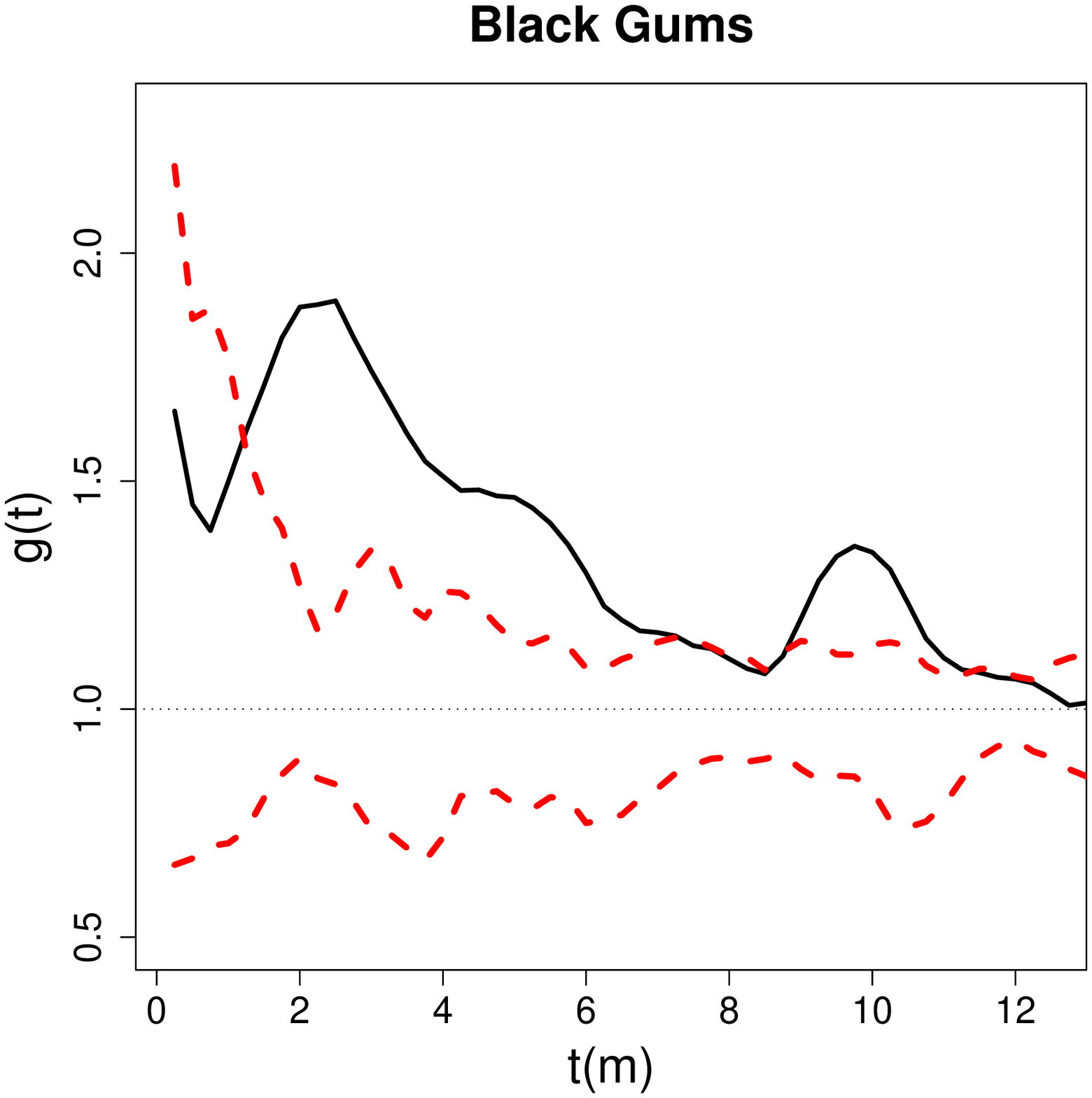} }}
\rotatebox{-90}{ \resizebox{2 in}{!}{\includegraphics{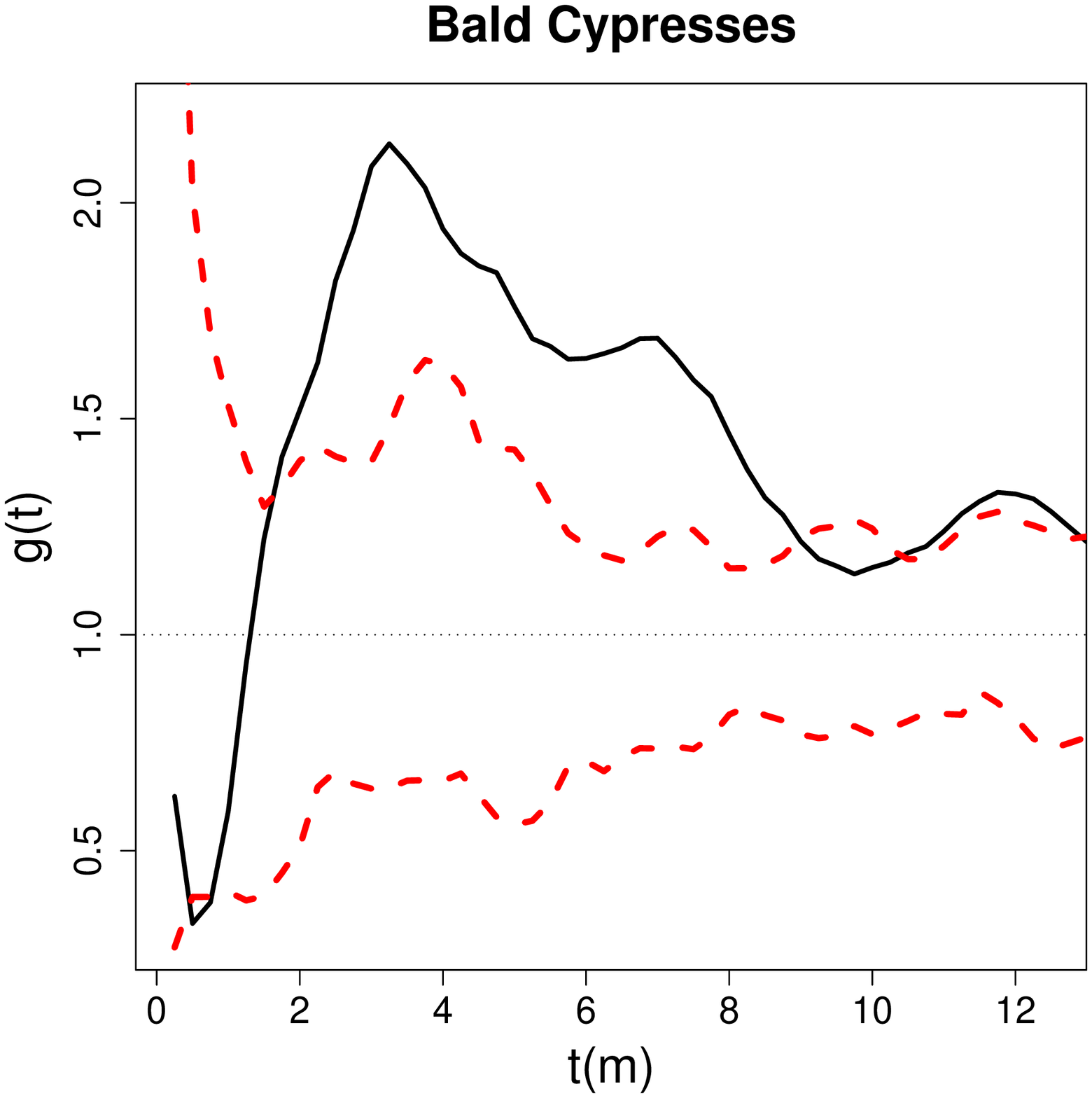} }}
\caption{
\label{fig:swamp-second-order-uni}
Ripley's univariate $L$-functions (top row) $\widehat{L}_{ii}(t)-t$ for $i=0,1,2$,
where $i=0$ stands for all data combined, $i=1$ for black gums,
and $i=2$ for bald cypresses;
and pair correlation functions $g(t)$ for all trees combined and for each species (bottom row).
Wide dashed lines are the upper and lower (pointwise) 95 \% confidence bounds for the
functions based on Monte Carlo simulation under the CSR independence pattern.}
\end{figure}

The results based on NNCT-tests
pertain to small scale interaction at about the average NN distances.
We might also be interested in the causes of the segregation
and the type and level of interaction between the tree species
at different scales (i.e., distances between the trees).
To answer such questions, we also present the second-order
analysis of the swamp tree data (\cite{diggle:2003})
using the functions (or some modified version of them) provided
in spatstat package in R (\cite{baddeley:2005}).
We use Ripley's univariate and bivariate $L$-functions
which are modified versions of his $K$-functions.
The estimator $\widehat{K}(t)$ of $K(t)$ is approximately unbiased for $K(t)$ at each fixed $t$.
Bias depends on the geometry of the study area and increases with $t$.
For a rectangular region
it is recommended to use $t$ values up to 1/4 of the smaller side length of the rectangle.
So we take the values $t \in [0,12.5]$ in our analysis,
since the rectangular region is $50 \times 200$ m.
But Ripley's $K$-function is cumulative,
so interpreting the spatial interaction at larger distances
is problematic (\cite{wiegand:2007}).
The (accumulative) pair correlation function $g(t)$
is better for this purpose (\cite{stoyan:1994}).
The pair correlation function of a (univariate)
stationary point process is defined as
$$g(t) = \frac{K'(t)}{2\,\pi\,t}$$
where $K'(t)$ is the derivative of $K(t)$.
However if $g(t)>0$, the pair correlation function estimates might have critical behavior
for small $t$ since the estimator of variance and hence
the bias are considerably large.
This problem gets worse especially in cluster processes (\cite{stoyan:1996}).
See for example Figure \ref{fig:swamp-second-order-uni}
where the confidence bands for smaller $t$ values are much wider compared
to those for larger $t$ values.
So pair correlation function analysis is more reliable for larger distances.
In particular, it is safer to use $g(t)$ for distances
larger than the average NN distance in the data set.
We can use Ripley's $L$-function for distances up to the average NN distance,
or use NNCT-tests for about the average NN distance.

Ripley's univariate $L$-functions and the pair correlation functions for both species
combined and each species for the swamp tree data are
presented in Figure \ref{fig:swamp-second-order-uni}.
The average NN distance in the swamp tree data
is $3.08 \pm 1.70$ m (mean$\pm$standard deviation).
So Ripley's $L$-function is reliable for up to about 3 meters,
where we see that all trees combined do not significantly deviate from CSR,
however, black gums seem to be significantly aggregated for distances about
$[2,3]$ meters and bald cypresses are significantly aggregated for distances about 3 meters.
For other distances in $[0,3]$ meters,
the pattern is not significantly different from CSR.
Hence, segregation of the species detected by the NNCT-tests might be due to different
levels and types of aggregation of the species in the study region.
The pair correlation function is more reliable for distances larger than 3 meters.
Then we observe that all trees are significantly aggregated for distances about
$[3,5]$ meters;
black gums are significantly aggregated for about $[3,7]$ and $[9,11]$ meters;
and bald cypresses are significantly aggregated for about $[3,9]$ and $[10,12.5]$ meters.
For other distances in $[3,12.5]$ meters,
the pattern is not significantly different from CSR.

We also calculate Ripley's bivariate $L$-function $\widehat{L}_{ij}(t)$.
By construction, $L_{ij}(t)$ is symmetric in $i$ and $j$ in theory,
that is, $L_{ij}(t)=L_{ji}(t)$ for all $i,j$.
But in practice edge corrections will render it slightly asymmetric,
i.e., $\widehat{L}_{ij}(t)\not=\widehat{L}_{ji}(t)$.
The corresponding estimates are pretty close in our example,
so only one bivariate plot is presented.
Under CSR independence, we have $L_{ij}(t)-t=0$.
If the bivariate pattern is segregation,
then $L_{ij}(t)-t$ tends to be negative,
if it is association then $L_{ij}(t)-t$ tends to be positive.
See (\cite{diggle:2003}) for more detail.
The same definition of the pair correlation function
can be applied to Ripley's bivariate $K$ or $L$-functions as well.
The benchmark value of $K_{ij}(t) = \pi \, t^2$ corresponds to $g(t) = 1$;
$g(t) < 1$ suggests segregation of the species;
and $g(t) > 1$ suggests association of the species.

\begin{figure}[hbp]
\centering
\rotatebox{-90}{ \resizebox{2 in}{!}{\includegraphics{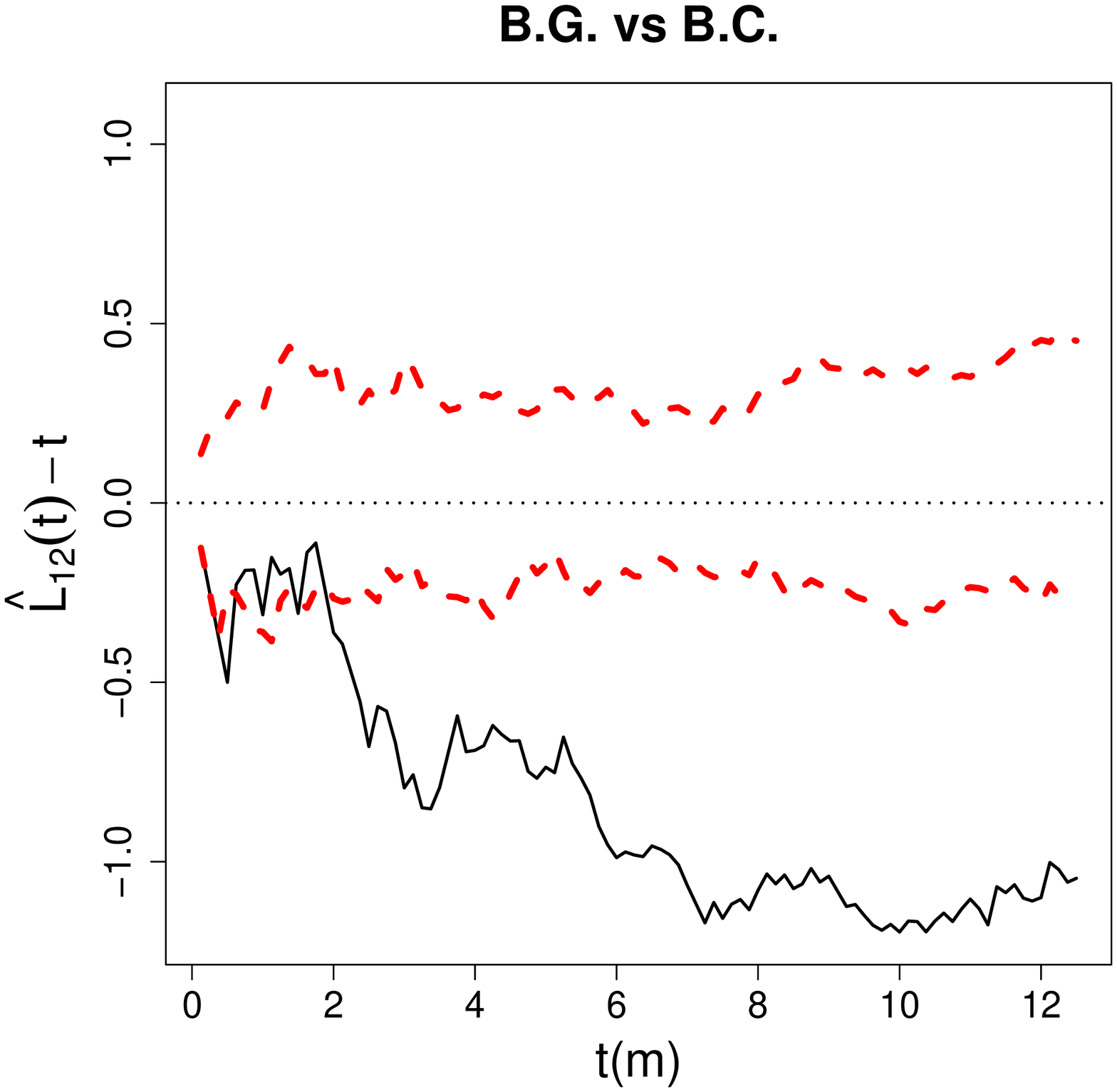} }}
\rotatebox{-90}{ \resizebox{2 in}{!}{\includegraphics{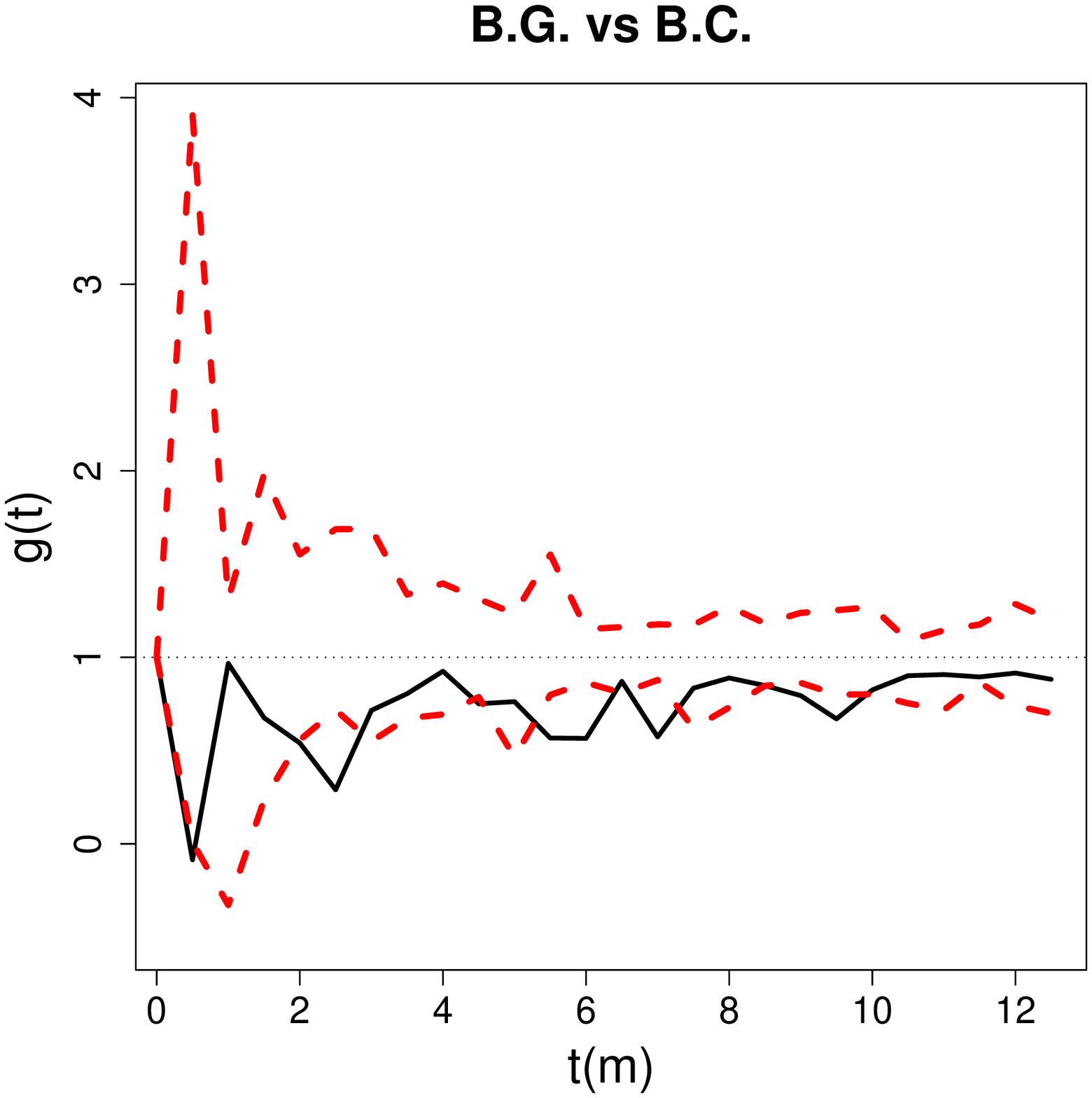} }}
\caption{
\label{fig:swamp-second-order-multi}
Ripley's bivariate $L$-function $\widehat{L}_{12}(t)-t$ (left)
and pair correlation function $g(t)$ (right) for the swamp tree data.
Wide dashed lines are the upper and lower (pointwise) 95 \% confidence bounds for the
functions based on Monte Carlo simulations under the CSR independence pattern.
B.G. = black gums and B.C. = bald cypresses.}
\end{figure}

Ripley's bivariate $L$-function and the bivariate pair correlation function for the
species in swamp tree data are plotted in Figure \ref{fig:swamp-second-order-multi}.
For 0-3 meter distances,
Ripley's bivariate $L$-function suggests that
the tree species are significantly segregated for distances
about 0.5 and $[1.8,3]$ meters,
and do not significantly deviate from CSR for other distances.
For 3-12.5 meters,
the pair correlation function suggests that the tree species
are significantly segregated for distances about $[5,7]$ and 9 meters,
and do not significantly deviate from CSR for other distances.

Considering Figures \ref{fig:swamp-second-order-uni}
and \ref{fig:swamp-second-order-multi},
we observe that Ripley's $L$ and pair correlation functions
usually detect the same large-scale pattern but at different ranges of distance values.
Ripley's $L$ suggests that the particular pattern is significant
for a wider range of distance values compared to $g(t)$,
since values of $L$ at small scales confound the values of $L$
at larger scales where $g(t)$ is more reliable to use
(\cite{wiegand:2004} and \cite{loosmore:2006}).

\subsection{Pyramidal Neuron Data}
\label{sec:benes-data}
This data set consists of the $(x,y)$-coordinates of pyramidal neurons in area 24, layer 2
of the cingulate cortex.
The data are taken from a unit square region
(unit of measurement unknown) in each of 31 subjects, grouped as
follows: controls consists of 12 subjects and correspond to cell numbers 1--655,
schizoaffectives consists of 9 subjects and correspond cell numbers 656--1061,
and schizophrenics consists of 10 subjects and correspond cell numbers 1062--1400.
Controls are the subjects with no previous history of any mental disorder,
schizoaffective disorder is a psychiatric disorder
where both the symptoms of mood disorder and psychosis occur,
and schizophrenia is a psychotic disorder
characterized by severely impaired thinking, emotions, and behavior.
\cite{diggle-benes:1991} applied several methods for the analysis of
the spatial distributions of pyramidal neurons in the cingulate cortex of human subjects
in three diagnostic groupings.
With a scaled Poisson analysis they found significant differences
between the groups in the mean numbers of neurons
in the sampled region, as well as a high degree of extra-Poisson
variation in the distribution of cell counts within these groups.
They employed two different functional descriptors of spatial
pattern for each subject to investigate departures from
completely random patterns, both between subjects and between groups,
while adjusting for cell count differences.
Since the distributions of their main functional pattern descriptor and
of their derived test statistic are unknown,
they applied a bootstrap procedure to attach $p$-values to their findings.

Since the definition of the rectangular domain for identifying neuron positions
is independent of neuronal cell density or the pattern
and this sampling domain is almost identical for each subject,
\cite{diggle-benes:1991} merged (i.e., pooled) the data for each group.
That is, the pyramidal neuron locations from control subjects
were pooled into one group, from schizoaffective subjects into another,
and schizophrenic subjects into another.
Although, the spatial distributions between subjects are not the same,
we think pooling the data by group might reveal more than what might be concealed.
\cite{diggle-benes:1991} computed and compared Ripley's univariate $K$-functions
to detect differences between patterns across the three groups.
Pattern analysis of the cellular arrangements demonstrated
significant deviation from CSR in favor of spatial regularity for each group.
On the pooled data, \cite{ceyhan:class2009} applied
a $3 \times 3$ NNCT-analysis and Ripley's $L$-functions
and found that deviation is toward association
of controls with schizoaffectives and vice versa.
In this article we will only consider the pyramidal neurons of
controls and schizoaffectives.

\begin{figure}[hbp]
\centering
\rotatebox{-90}{ \resizebox{3.5 in}{!}{\includegraphics{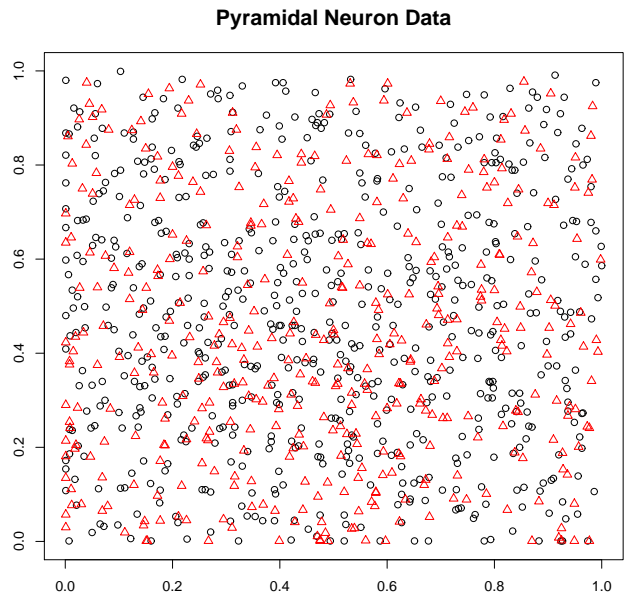} }}
\caption{
\label{fig:benes}
The scatter plots of the locations of neurons of controls (circles $\circ$),
and schizoaffectives (triangles $\triangle$) in the pyramidal neuron data.}
\end{figure}

We plot the locations of these points in the study region
in Figure \ref{fig:benes} and provide the corresponding NNCT together with
percentages based on row and column sums in Table \ref{tab:NNCT-benes}.
Observe that the pyramidal neuron locations appear to be somewhat regularly spaced.
Also the percentages are slightly smaller for the diagonal cells,
compared to the marginal (row or column) percentages,
which might be interpreted as presence of a deviation from CSR independence
in favor of association.

\begin{table}[ht]
\centering
\begin{tabular}{cc}

\begin{tabular}{cc|cc|c}
\multicolumn{2}{c}{}& \multicolumn{2}{c}{NN}& \\
\multicolumn{2}{c}{}&    Ctrl &  S.A.   &   sum  \\
\hline
& Ctrl &    368  &   288    &   656  \\
\raisebox{1.5ex}[0pt]{base}
& S.A. &    273 &  136    &   409  \\
\hline
&sum     &  641   & 424    &  1065  \\
\end{tabular}
&
\begin{tabular}{cc|cc|c}
\multicolumn{2}{c}{}& \multicolumn{2}{c}{NN}& \\
\multicolumn{2}{c}{}&   Ctrl &  S.A.   &    \\
\hline
& Ctrl &   56 \%  &    44 \%    &   62 \%   \\
& S.A. &    67 \%  &  33 \%     &   38 \%   \\
\hline
&    &    60 \%    & 40\%      &  100 \%   \\
\end{tabular}

\end{tabular}
\caption{\label{tab:NNCT-benes}
The NNCT for the pyramidal neuron data (left)
and the corresponding percentages (right).
Ctrl = Control and S.A. = Schizoaffective.}
\end{table}

The locations of the pyramidal neurons can be viewed a priori resulting
from different processes, so the more appropriate null hypothesis is the CSR independence pattern.
We calculate $Q=668$ and $R=668$ for this data set.
In Table \ref{tab:Benes-test-stat},
the cell-specific and the directional test statistics
and the associated nd the associated $p$-values.
Observe that $\pasy$, $\pmc$, and $\prand$ values
are similar for each test.
The tests are significant except for $Z^D_{22}$ at .05 level,
implying significant deviation from CSR independence.
Based on the directional tests,
this deviation is toward association of controls with schizoaffectives and vice versa,
since all tests are significant against the left-sided alternative.

\begin{table}[ht]
\centering
\begin{tabular}{|c||c|c|c|c|c|c|c|c|c|c|}
\hline
\multicolumn{11}{|c|}{Test statistics and the associated $p$-values for the pyramidal neuron data} \\
\hline
 & $Z^D_{11}$ & $Z^D_{22}$ &  $Z^C_{11}$ &  $Z^C_{22}$ &  $Z_P$ & $Z_{mc}$ & $Z^a_{mc}$ & $Z^s_{mc}$ &  $Z_I$ &  $Z_{II}$ \\
\hline
test statistics & -2.86 & -1.90 &  -2.70 &  -2.70 & -3.45 & -2.70 & -2.68 & -2.66 & -2.68 & -2.70 \\
\hline
\multicolumn{11}{|c|}{against the two-sided alternative} \\
\hline
$\pasy$ & .0042 & .0575 & .0069 & .0069 & .0006 & .0068 & --- & --- & .0073 & .0069 \\
\hline
$\pmc$ & ???.0042 & .0575 & .0003 & .0003 & .0006 & .0068 & --- & --- & .0073 & .0069 \\
\hline
$\prand$ & .0048 & .0633 & .0080 & .0062 & --- & --- & --- & --- & .0064 & .0063 \\
\hline
\multicolumn{11}{|c|}{against the right-sided (i.e., segregation) alternative} \\
\hline
$\pasy$ & .9979 & .9713 & .9965 & .9965 & .9997 & --- & --- & .9961 & .9964 & .9965 \\
\hline
$\pmc$ & ???.9979 & .9713 & .0001 & .0002 & .9997 & --- & --- & .9961 & .9964 & .9965 \\
\hline
$\prand$ & .9979 & .9613 & .9975 & .9975 & --- & --- & --- & --- & .9975 & .9975 \\
\hline
\multicolumn{11}{|c|}{against the left-sided (i.e., association) alternative} \\
\hline
$\pasy$  & .0021 & .0287 & .0035 & .0035 & .0003 & --- & .0037 & --- & .0036 & .0035 \\
\hline
$\pmc$ & ???.9979 & .9713 & .0001 & .0002 & .9997 & --- & --- & .9961 & .9964 & .9965 \\
\hline
$\prand$ & .0021 & .0397 & .0025 & .0025 & --- & --- & --- & --- & .0025 & .0025 \\
\hline
\end{tabular}
\caption{ \label{tab:Benes-test-stat}
Test statistics and the associated $p$-values for the two-sided and directional alternatives
for the pyramidal neuron data.
The labeling of the $p$-values are as in Table \ref{tab:swamp-test-stat}.
}
\end{table}

\begin{figure}[hbp]
\centering
\rotatebox{-90}{ \resizebox{2 in}{!}{\includegraphics{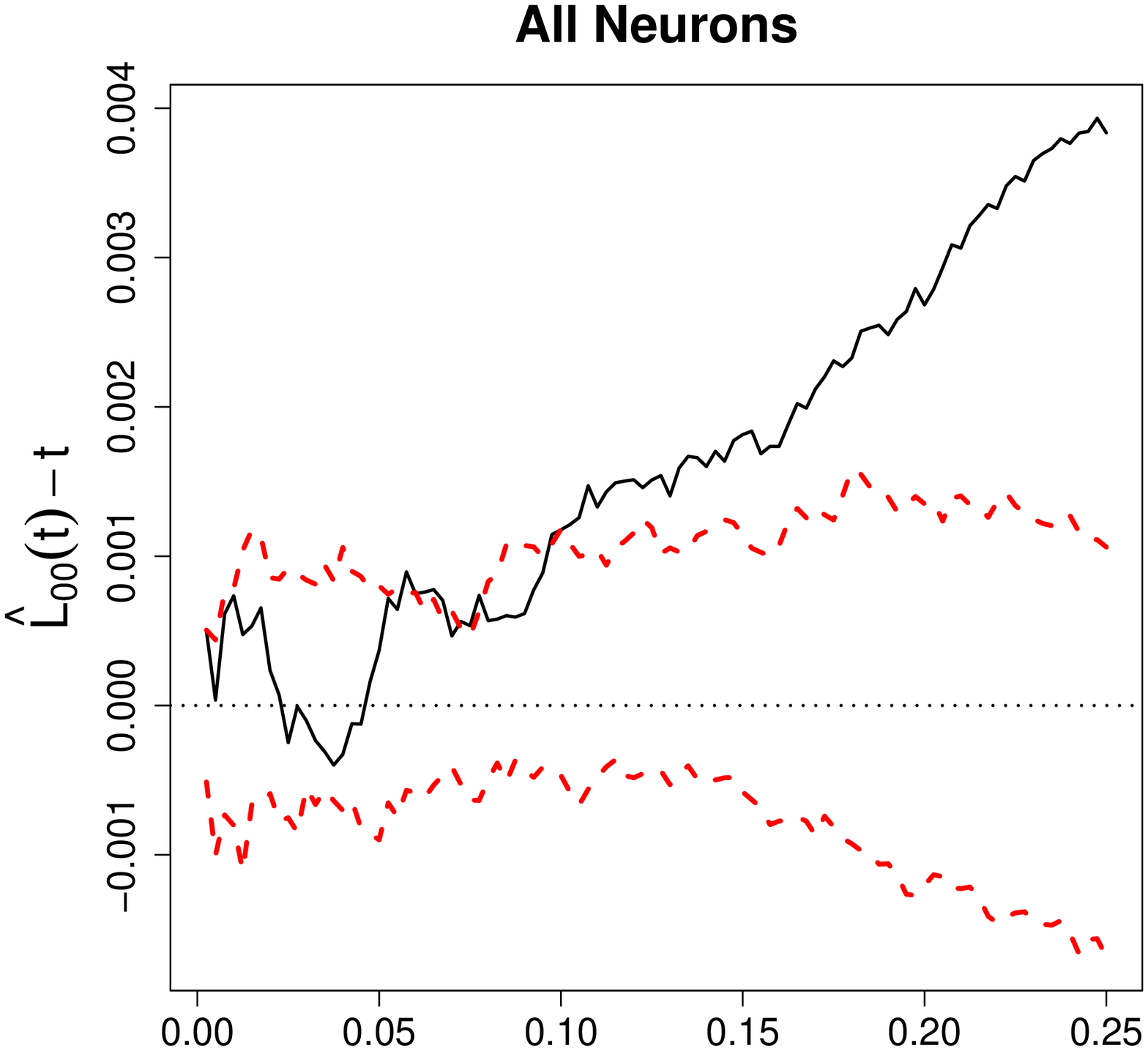} }}
\rotatebox{-90}{ \resizebox{2 in}{!}{\includegraphics{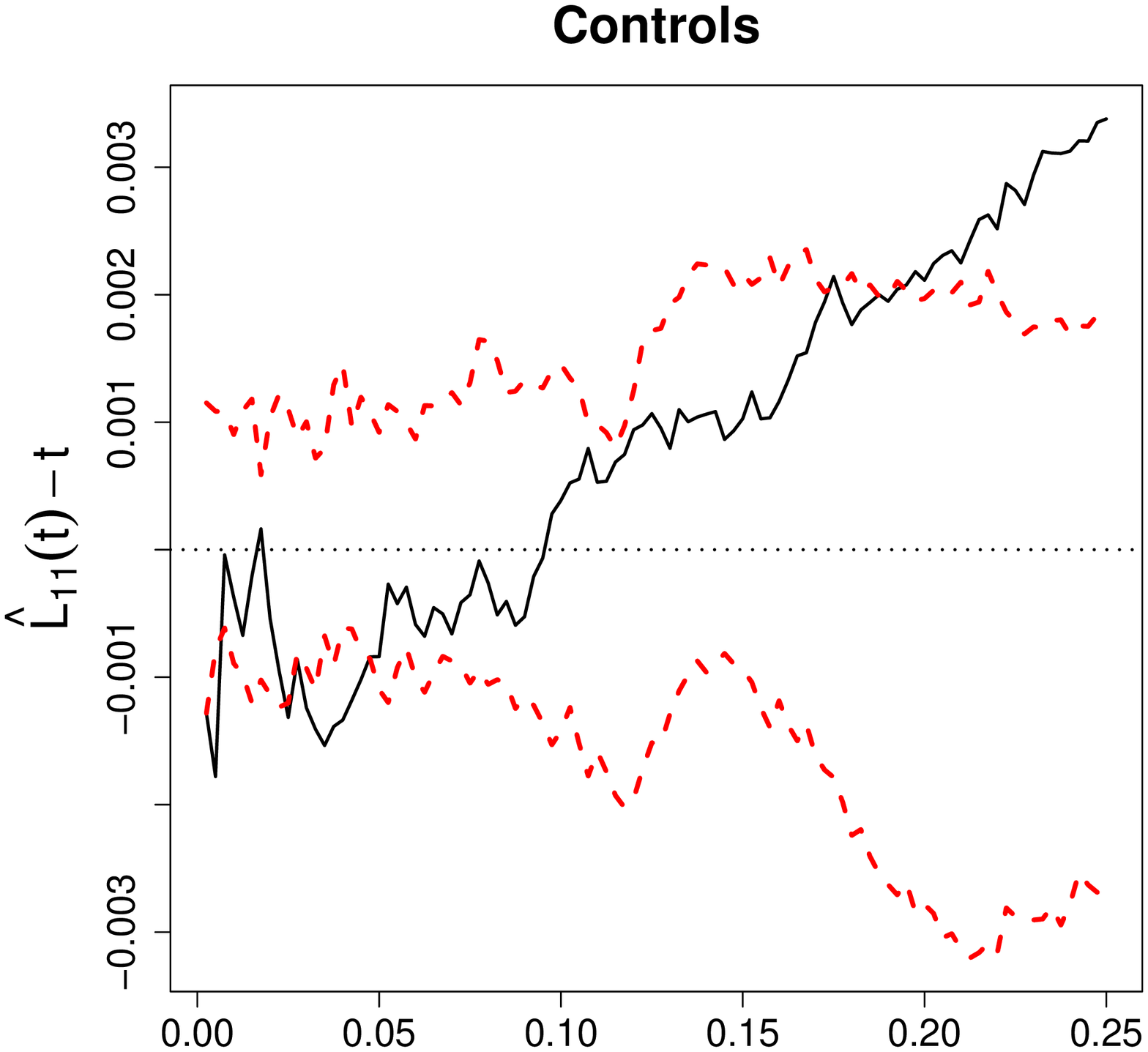} }}
\rotatebox{-90}{ \resizebox{2 in}{!}{\includegraphics{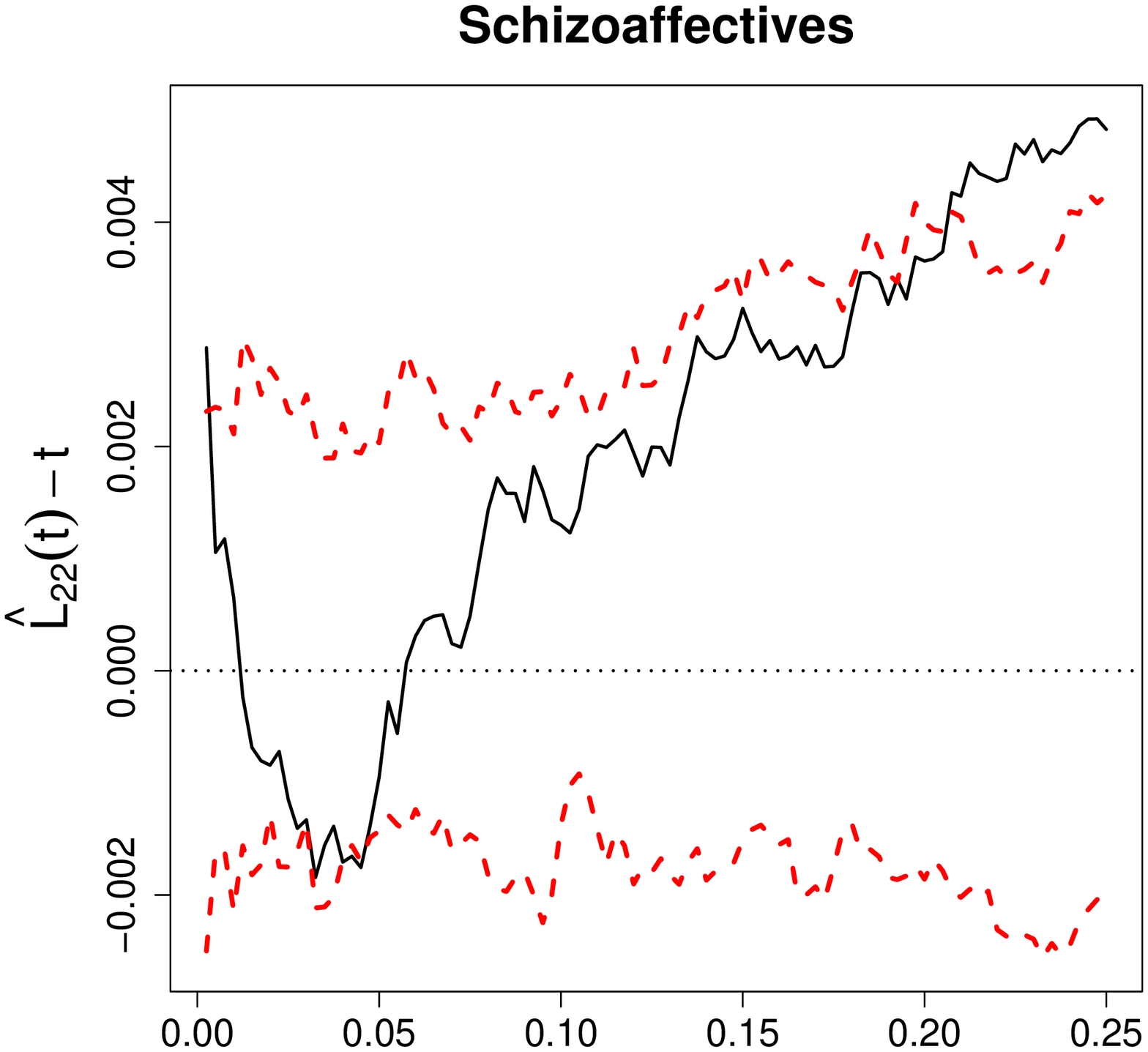} }}
\rotatebox{-90}{ \resizebox{2 in}{!}{\includegraphics{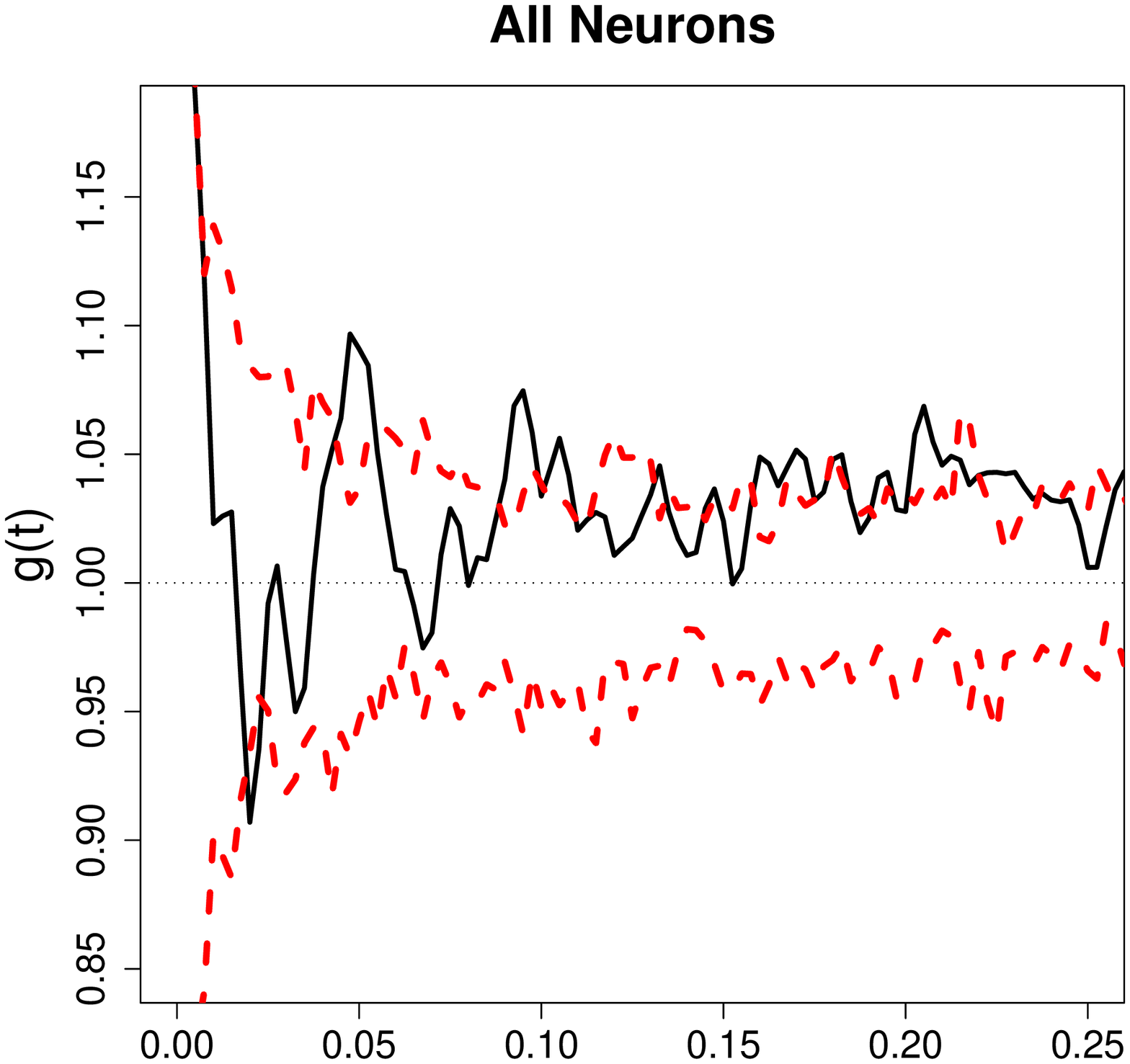} }}
\rotatebox{-90}{ \resizebox{2 in}{!}{\includegraphics{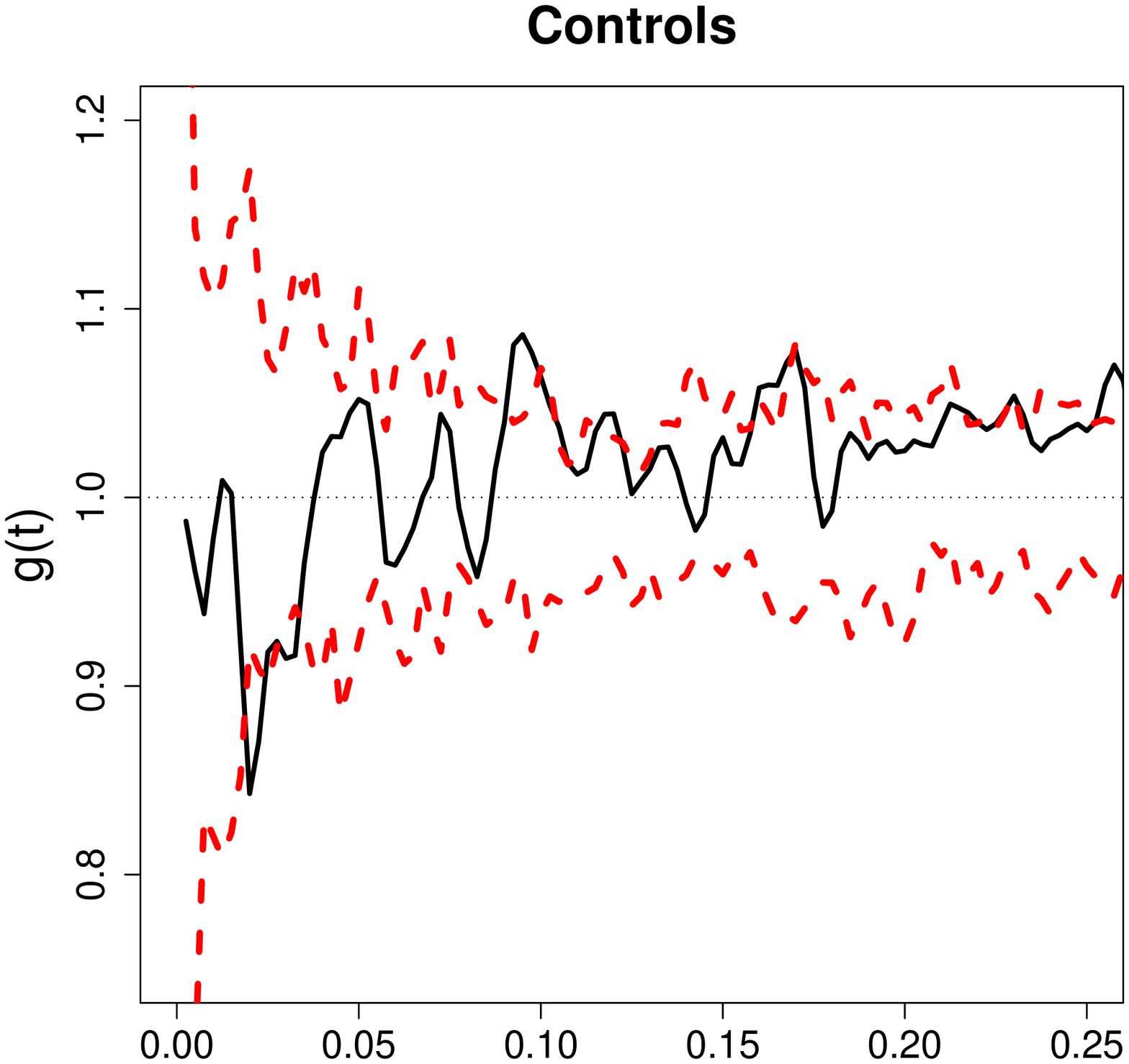} }}
\rotatebox{-90}{ \resizebox{2 in}{!}{\includegraphics{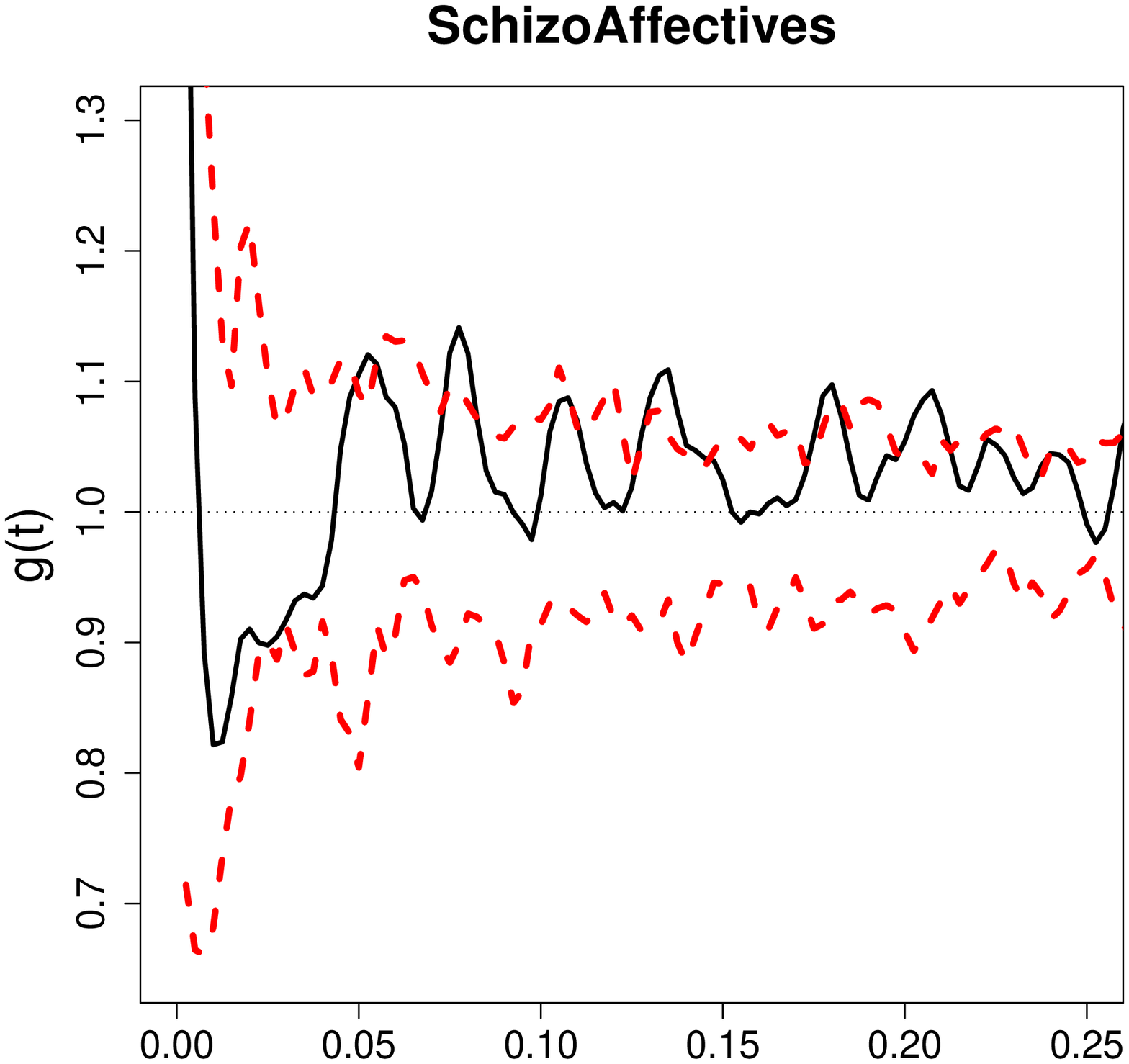} }}
\caption{
\label{fig:Benes-second-order-uni}
Ripley's univariate $L$-functions (top row) $\widehat{L}_{ii}(t)-t$ for $i=0,1,2$,
where $i=0$ stands for all data combined, $i=1$ for controls and $i=2$ for schizoaffectives;
and pair correlation functions $g(t)$ for all neurons combined and for each class (bottom row).
Wide dashed lines are the upper and lower (pointwise) 95 \% confidence bounds for the
functions based on Monte Carlo simulation under the CSR independence pattern.}
\end{figure}

To find out what might be causing the association,
and what is the type and level of interaction at different scales
we plot Ripley's (univariate) $L$-function and pair correlation function
for all data combined and for each (pooled) group in Figure \ref{fig:Benes-second-order-uni}
where the upper and lower 95 \% confidence bounds are also provided.
The average NN distance for this data set is 0.0155 ($\pm$ 0.0086)),
so we only consider distances up to 0.02 for Ripley's $L$-function.
For this range of distances Ripley's univariate $L$-function suggests
no significant deviation from CSR pattern for all data combined and schizoaffectives,
but it indicates significant regularity for controls at $t \approx .01$.
For $t \in [.02,.25]$, the pair correlation function suggests that
all neurons are significantly aggregated for distances about .04, .08, .09, .15, .20, and .22;
control neurons are significantly aggregated at about .08, .10, and .25;
and schizoaffectives are significantly aggregated at about .05, .07, .13 and .18.
At other distances, the neurons do not significantly deviate from CSR.
This is along the lines of the NNCT analysis results, which indicate
deviation from CSR independence at smaller scales.
The significant spatial regularity of the controls might explain the
association of neurons of controls and schizoaffectives.

\begin{figure}[hbp]
\centering
\rotatebox{-90}{ \resizebox{2 in}{!}{\includegraphics{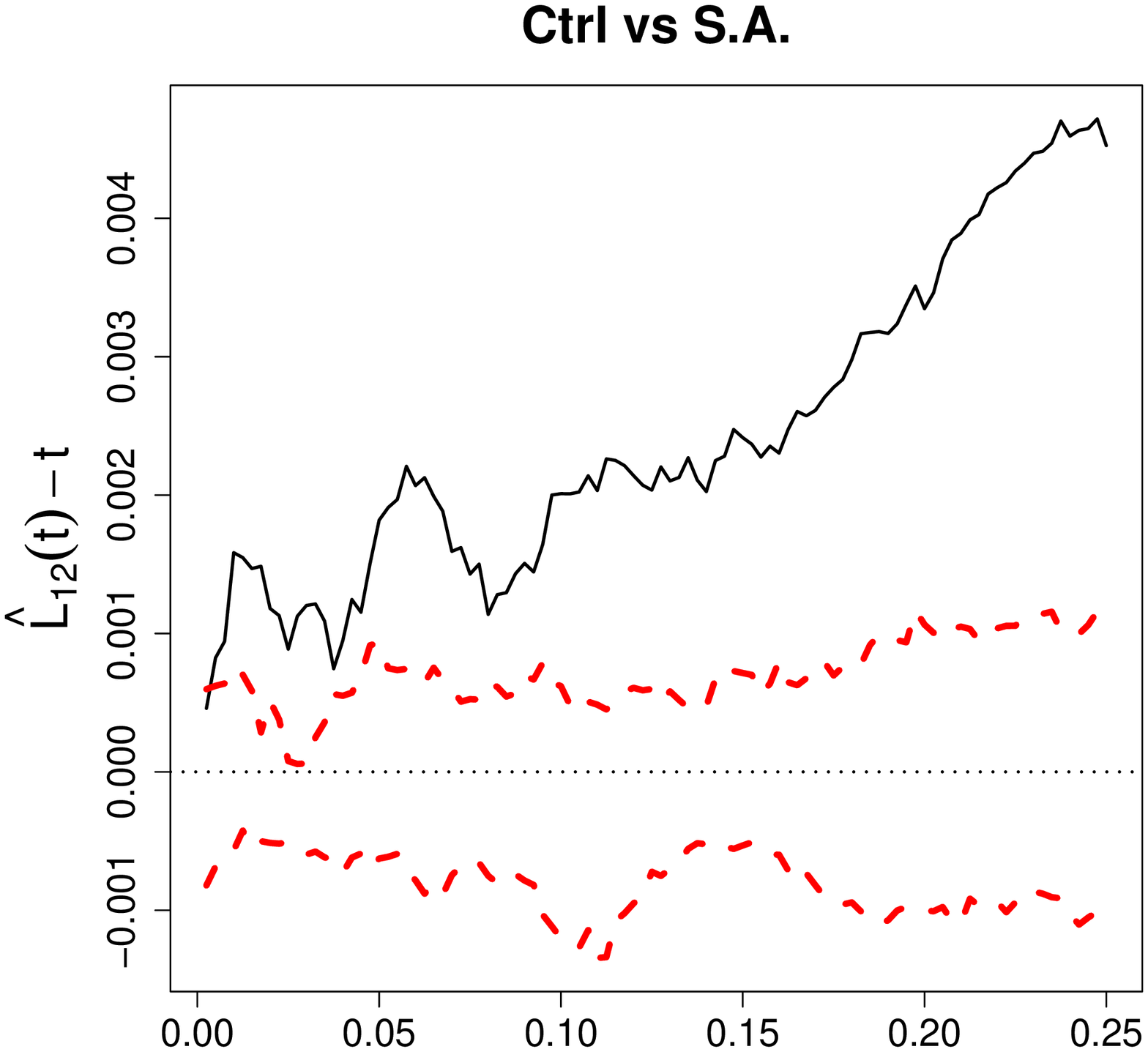} }}
\rotatebox{-90}{ \resizebox{2 in}{!}{\includegraphics{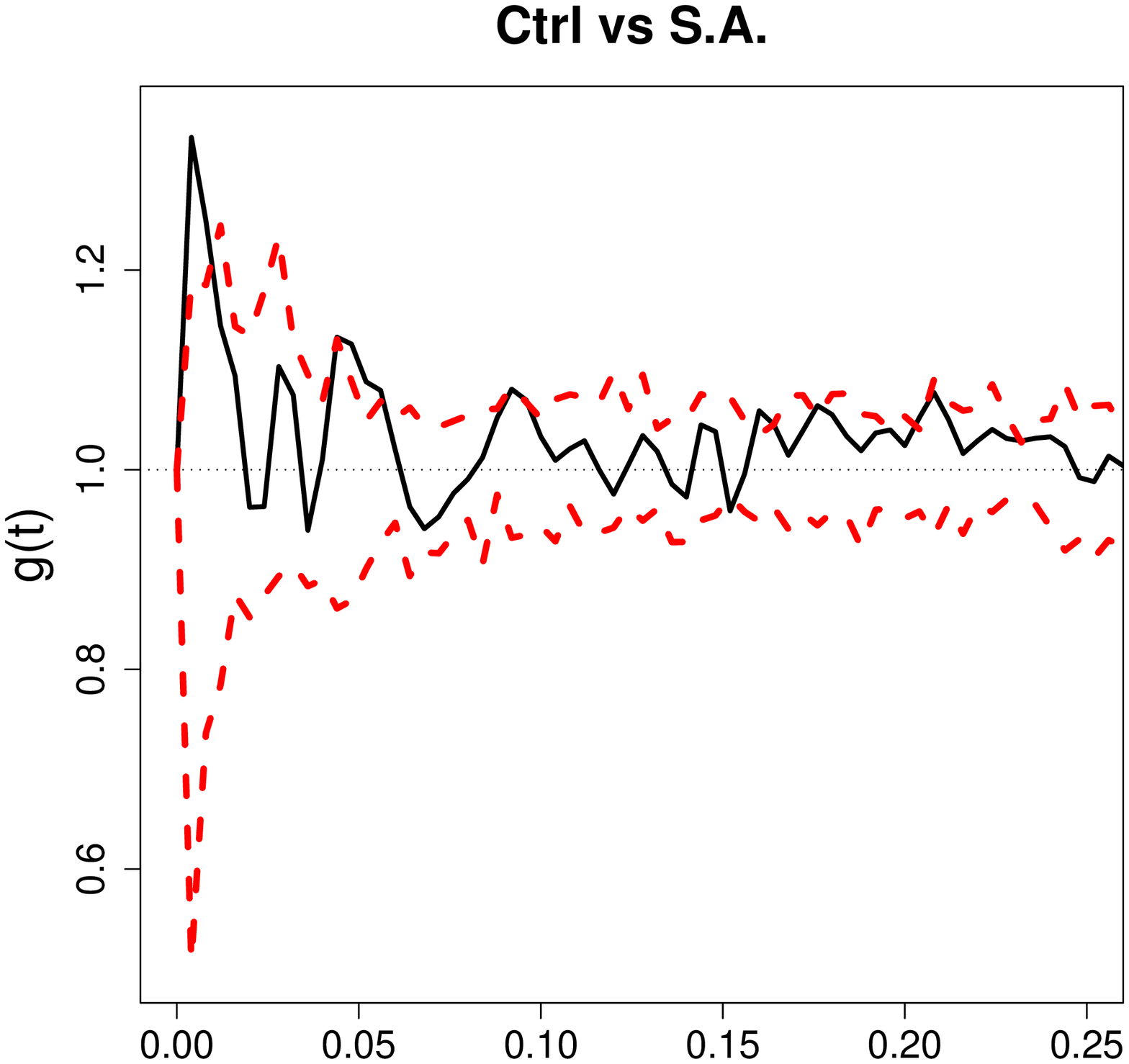} }}
\caption{
\label{fig:Benes-second-order-multi}
Ripley's bivariate $L$-function $\widehat{L}_{12}(t)-t$ (left)
and pair correlation function $g(t)$ (right) for the pyramidal neuron data.
Wide dashed lines are the upper and lower (pointwise) 95 \% confidence bounds for the
functions based on Monte Carlo simulations under the CSR independence pattern.
Ctrl = Control and S.A. = Schizoaffective.}
\end{figure}

We also plot Ripley's bivariate $L$-function and pair correlation function
together with the upper and lower 95 \% confidence bounds in Figure \ref{fig:Benes-second-order-multi}.
For distances up to 0.02 for Ripley's bivariate $L$-function
suggests that control and schizoaffective neurons are significantly associated
with each other.
For $t \in [.02,.25]$, the pair correlation function suggests that
control and schizoaffective neurons are significantly associated at about .04 only.
At other distances, the neurons do not significantly deviate from CSR independence.
So at smaller scales (i.e., $t \lesssim 0.02$) the univariate and
bivariate $L$-functions seem to be in agreement with the NNCT results
which indicate the association of controls and schizoaffectives.

\subsection{Artificial Data}
\label{sec:arti-data}
\begin{figure}[hbp]
\centering
\rotatebox{-90}{ \resizebox{3. in}{!}{ \includegraphics{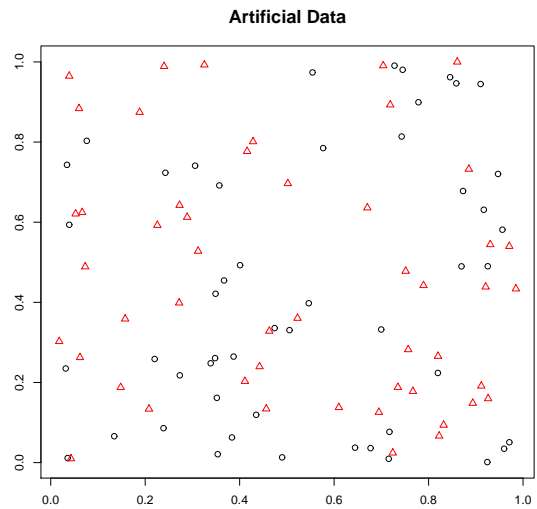} }}
\caption{
\label{fig:Arti}
The scatter plots of the locations of $X$ points (circles $\circ$) and $Y$
points (triangles $\triangle$).}
\end{figure}

In this section, we provide an artificial example, a
random sample of size 100 (with $50$ $X$-points and $50$ $Y$-points
uniformly generated on the unit square).
The question of interest is
the spatial interaction between $X$ and $Y$ classes.
We plot the locations of these points in the study region
in Figure \ref{fig:Arti} and the
corresponding NNCT together with percentages are provided
in Table \ref{tab:NNCT-arti}.
Observe that the percentages are slightly
larger for the diagonal cells, which might be interpreted as
presence of mild (not necessarily significant) segregation for both classes.

\begin{table}[ht]
\centering
\begin{tabular}{cc}

\begin{tabular}{cc|cc|c}
\multicolumn{2}{c}{}& \multicolumn{2}{c}{NN}& \\
\multicolumn{2}{c}{}&    $X$ &  $Y$   &   sum  \\
\hline
& $X$ &    29  &   21    &   50  \\
\raisebox{1.5ex}[0pt]{base}
& $Y$ &    20 &  30    &   50  \\
\hline
&sum     &    49   & 51            &  100  \\
\end{tabular}
&
\begin{tabular}{cc|cc|c}
\multicolumn{2}{c}{}& \multicolumn{2}{c}{NN}& \\
\multicolumn{2}{c}{}&    $X$ &  $Y$   &    \\
\hline
& $X$ &   58 \%  &    42 \%    &   50 \%   \\
& $Y$ &   40 \%  &  60 \%     &   50 \%   \\
\hline
&    &    49 \%    & 51\%      &  100 \%   \\
\end{tabular}

\end{tabular}
\caption{ \label{tab:NNCT-arti} The NNCT for the artificial data and
the corresponding percentages (in parenthesis).}
\end{table}

\begin{table}[ht]
\centering
\begin{tabular}{|c||c|c|c|c|c|c|c|c|c|c|}
\hline
\multicolumn{11}{|c|}{Test statistics and the associated $p$-values for the artificial data} \\
\hline
 & $Z^D_{11}$ & $Z^D_{22}$ &  $Z^C_{11}$ &  $Z^C_{22}$ &  $Z_P$ & $Z_{mc}$ & $Z^a_{mc}$ & $Z^s_{mc}$ &  $Z_I$ &  $Z_{II}$ \\
\hline
test statistics  & 1.13 & 1.40 & 1.49 & 1.49 & 1.80 & 1.41 & 1.34 & 1.46 & 1.49 & 1.49 \\
\hline
\multicolumn{11}{|c|}{against the two-sided alternative} \\
\hline
$\pasy$ & .2570 & .1615 & .1365 & .1356 & .0718 & .1586 & --- & --- & .1360 & .1360 \\
\hline
$\pmc$ & .2593 & .1614 & .1404 & .1371 & --- & --- & --- & --- & .1414 & .1380 \\
\hline
$\prand$ & .2842 & .1822 & .1406 & .1431 & --- & --- & --- & --- & .1464 & .1368 \\
\hline
\multicolumn{11}{|c|}{against the right-sided (i.e., segregation) alternative} \\
\hline
$\pasy$ & .1285 & .0808 & .0682 & .0678 & .0359 & --- & --- & .0726 & .0680 & .0680 \\
\hline
$\pmc$ & .1323 & .0801 & .0716 & .0701 & --- & --- & --- & --- & .0722 & .0717 \\
\hline
$\prand$ & .1578 & .1002 & .0723 & .0671 & --- & --- & --- & --- & .0787 & .0787 \\
\hline
\multicolumn{11}{|c|}{against the left-sided (i.e., association) alternative} \\
\hline
$\pasy$  & .8715 & .9192 & .9318 & .9322 & .9641 & --- & .9106 & --- & .9320 & .9320 \\
\hline
$\pmc$   & .8695 & .9212 & .9287 & .9302 & --- & --- & --- & --- & .9281 & .9286 \\
\hline
$\prand$ & .9013 & .9389 & .9316 & .9368 & --- & --- & --- & --- & .9252 & .9252 \\
\hline
\end{tabular}
\caption{ \label{tab:arti-test-stat}
Test statistics and the associated $p$-values for the two-sided and directional alternatives
for the artificial data.
The labeling of the $p$-values are as in Table \ref{tab:swamp-test-stat}.
}
\end{table}

Observe that in Table \ref{tab:arti-test-stat},
$p$-value for the two-sided alternative of the directional test $Z_P$ is almost
significant ($p=.0718$),
and for the right-sided alternative i.e. for segregation it is significant ($p=.0359$),
which might be interpreted as evidence of deviation from CSR independence,
but given the reservations on the
appropriateness for testing CSR independence or RL, other tests are more reliable,
and they all give insignificant $p$-values.
Furthermore, the plot in Figure \ref{fig:Arti} is not
suggestive of any deviation from CSR independence, and the dependence
between cell counts confounds the conclusion based on $Z_P$.

\begin{figure}[hbp]
\centering
\rotatebox{-90}{ \resizebox{2 in}{!}{\includegraphics{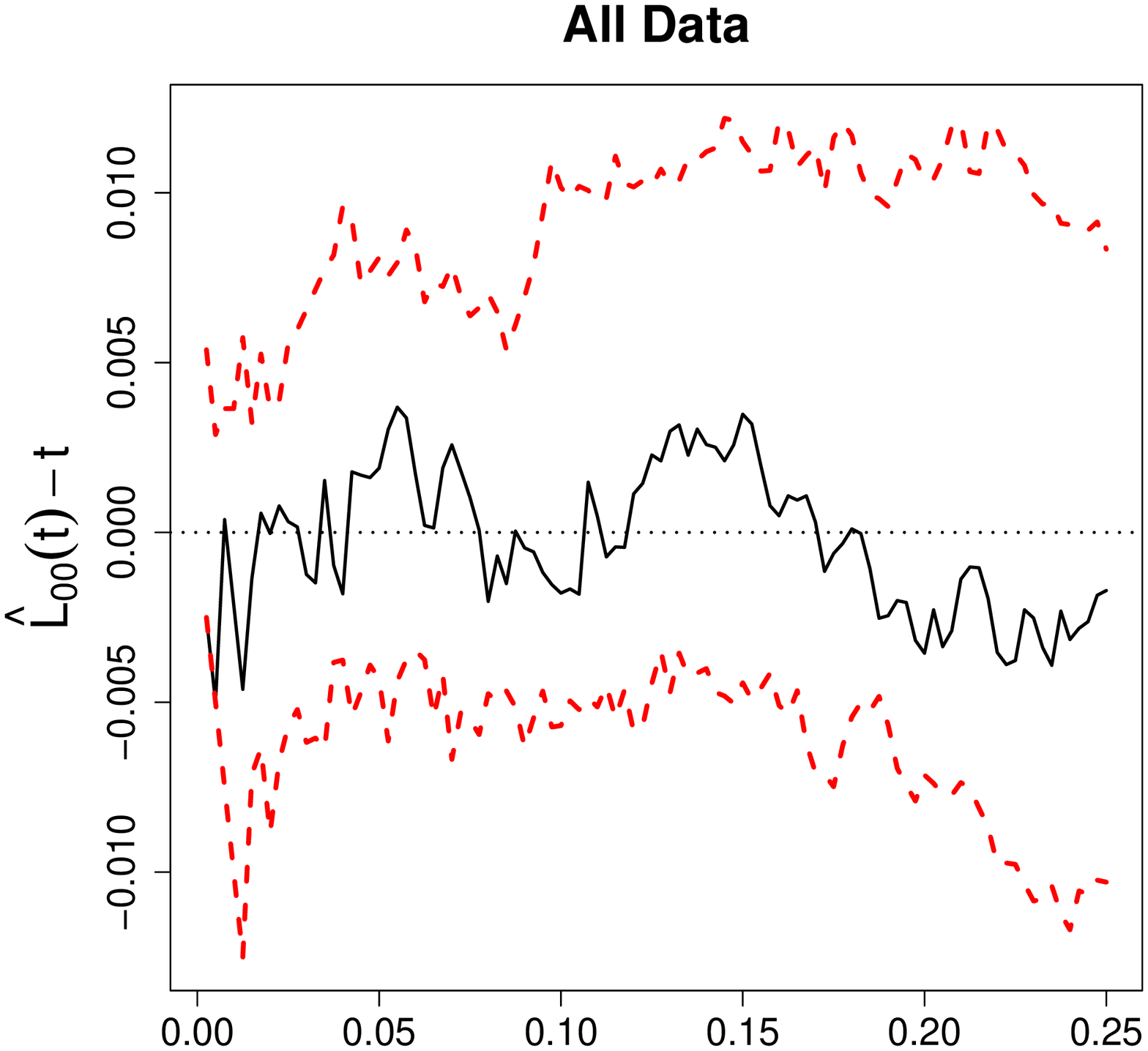} }}
\rotatebox{-90}{ \resizebox{2 in}{!}{\includegraphics{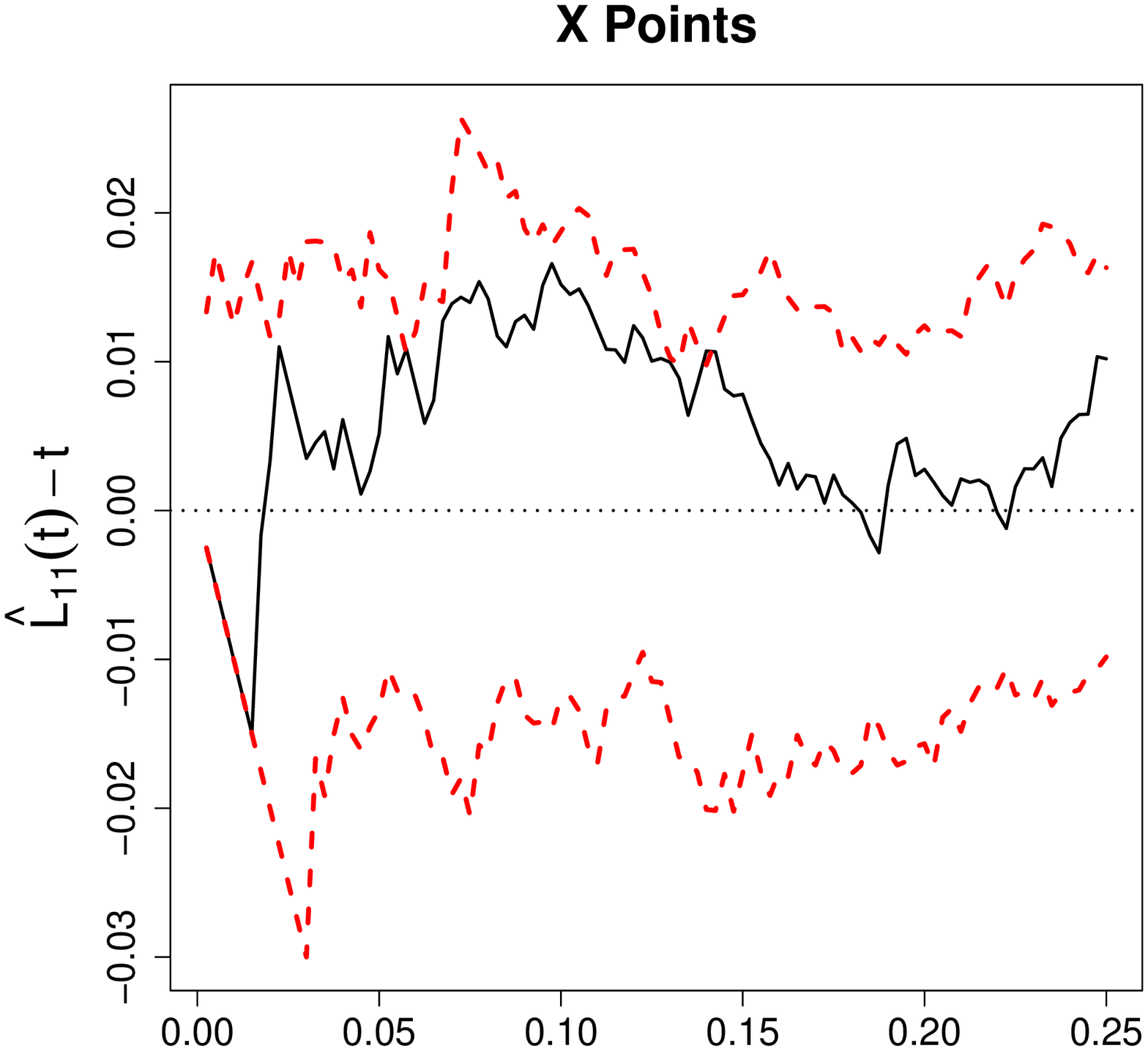} }}
\rotatebox{-90}{ \resizebox{2 in}{!}{\includegraphics{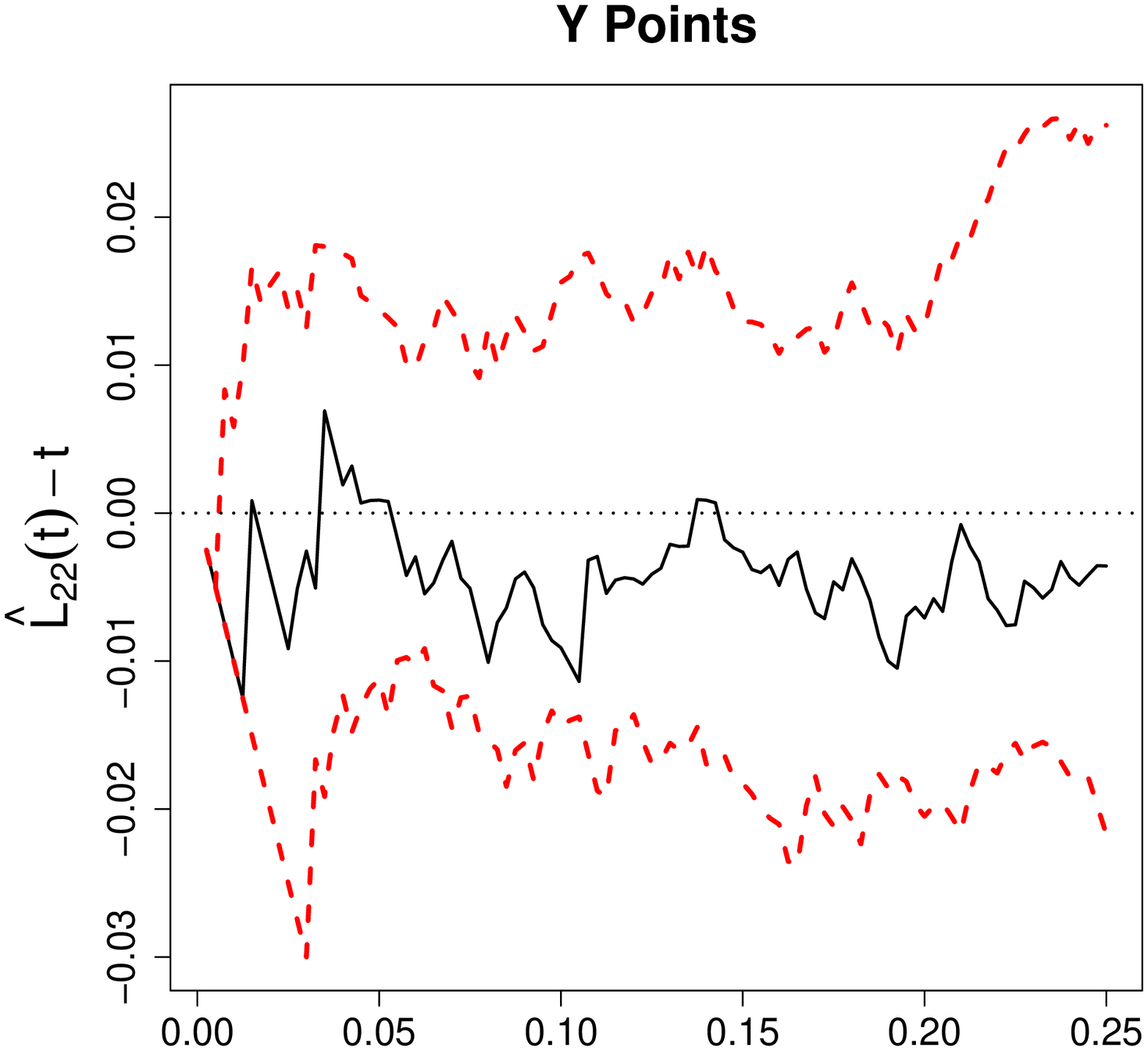} }}
\rotatebox{-90}{ \resizebox{2 in}{!}{\includegraphics{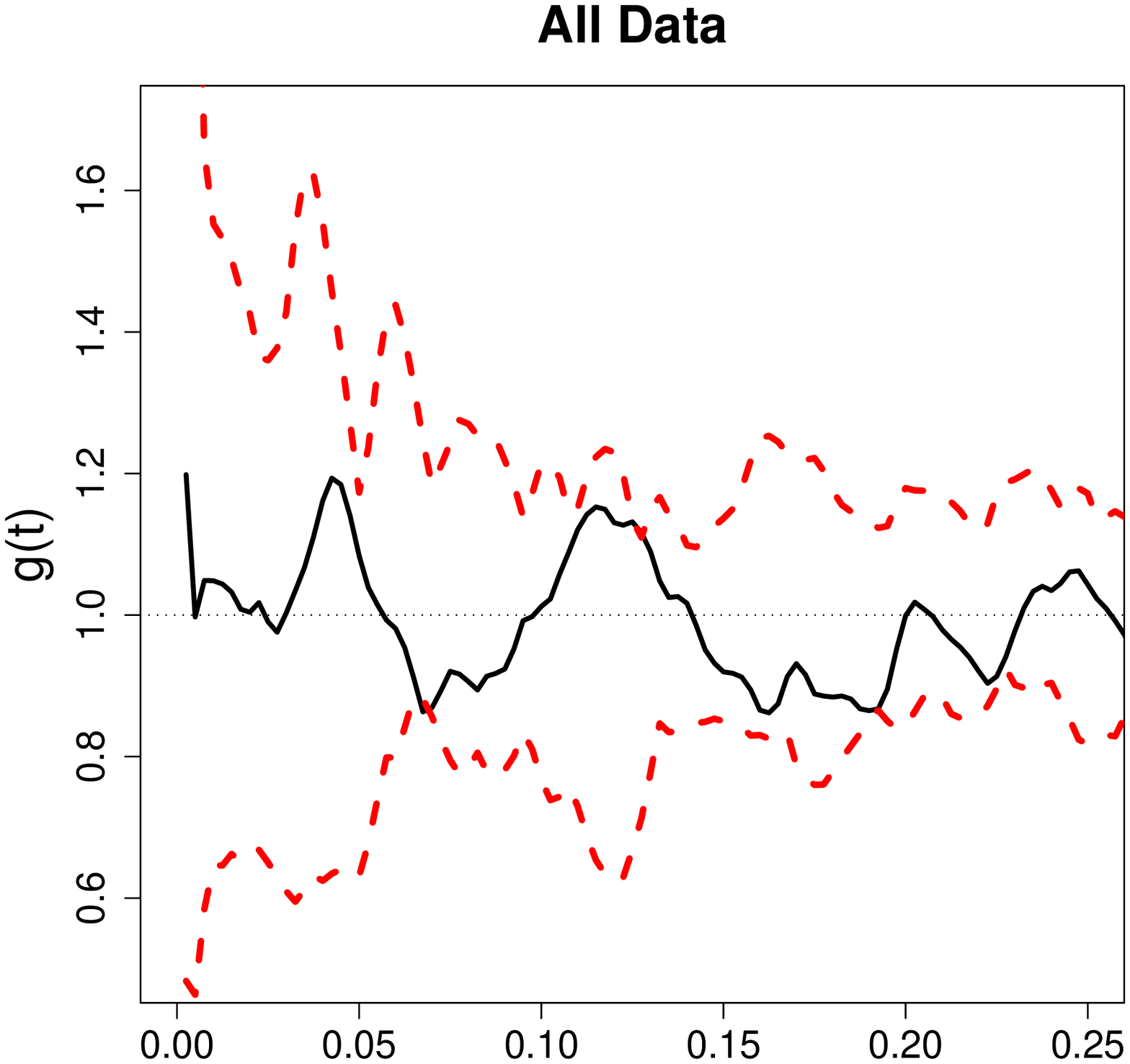} }}
\rotatebox{-90}{ \resizebox{2 in}{!}{\includegraphics{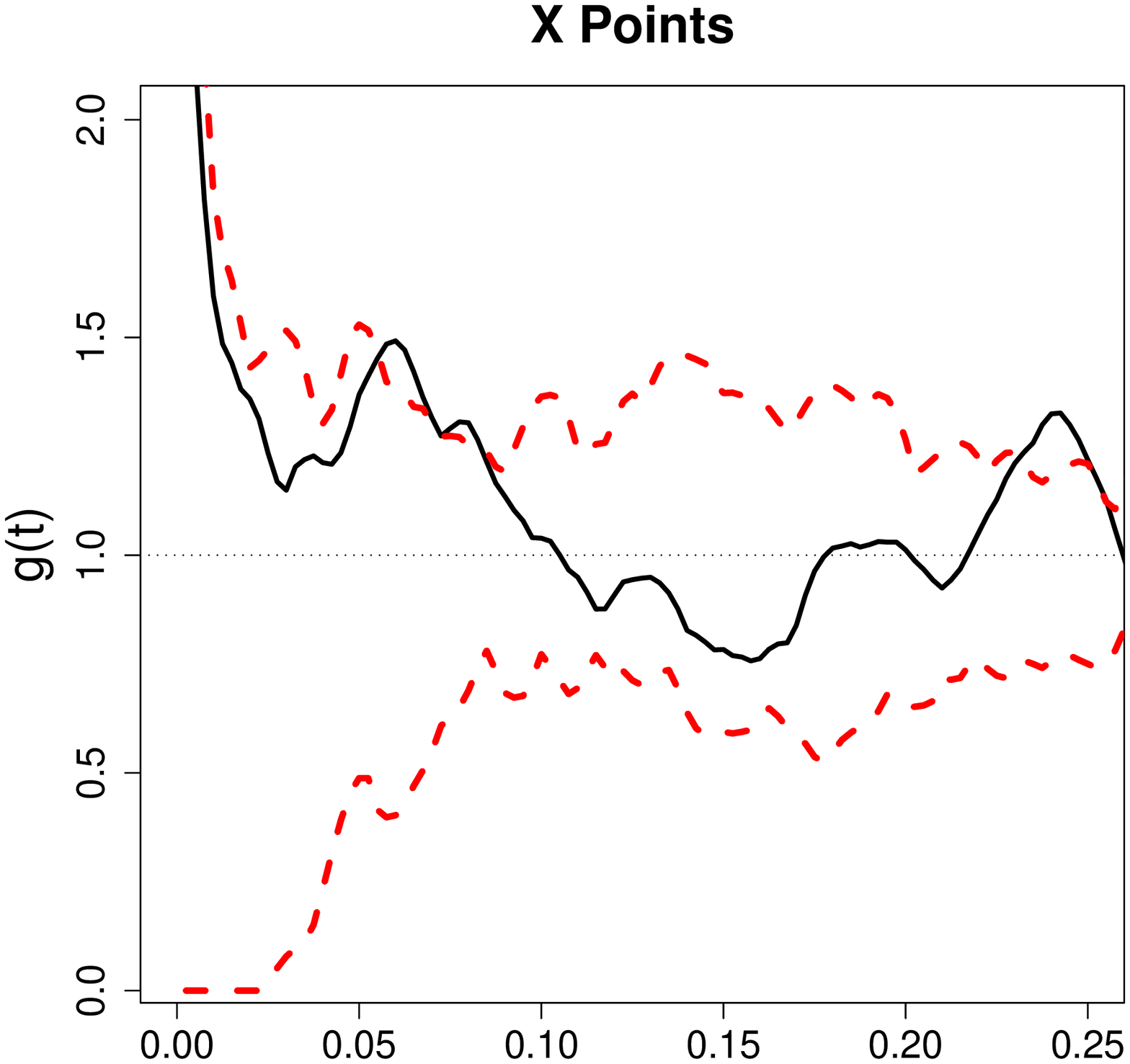} }}
\rotatebox{-90}{ \resizebox{2 in}{!}{\includegraphics{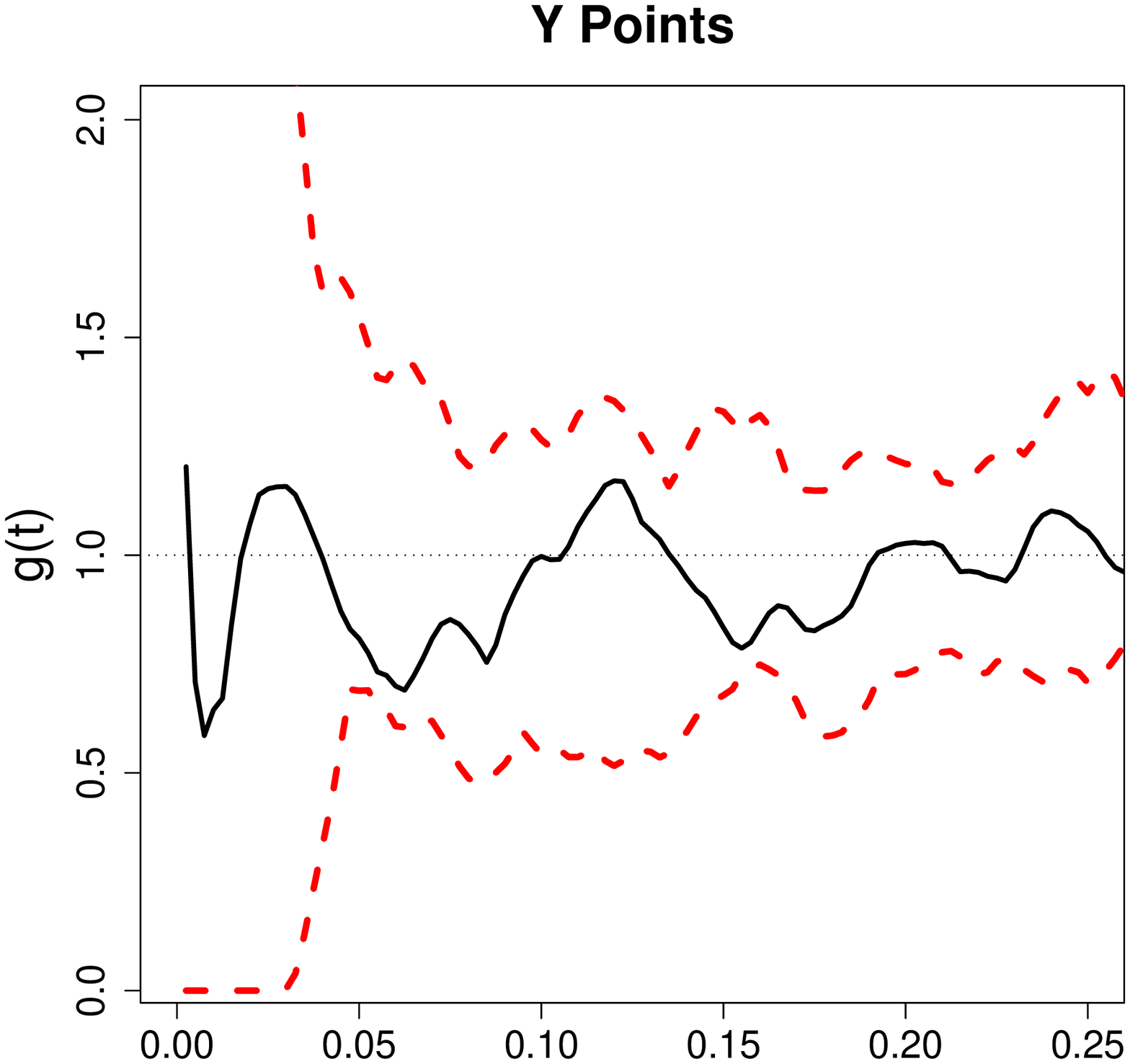} }}
\caption{
\label{fig:arti-second-order-uni}
Ripley's univariate $L$-functions (top row) $\widehat{L}_{ii}(t)-t$ for $i=0,1,2$,
where $i=0$ stands for all data combined, $i=1$ for $X$ points and $i=2$ for $Y$ points;
and pair correlation functions $g(t)$ for all data combined and for each class (bottom row).
Wide dashed lines are the upper and lower (pointwise) 95 \% confidence bounds for the
functions based on Monte Carlo simulation under the CSR independence pattern.}
\end{figure}

We also plot Ripley's (univariate) $L$-function and pair correlation function
for all data combined and for each group in Figure \ref{fig:arti-second-order-uni}.
The average NN distance for this data set is 0.05 ($\pm$ 0.03)),
so we only consider distances up to 0.05 for Ripley's $L$-function.
For this range of distances Ripley's univariate $L$-function suggests
no significant deviation from CSR pattern for each plot.
For $t \in [.05,.25]$, the pair correlation function suggests that
all data points do not significantly deviate from CSR;
neither do the $Y$ points;
but $X$ points are significantly aggregated at about .05 and $[.22,.24]$
and at other distances $X$ points do not significantly deviate from CSR.
This is along the lines of the NNCT analysis results, which indicate
no significant deviation from CSR independence at smaller scales.

\begin{figure}[hbp]
\centering
\rotatebox{-90}{ \resizebox{2 in}{!}{\includegraphics{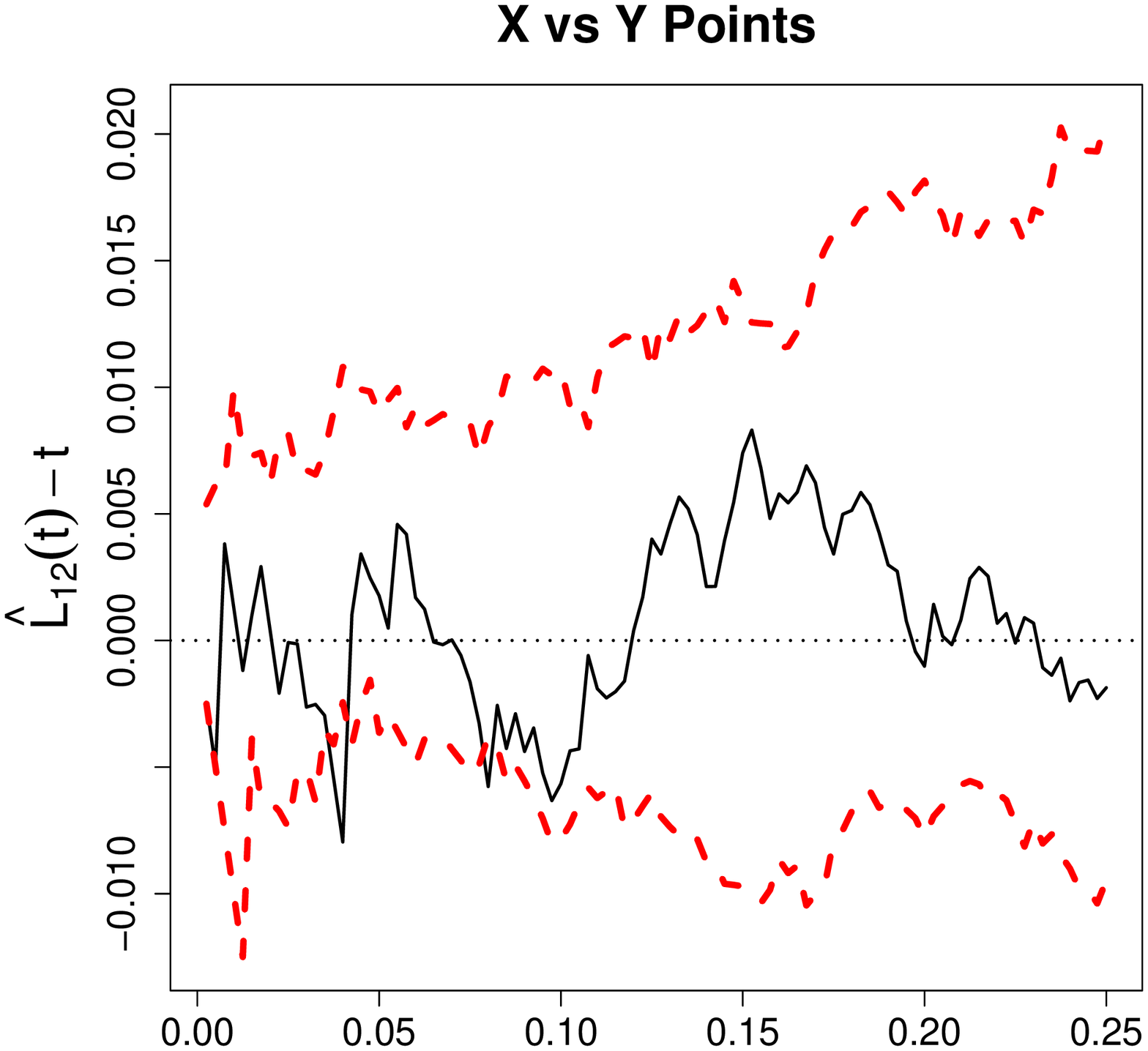} }}
\rotatebox{-90}{ \resizebox{2 in}{!}{\includegraphics{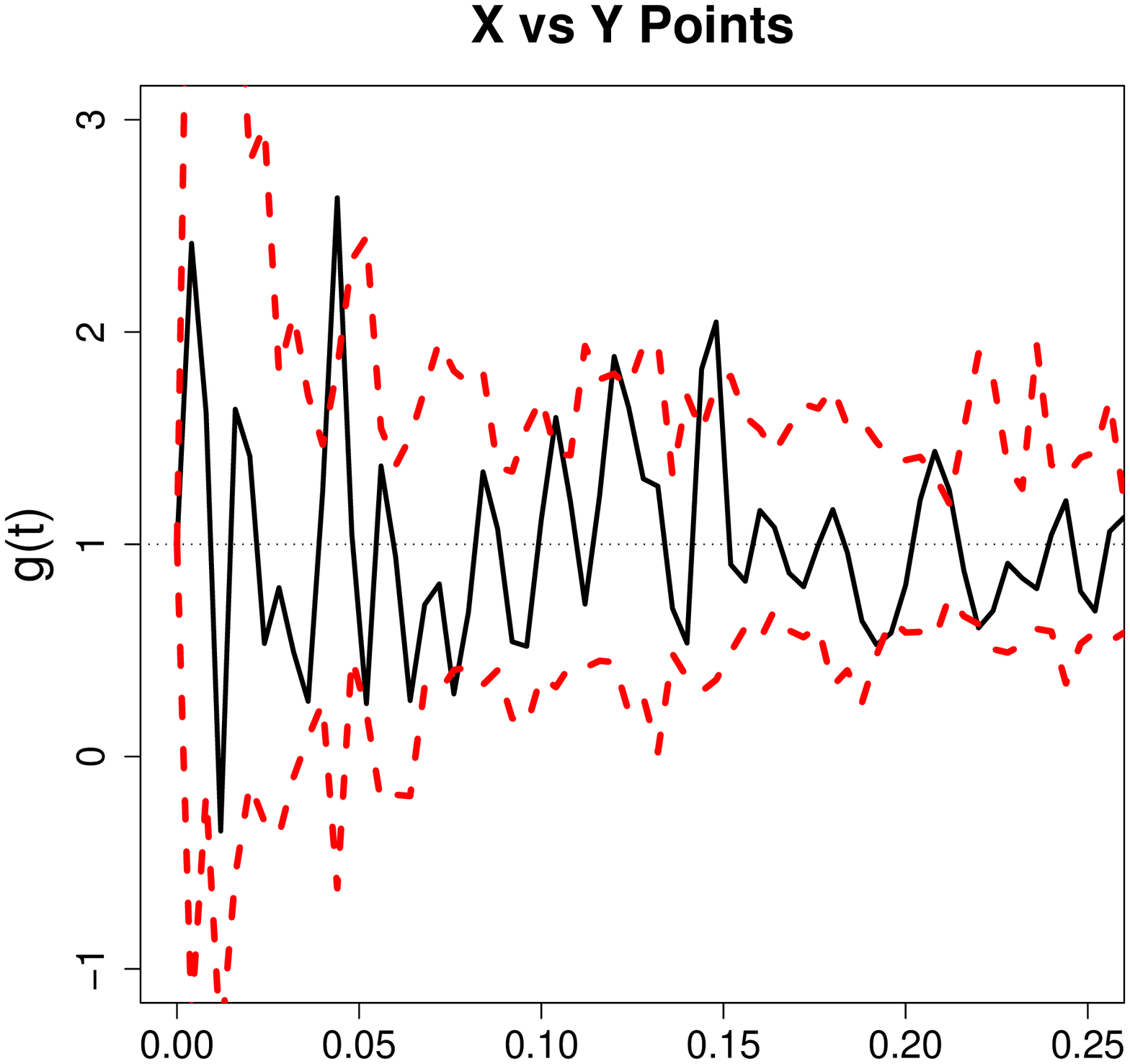} }}
\caption{
\label{fig:arti-second-order-multi}
Ripley's bivariate $L$-function $\widehat{L}_{12}(t)-t$ (left)
and pair correlation function $g(t)$ (right) for the artificial data.
Wide dashed lines are the upper and lower (pointwise) 95 \% confidence bounds for the
functions based on Monte Carlo simulations under the CSR independence pattern.
Ctrl = Control and S.A. = Schizoaffective.}
\end{figure}

We also plot Ripley's bivariate $L$-function and pair correlation function
in Figure \ref{fig:arti-second-order-multi}.
For distances up to 0.05 for Ripley's bivariate $L$-function
suggests that $X$ and $Y$ points are significantly segregated at about $t=.04$.
For $t \in [.05,.25]$, the pair correlation function suggests that
$X$ and $Y$ points are significantly segregated at about .14 and .20 only.
At other distances, the points do not significantly deviate from CSR independence.
So at smaller scales (i.e., $t \lesssim 0.02$) the univariate and
bivariate $L$-functions seem to be in agreement with the NNCT results
which indicate no deviation from CSR independence.

\section{Discussion and Conclusions}
\label{sec:disc-conc}

Pielou's and Dixon's segregation tests based on nearest neighbor contingency tables (NNCTs)
are $\chi^2$-tests, hence are used for two-sided alternatives.
That is, when the null patterns of CSR independence or RL are rejected,
these tests do not indicate the direction of the alternative pattern,
which consist of segregation or association patterns.
In this article, we discuss directional (i.e., one-sided)
tests of segregation based on NNCTs.
We propose a directional version of Pielou's test
by partitioning the $\chi^2$ test statistic in the usual fashion (\cite{bickel:1977}).
However, the problem that confounds Pielou's test
(i.e., the problem of dependence between cell counts)
is inherited by the directional versions also.
This makes the directional version of Pielou's test liberal
in rejecting the null case of CSR independence or RL.
We also consider the directional versions of the cell-specific tests
due to \cite{dixon:1994} and \cite{ceyhan:cell2008}.
Furthermore, we introduce two new directional tests of segregation.

We discuss the differences in these NNCT-tests,
compare the tests using extensive Monte Carlo simulations
under RL and CSR independence and under various segregation and association alternatives.
We also illustrate the tests on four examples and compare them with Ripley's
$L$-function (\cite{ripley:2004}).
We demonstrate that under the CSR independence pattern,
NNCT-tests are conditional on $Q$ and $R$,
but not under RL.

Based on our Monte Carlo simulations,
we conclude that the asymptotic approximation for the cell-specific
and the directional tests is appropriate only
when the corresponding cell count in the NNCT is larger than 10.
When a cell count is less than 10,
we recommend the Monte Carlo randomization of these tests.
Type I error rates (empirical significance levels)
of Ceyhan's cell-specific and of the new directional
tests are more robust to the differences in sample sizes
(i.e., differences in relative abundances).
Considering the empirical significance levels
and power estimates of the tests,
we recommend version II of the new tests (defined in Equation \eqref{eqn:new-versions})
for the two-sided alternatives
provided the sample sizes are not very different.

The CSR independence pattern assumes that the study region
is unbounded for the analyzed pattern,
which is not the case in practice.
Edge effects are a constant problem in the analysis of empirical
(i.e., bounded) data sets and much effort has gone into the
development of edge correction methods (\cite{yamada:2003}).
So the edge (or boundary) effects might confound the test results
if the null pattern is the CSR independence.
Two correction methods for the edge effects on NNCT-tests,
namely buffer zone correction and toroidal correction,
are investigated in (\cite{ceyhan:cell-class-edge-correct}, \cite{ceyhan:overall}, and \cite{ECarXivCorrected:2008})
where it is recommended that inner or outer buffer zone correction for NNCT-tests
could be used with the width of the buffer area being about the average NN distance.
But larger buffer areas are not recommended
since they are wasteful with little additional gain.
On the other hand, toroidal edge correction is recommended
with points within the average NN distance in the additional copies
around the study region.
For larger distances, the gain might not be worth the effort.
We extend these recommendations for the new directional tests also.

NNCT-tests summarize the pattern in the data set for small scales,
more specifically, they provide information on the pattern
around the average NN distance between all points.
On the other hand, pair correlation function $g(t)$
and Ripley's classical $K$ or $L$-functions and other variants provide
information on the pattern at various scales.
However, the classical $L$-function is not appropriate for the null pattern of RL
when locations of the points have spatial inhomogeneity.
For such cases, Diggle's $D$-function (\cite{diggle:2003} p. 131)
is more appropriate in testing the bivariate spatial clustering at various scales.
Our example illustrates that for distances around the average NN distance,
NNCT-tests and Ripley's bivariate $L$-function yield similar results.

If significant,
the cell-specific test and the new tests (for the two-sided alternative)
imply significant deviation from the null pattern.
Furthermore, the sign of the test statistic will be suggestive of segregation
(if positive) and association (if negative).
But these tests are more powerful against the one-sided alternatives.
For a data set for which CSR independence is the reasonable null pattern,
we recommend the NNCT-tests
if the question of interest is the spatial interaction at small scales
(i.e., about the mean NN distance).
One can also perform Ripley's $K$ or $L$-function
and only consider distances up to around the average NN distance
and compare the results with those of NNCT analysis.
If the spatial interaction at higher scales is of interest,
pair correlation function is recommended (\cite{loosmore:2006}).
On the other hand, if the RL pattern is the reasonable null pattern for the data,
we recommend the NNCT-tests if the small-scale interaction is of interest
and Diggle's $D$-function if the spatial interaction at higher scales is also of interest.

\section*{Acknowledgments}
Some of the Monte Carlo simulations presented in this article
were executed on the Hattusas cluster of
Ko\c{c} University High Performance Computing Laboratory.



\end{document}